\documentclass[%
 reprint,
superscriptaddress,
 amsmath,amssymb,na
 aps,
prd,
]{revtex4-1}

\pdfoutput=1

\usepackage{braket}				% Allows Dirac notation
\usepackage{graphicx}			% Include figure files
\usepackage{bm}				% bold math
\usepackage[usenames, dvipsnames]{color}	
\usepackage{hyperref}			% add hypertext capabilities
\usepackage{mathtools}
\usepackage[normalem]{ulem} %strikethrough text
\usepackage[caption=false]{subfig}
\usepackage{float}

\DeclareMathOperator{\tr}{tr} 			% for trace
\DeclareMathOperator{\PV}{PV} 			% for trace

\renewcommand{\Re}{\mathop{\rm Re}}	% for the real part
\DeclareMathOperator\sgn{sgn}		% for the algebraic sign
\DeclareMathOperator\erf{erf}			% for the error function
\newcommand{\nn}{\nonumber}		% for no numbering

\usepackage{colonequals} %Get a nice looking :=
\newcommand{\ce}{\colonequals}

\usepackage[usenames, dvipsnames]{xcolor}
\newcommand{\tcb}{\textcolor{blue}}			

\newcommand{\tcp}{\textcolor{purple}}			

\definecolor{capri}{rgb}{0.0, 0.75, 1.0}

\begin{document}

\title{Entangling detectors in anti-de Sitter space}

\author{Laura J. Henderson}
\email[]{l7henderson@uwaterloo.ca}
\affiliation{Institute for Quantum Computing, University of Waterloo, Waterloo, Ontario, Canada, N2L 3G1}
\affiliation{Department of Physics and Astronomy, University of Waterloo, Waterloo, Ontario, Canada, N2L 3G1}

\author{Robie A. Hennigar}
\email[]{rhennigar@uwaterloo.ca}
\affiliation{Department of Mathematics and Statistics, Memorial University of Newfoundland, St. John’s, Newfoundland and Labrador, A1C 5S7, Canada}
%\affiliation{Department of Physics and Astronomy, University of Waterloo, Waterloo, Ontario, Canada, N2L 3G1}

\author{Robert~B.~Mann}
\email[]{rbmann@uwaterloo.ca}
\affiliation{Institute for Quantum Computing, University of Waterloo, Waterloo, Ontario, Canada, N2L 3G1}
\affiliation{Department of Physics and Astronomy, University of Waterloo, Waterloo, Ontario, Canada, N2L 3G1}
\affiliation{Perimeter Institute for Theoretical Physics, 31 Caroline St. N., Waterloo, Ontario, Canada, N2L 2Y5}

\author{Alexander R. H. Smith}
\email[]{alexander.r.smith@dartmouth.edu}
\affiliation{Department of Physics and Astronomy, Dartmouth College, Hanover, New Hampshire 03755, USA}

\author{Jialin Zhang}
\email[]{jialinzhang@hunnu.edu.cn \\ }
\affiliation{Department of Physics and Synergetic Innovation Center for Quantum Effects and Applications, Hunan Normal University, Changsha, Hunan 410081, China}

\date{\today}

\begin{abstract}
We examine in (2+1)-dimensional anti-de Sitter (AdS) space the phenomena of entanglement harvesting\,---\,the process in which a pair of detectors (two-level atoms) extract entanglement from a quantum field through local interactions with the field. We begin by reviewing the Unruh-DeWitt detector and its interaction with a real scalar field in the vacuum state, as well as the entanglement harvesting protocol in general. We then examine how the entanglement harvested by a pair of such detectors depends on their spacetime trajectory, separation, spacetime curvature, and boundary conditions satisfied by the field. The harvested entanglement is interpreted as an indicator of field entanglement between the localized regions where the detectors interact with the field, and thus this investigation allows us to probe indirectly the entanglement structure of the AdS vacuum. We find an island of separability for specific values of the detectors' energy gap and separation at intermediate values of the AdS length for which entanglement harvesting is not possible; an analogous phenomena is observed in AdS$_4$, to which we compare and contrast our  results. In the process we examine how the transition probability of a single detector, as a proxy for local fluctuations of the field, depends on spacetime curvature, its location in AdS space, and boundary conditions satisfied by the field.
\end{abstract}

\maketitle

%========================================
%========================================

\section{Introduction}

The theory of quantum fields propagating on curved spacetimes investigates situations where both quantum and gravitational effects are important, but the gravitational field can be treated classically and the backreaction of the fields on the spacetime can be ignored \cite{Birrell:1982, Wald:1994}. This area of research has given us insights into the early Universe and black hole physics, leading to valuable clues of what we should expect from a full-fledged quantum theory of gravity.

Physically motivated detector models, such as the Unruh-DeWitt detector~\cite{Unruh:1976, DeWitt:1979}, provide an operational way in which to probe properties of quantum fields on curved spacetimes. These detectors move along a classical trajectory through a given spacetime while their internal degrees of freedom interact locally with a quantum field living on the spacetime. Most often the interaction coupling the detector and field is inspired by the light-matter interaction widely used in the field of quantum optics~\cite{Mandel:1995,Alvaro,Martin-Martinez2013, Pozas-Kerstjens:2016}. By measuring the internal degrees of freedom of a detector, or collection of detectors, after their interaction with the field has ceased, one effectively performs a measurement on the quantum field itself, allowing one to probe properties of the field such as local fluctuations and their correlations.

Entanglement is a particularly interesting property of a quantum field theory that is well suited to being probed by detectors, and is now finding applications in disparate areas of physics such as the study of critical phenomena in condensed matter systems~\cite{Osterloh:2002, Vidal:2003, Amico:2008}, in the description of non-classical states of light within the field of quantum optics~\cite{Mandel:1995, Quantum-Information:2011}, and offering an explanation for the origin of black hole entropy~\cite{Bombelli:1986, Callan:1994, Srednicki:1993}. It was first discovered within the context of algebraic quantum field theory that the vacuum state of a free quantum field is entangled as seen by the local inertial observers in flat spacetime, even if these observers are localized in spacelike-separated regions~\cite{Summers:1985,Summers:1987,Summers:1987fn}. This result is surprising because it suggests that observers can violate a Bell-like inequality by simply observing local vacuum fluctuations of a quantum field\,---\,the vacuum is a resource for entanglement.

It was later pointed out that this property of the electromagnetic vacuum could in principle be exploited to generate entanglement between a pair of atoms~\cite{Valentini:1991}. Extensive research has since investigated this process of localized detectors (two-level atoms) extracting entanglement/non-local correlations from the vacuum state of a quantum \mbox{field~\cite{Reznik:2002fz,Reznik:2005,Steeg:2009, Olson:2011, Hu:2012jr, Smith:2016a, Martin-MartinezSUS, Martin-Martinez:2016, Pozas-Kerstjens:2015,Pozas-Kerstjens:2016, Lin:2016}}, which has become known as the entanglement harvesting protocol~\cite{Salton:2014jaa}. The harvested entanglement can depend on the properties of the detectors used, however, it also gives an indication of the entanglement structure of the field itself.

A particular interesting application of the entanglement harvesting protocol is to examine how vacuum entanglement is affected by spacetime structure, such as the expansion rate of the Universe~\cite{Steeg:2009} and global spacetime topology~\cite{Smith:2016a}. However, apart from these investigations, little is known about entanglement harvesting in curved spacetimes. It has been shown that a weakly curved background can either increase or decrease harvested entanglement depending on the choice of vacuum state~\cite{Cliche:2011}, and recently the entanglement harvesting protocol was investigated in a black hole spacetime~\cite{Henderson:2017}, illustrating substantive differences differences from the corresponding situation in Minkowski space. The presence of the black hole was found to inhibit entanglement: as two detectors of fixed proper separation approach the black hole, their concurrence decreases to zero at a finite distance away from the horizon, this distance depends on the energy gaps of the detectors, the mass of the black hole, and the curvature of the spacetime.

Motivated by the above, in this article we present a detailed study of entanglement harvesting in anti-de Sitter (AdS) space. Quantum fields on AdS space are of particular interest because of the rapid development of the anti-de Sitter/conformal field theory correspondence which posits a connection between conformal field theories and the bulk geometry of AdS \cite{Maldacena:1998}. We   note that entanglement harvesting in the bulk may have implications for entanglement entropy in the context of the AdS/CFT correspondence conjecture.
%{\bf Alternate presentation (motivated by Alex's request for more clarity)}:
If one considers quantum corrections to the well-known Ryu-Takayanagi prescription~\cite{Ryu:2006} then the entanglement of quantum fields in the bulk across the Ryu-Takayanagi surface contributes to the entanglement entropy of the boundary CFT~\cite{Faulkner:2013ana}. Understanding the nature of entanglement in the bulk is likely to be important for understanding the quantum corrections to the entanglement in the boundary.

 We shall investigate how entanglement harvesting depends on the AdS length and the influence of the boundary conditions satisfied by the field at infinity, comparing our results to the flat spacetime counterpart and to the results obtained for the BTZ black hole \cite{Henderson:2017}. For this reason we shall consider $(2+1)$-dimensional AdS spacetime.  This lower-dimensional setting has the advantage of yielding insight into the significant physical processes with relative computational ease and efficiency.  We find a number of novel features that are particularly visible in the static case, most prominently an island of separability in parameter space at intermediate values of the AdS length, whose origin remains to be understood.  This same
phenomenon has been observed in a concurrent study in $(3+1)$-dimensional AdS spacetime \cite{KRE}, and we shall also, where appropriate,  compare our results to those obtained in that setting.

We begin in Sec.~\ref{sec2} by reviewing the basic formalism of the field-detector coupling and describing in general the entanglement harvesting protocol in curved spacetimes. In Sec.~\ref{sec3} we investigate the influence of the structure of AdS space on the entanglement harvesting protocol, including the AdS length and boundary conditions satisfied by the field. We consider two different arrangements of detectors: a pair of static non-geodesic detectors and a pair of detectors undergoing circular geodesic orbits. This allows for a comparison between entanglement harvesting along geodesic and non-geodesic detector trajectories. In addition the transition probability of a single detector will be derived in all cases. Finally, we conclude in Sec.~\ref{sec-conc} by summarizing the results presented and outlining prospects for future research.

Throughout this article we adopt the mostly positive convention for the metric signature $(-1,1,1)$ and employ natural units $\hbar = c = 8G = 1$.

%========================================
%========================================
\section{Unruh-DeWitt Detectors in AdS$_3$}
\label{sec2}

In this section we introduce the Unruh-DeWitt detector as a simplified model of a two-level atom interacting locally with a quantum field. We review the derivation of the joint state of two such detectors after their interaction with the field has ceased, stating explicitly the matrix elements appearing in this state to leading order in the interaction strength in terms of the vacuum Wightman function. We then introduce scalar field theory on (2+1)-AdS space, stating explicitly the associated vacuum Wightman function.

%========================================
%========================================
\subsection{The Unruh-DeWitt detector}

Consider a detector moving along the spacetime trajectory $x_D(\tau)$, parametrized in terms of the detectors proper time $\tau$, with its ground state $\ket{0}_D$ and excited state $\ket{1}_D$ separated by an energy gap $\Omega_D$. As a simplified model of the light-matter interaction \cite{Martin-Martinez2013, Alvaro}, suppose the detector couples to a real scalar field $\phi(x)$ through the interaction Hamiltonian
\begin{align}
H_{D}(\tau)=\lambda\chi_D(\tau)\Big[e^{i\Omega_D \tau}\sigma^+ + e^{-i\Omega_D \tau}\sigma^-\Big] \phi[x_D(\tau)],
\label{HD}
\end{align}
where $\lambda$ is the strength of the interaction, $\chi_D(\tau)$ is a switching function specifying the duration of the detector's interaction with the field, and $\sigma^\pm$ denote $SU(2)$ ladder operators acting on the detector's Hilbert space.

Consider now two such detectors, labeled by $A$ and $B$, which both begin ($\tau_A,\, \tau_B \to -\infty$) in their ground state and then couple to the same scalar field, which begins in an appropriately defined vacuum state $\ket{0}$. In such a scenario the initial state of the detectors and field together is $\ket{\Psi_i} \ce \ket{0}_A\ket{0}_B \ket{0}$. In the far future ($\tau_A,\, \tau_B \to \infty$), after the interactions between the field and detectors has ceased, the joint state of the detectors and field together is $\ket{\Psi_f} = U \ket{\Psi_i}$, where the evolution operator $U$ is given by
\begin{align}
U \ce \mathcal{T} \exp\left[  -i \int  dt\,  \sum_{D \in \{A,B\}}  \frac{d \tau_D}{dt} H_{D}(\tau_D) \right],
\end{align}
where $\mathcal{T}$ denotes the time ordering operator and $t$ is the time coordinate with respect to which the vacuum state of the field is defined. The final state of the detectors alone is obtained by tracing out the field degrees of freedom
 \begin{align}
 \rho_{AB} &:=\tr_\phi \left( U\ket{\Psi_i}\bra{\Psi_i}U^{\dagger} \right) \nn \\
 &= \begin{pmatrix}
1 - P_A - P_B  & 0 & 0 & X \\
0 & P_B  & C & 0 \\
0 & C^* & P_A & 0 \\
X^* & 0 & 0 & 0
\end{pmatrix} + \mathcal{O}\!\left(\lambda^4\right),
\label{FinsalState2}
 \end{align}
which we have expressed in the computational basis $\{ \ket{0}_A \ket{0}_B, \ket{0}_A \ket{1}_B, \ket{1}_A \ket{0}_B, \ket{1}_A \ket{1}_B \}$, and
\begin{widetext}
\begin{align}
P_D &:= \lambda^2 \int d\tau_D  d \tau_D' \, \chi_D(\tau_D) \chi_D(\tau_D') e^{-i \Omega_D \left(\tau_D-\tau_D'\right)} W\!\left(x_D(t) ,x_D(t')\right) \label{PJ}  \quad \mbox{for} \quad D \in \{A,B\},
 \\
C&:= \lambda^2 \int d\tau_A d\tau_B   \,  \chi_A(\tau_A) \chi_B(\tau_B) e^{- i \left( \Omega_A \tau_A - \Omega_B \tau_B \right)} W\!\left(x_A(t) , x_B(t')\right) \label{defC}, \\
X &:=-\lambda^2  \int  d\tau_A d\tau_B   \, \chi_A(\tau_A) \chi_B(\tau_B)  e^{-i\left( \Omega_A  \tau_A + \Omega_B  \tau_B\right)} \Big[ \theta(t'-t) W\!\left(x_A(t), x_B(t')\right)
+ \theta(t-t') W\!\left(x_B(t'),x_A(t) \right)  \Big] , \label{defX}
\end{align}
\end{widetext}
where $W(x,x'):=\bra{0} \phi(x) \phi(x') \ket{0}$ is the Wightman function associated with the field, and $P_D$ is the probability that a detector has transitioned from its ground to excited state as a result of its interaction with the field; in some contexts this probability is referred to as the response function of the detector~\cite{Birrell:1982}.  For spatially fixed  detectors $A$ and $B$ we have  $\tau_A = \gamma_A t$ and $\tau_B = \gamma_B t$, where the quantities $\gamma_D$ will be functions of their spatial location. In what follows we will assume both detectors have the same proper energy gap $\Omega$ and switching parameter $\sigma$ in their own rest frame, i.e, $\Omega_D = \Omega_A=\Omega_B=\Omega$ and $\sigma_D=\sigma_A=\sigma_B=\sigma$.

To quantify the entanglement among the detectors resulting from their interaction with the field we employ the concurrence as a measure of entanglement \cite{Wootters:1997id}. The concurrence evaluated for the final state of the detectors given in Eq.~\eqref{FinsalState2} yields \cite{Smith:2017b,Smith:2016a}
\begin{align}\label{concurrence-eq}
\mathcal{C} \left[ \rho_{AB} \right] = 2\max \left[0,|X|-\sqrt{P_AP_B}\right] + \mathcal{O}\!\left(\lambda^4\right).
\end{align}
From Eq.~\eqref{concurrence-eq} we see that concurrence, and consequently the entanglement harvested, is a competition between two quantities: the off diagonal matrix element $|X|$, which leads to non-local correlations between the detectors, and the geometric mean of the detector's transition probabilities $P_A$ and $P_B$ (the ``noise'').

This process of detectors extracting entanglement from the vacuum through the local interactions described above is known as the entanglement harvesting protocol. Assuming that any interaction between the detectors mediated by the field is negligible, which is certainly the case if the detectors are spacelike separated, the resulting entanglement of the final state of the detectors in Eq.~\eqref{FinsalState2} is a consequence of existing entanglement between the regions of the field in which the detectors interacted, some of which has been transferred to the detectors. In this way, the entanglement harvesting protocol, specifically quantifying the entanglement of the state given in  Eq.~\eqref{FinsalState2} through an entanglement measure like the concurrence, provides an indication of the entanglement structure of the vacuum state. We will exploit this fact in Sec.~\ref{sec3} to examine the entanglement structure of the AdS vacuum and its dependence on spacetime curvature and boundary conditions satisfied by the field as seen by a pair of detectors.

%========================================
%========================================
\subsection{The AdS$_3$ Wightman function}

AdS space is a maximally symmetric spacetime of constant negative curvature that solves Einstein's equations with negative cosmological constant $\Lambda \ce -1/\ell^2$. The (2+1)-dimensional AdS (AdS$_3$) geometry can be described by a $3$-dimensional hyperboloid
\begin{align}
X_1^2+X_2^2-T_1^2-T_2^2=-\ell^2, \label{hyp1}
\end{align}
embedded in a flat 4-dimensional geometry
\begin{align}\label{ds4}
dS^2=  dX_1^2 + dX_2^2  -dT_1^2 - dT_2^2,
\end{align}
with coordinates $(X_1, X_2, T_1, T_2)$~\cite{Carlip:2003}.

Global coordinates $(\bar{\tau}, \rho, \phi)$ covering AdS$_3$ are obtained via the transformation
\begin{align}
X_1 &=\ell\sinh\rho\sin\phi, &X_2&=\ell\sinh\rho\cos\phi, \nn \\
T_1&=\ell\cosh\rho\cos\bar{\tau}, &T_2&=\ell\cosh\rho\sin\bar{\tau}, \label{coordinate1}
\end{align}
in which the induced metric on AdS$_3$ becomes
\begin{align}
ds^2 = \ell^2 \left( -\cosh^2\rho \, d\bar{\tau}^2+d\rho^2+\sinh^2\rho \, d\phi^2 \right),
\end{align}
where $\bar{\tau} \in \mathbb{R}$, $\rho \in \mathbb{R}^+$, and $\phi \in [0,2\pi)$\footnote{ Strictly speaking, when obtained from this embedding picture, the time coordinate $\bar\tau$ is periodic with period $2 \pi$. By allowing $\bar\tau$ to take on all real values we are implicitly moving to the universal cover of the space. It is this universal cover that is customarily (and also in this work) referred to as AdS.}. Under the coordinate transformation $t:=\ell \bar{\tau}$ and $r:=\ell \sinh \rho$, the AdS$_3$ metric may be cast in the familiar form
\begin{align}\label{metric-ads}
ds^2=-\Big(1+\frac{r^2}{\ell^2}\Big) dt^2+\frac{dr^2}{1+r^2/\ell^2}+r^2d\phi^2.
\end{align}
It will be useful for what follows to introduce the function $d(R_1, R_2 )$ denoting the proper distance between the spacetime points $x_1=(t,R_1, \phi)$ and $x_2=(t,R_2, \phi)$,
\begin{align}
d(R_1, R_2 ) \ce \ell\ln \left[\frac{R_2+\sqrt{R_2^2 + \ell ^2}}{R_1 + \sqrt{R_1^2 + \ell^2}} \right] .\label{properdistance}
\end{align}

We shall consider  a massless conformally coupled real scalar field on AdS$_3$, since its vacuum Wightman function takes the particularly simple form~\cite{Carlip:2003}
\begin{align}
W_{Ads}^{(\zeta)}(x,x')=\frac{1}{4\pi\ell\sqrt{2}} \left(\frac{1}{\sqrt{\sigma(x,x')}}-\frac{\zeta}{\sqrt{\sigma(x,x')+2}}\right), \label{wightmanf}
\end{align}
where
\begin{align}
\sigma(x,x') &\ce \frac{1}{2\ell^2} \bigg[ \left(X_1-X_1' \right)^2- \left(T_1-T_1'\right)^2  \nn \\
 &\quad   +\left(X_2-X_2'\right)^2 - \left(T_2-T_2'\right)^2 \bigg], \label{deltasigma}
\end{align}
is the square distance between $x$ and $x'$ in the embedding space $\mathbb{R}^{2,2}$ and the parameter $\zeta \in \{1,0,-1\}$ specifies either Dirichlet ($\zeta = 1$), transparent ($\zeta = 0$), or Neumann ($\zeta = -1$) boundary conditions satisfied by the field at spatial infinity. Furthermore, it is the vacuum state described by this Wightman function from which the Hartle-Hawking vacuum on the BTZ black hole may be constructed~\cite{Lifschytz:1994}; the BTZ black hole being asymptotically AdS space. Thus our investigations in this paper will allow for a proper comparison between entanglement harvesting between detectors in AdS$_3$ space and detectors outside the BTZ black hole~\cite{Henderson:2017} to investigate the effects of spacetime horizons.

Although the general method for computing Wightman functions is to carry out a mode sum over basis functions,
the structure of the Wightman function in AdS$_3$ is simple enough that a combination of analytic and numerical  integration is possible.  Our methods are complementary to
a concurrent study of entanglement harvesting in AdS$_4$ \cite{KRE}, in which mode sums were carried out to evaluate the relevant quantities of interest.

%========================================
%========================================
\section{Static detectors}
\label{sec3}

In this section we evaluate the transition probability of a static  detector at different locations in AdS$_3$, and answer the question of how much entanglement two such detectors can harvest from the AdS$_3$ vacuum.

Suppose detectors $A$ and $B$ are kept at distinct fixed positions $R_A$ and $R_B$, respectively, along a common radial direction. The spacetime trajectories describing such detectors parameterized by their proper time $\tau_D$ is
\begin{align}\label{traj-static}
x_D(\tau_D) \ce \left\{ t =  \tau_D/\gamma_D,  \  r = R_D, \  \phi= \phi_0\right\},
\end{align}
where $\gamma_D:=\sqrt{(R_D/\ell)^2+1}$.  We take the switching functions of the detectors to be Gaussians described by
\begin{align}
  \chi_A(\tau_A) &= \exp\left(-\frac{\left(\tau_A+\gamma_At_0/2\right)^2}{2\sigma^2}\right), \nn\\
  \chi_B(\tau_B) &= \exp\left(-\frac{\left(\tau_B-\gamma_Bt_0/2\right)^2}{2\sigma^2}\right).
  \label{eq:TimeDelay}
\end{align}
where $t_0>0$ corresponds to detector $A$ interacting with the field before detector $B$ and vice versa for $t_0<0$,  where the time delay $t_0$ is defined in the field frame. Each detector interacts with the field for an approximate amount of proper time $\sigma$.

To compute the transition probability $P_D$ and matrix element $X$ we evaluate the Wightman function in Eq.~\eqref{wightmanf} along the static detectors' trajectories, substitute the resulting expression into Eqs. \eqref{PJ} and \eqref{defX}, and express the result in a form that lends itself to being evaluated numerically, all of which is done explicitly in Appendix~\ref{Derivation of PD and X}. The resulting expression for the transition probability is
\begin{align}
P_{D}&= \frac{\lambda^2\sigma }{2\sqrt{2\pi}}  \Bigg( - \PV \int_{0}^{\infty} dy \,  \frac{e^{-{a_D} y^2} \sin \left(\beta_D{y} \right )}{ \sqrt{2} \sin \left(y/2 \right)} \nn \\
&\quad +\frac{\pi}{\sqrt{2}} \sum_{n\in \mathbb{Z} } (-1)^n \cos \left(2n\pi\beta_D \right)e^{-4n^2\pi^2{a_D}}  \nn\\
&\quad  -\zeta  {\rm Re}  \sum_{n \geq 0 }  (-1)^{n} \int_{\theta_{D_n}}^{\theta_{D_{n+1}}} dy \, \frac{e^{-{a_D} y^2} e^{-i\beta_D{y}}}{\sqrt{\cos y + \alpha^{+}_D}}  \Bigg), \label{PAPB}
\end{align}
where we have defined
\begin{align}
&a_D \ce \gamma_D^2 \ell^2/4\sigma^2
\qquad \beta_D \ce \gamma_D\ell \Omega \nn\\
&\alpha^{+}_D \ce   \left[ - (R_D/\ell)^2 + 1\right]/\gamma_D^2 \nn\\
&\theta_{D_n} \ce \max \left[ 0, \, \arccos\alpha^{+}_D + (2n-1) \pi \right].
\end{align}
The resulting expression for the matrix element $X$ is
\begin{align}
  X &= - \frac{\lambda^2\sigma}{2\sqrt{\pi}} K_{X} \nn\\
  &\phantom{=}\times \sum_{n\ge0} (-1)^n\left(\int_{\theta_{X_n}^-}^{\theta_{X_{n+1}}^-} dy\frac{e^{-a_Xy^2}\cosh\big((\Delta_T+i\beta_X)y\big)}{\sqrt{\cos y +\alpha_X^-}}\right. \nn\\
  &\phantom{=} - \zeta \left.\int_{\theta_{X_n}^+}^{\theta_{X_{n+1}}^+} dy\frac{e^{-a_Xy^2}\cosh\big((\Delta_T+i\beta_X)y\big)}{\sqrt{\cos y +\alpha_X^+}}\right)
  \label{X-equation}
\end{align}
where
\begin{align}
&K_{X} \ce \sqrt{\frac{\gamma_A\gamma_B}{\gamma_A^2+\gamma_B^2}}\exp\left(-\frac{\Omega^2\sigma^2}{2}\frac{(\gamma_A+\gamma_B)^2}{\gamma_A^2+\gamma_B^2}\right) \nn\\
&\quad \times \exp\left(-\frac{t_0^2}{2\sigma^2}\frac{\gamma_A^2\gamma_B^2}{\gamma_A^2+\gamma_B^2}+i\frac{\Omega t_0}{2}\frac{(\gamma_A+\gamma_B)^2(\gamma_A-\gamma_B)}{\gamma_A^2+\gamma_B^2}\right) \nn\\
&{a_X} \ce \frac{\gamma_A^2 \gamma_B^2}{\gamma_A^2 + \gamma_B^2} \frac{\ell^2}{2 \sigma^2 } \nn \\
&\Delta_T \ce -\frac{t_0\ell}{\sigma^2}\frac{\gamma_A^2\gamma_B^2}{\gamma_A^2+\gamma_B^2} \label{Kxeq} \\
&\beta_X \ce \frac{\gamma_A \gamma_B \left( \gamma_A -\gamma_B\right)}{\gamma_A^2 + \gamma_B^2} \ell \Omega \nn\\
&\alpha^{\pm}_X \ce  \frac{1}{\gamma_A \gamma_B} \left(-\frac{R_AR_B}{\ell^2 } \pm 1 \right) \nn\\
&\theta^\pm_{X_n} \ce \max \left[ 0, \,  \arccos\alpha^{\pm}_X + (2n-1) \pi \right]. \nn
\end{align}

In what follows, both the transition probability and matrix element $X$ given above were evaluated numerically in Mathematica to a precision of $ 10^{-17}$ using the double exponential and double exponential oscillatory integrations methods.

%----------------------------------------------------------------------
\begin{figure*}[t]
\subfloat[$d(0, R_D)/\sigma=0$]{%
  \includegraphics[width=.3\linewidth]{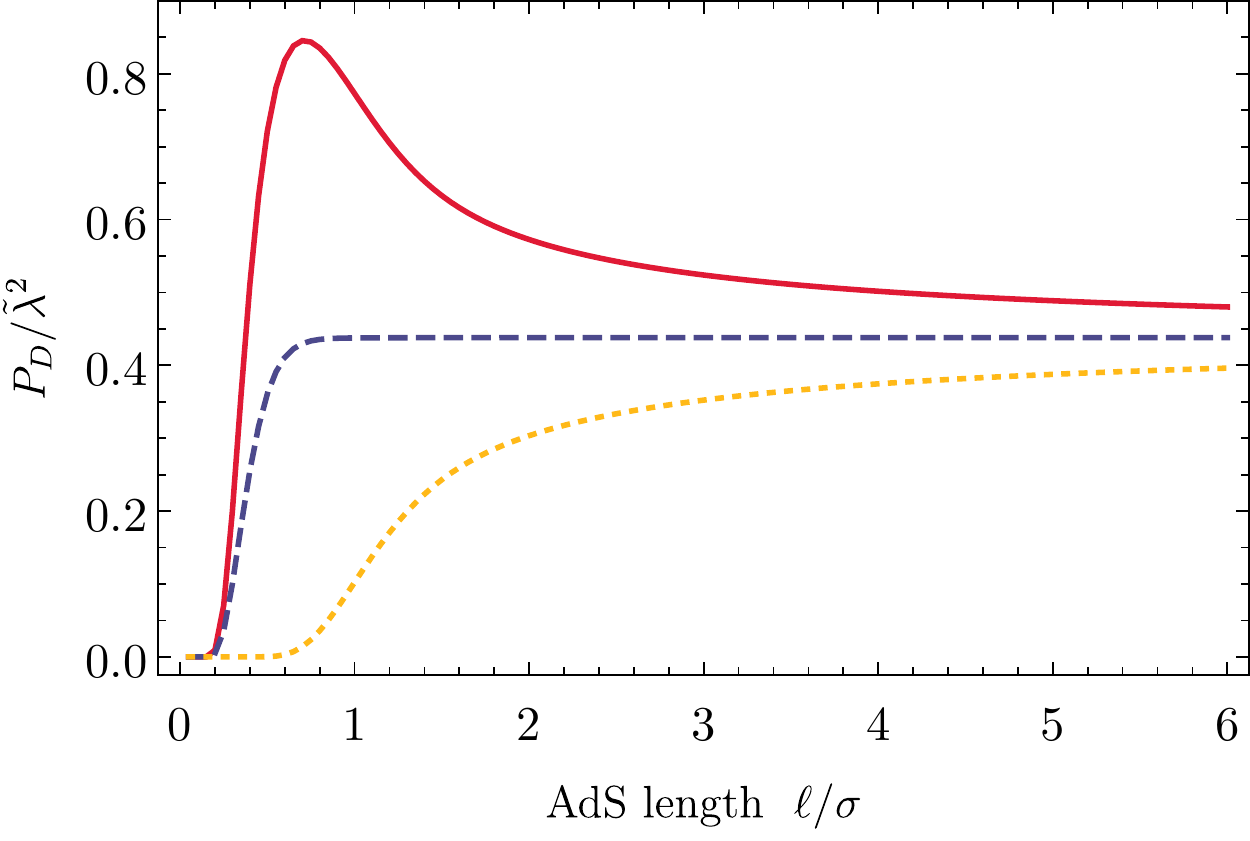}%
}%
\quad
 \subfloat[$d(0,R_D)/\sigma=1/2$]{%
  \includegraphics[width=.3\linewidth]{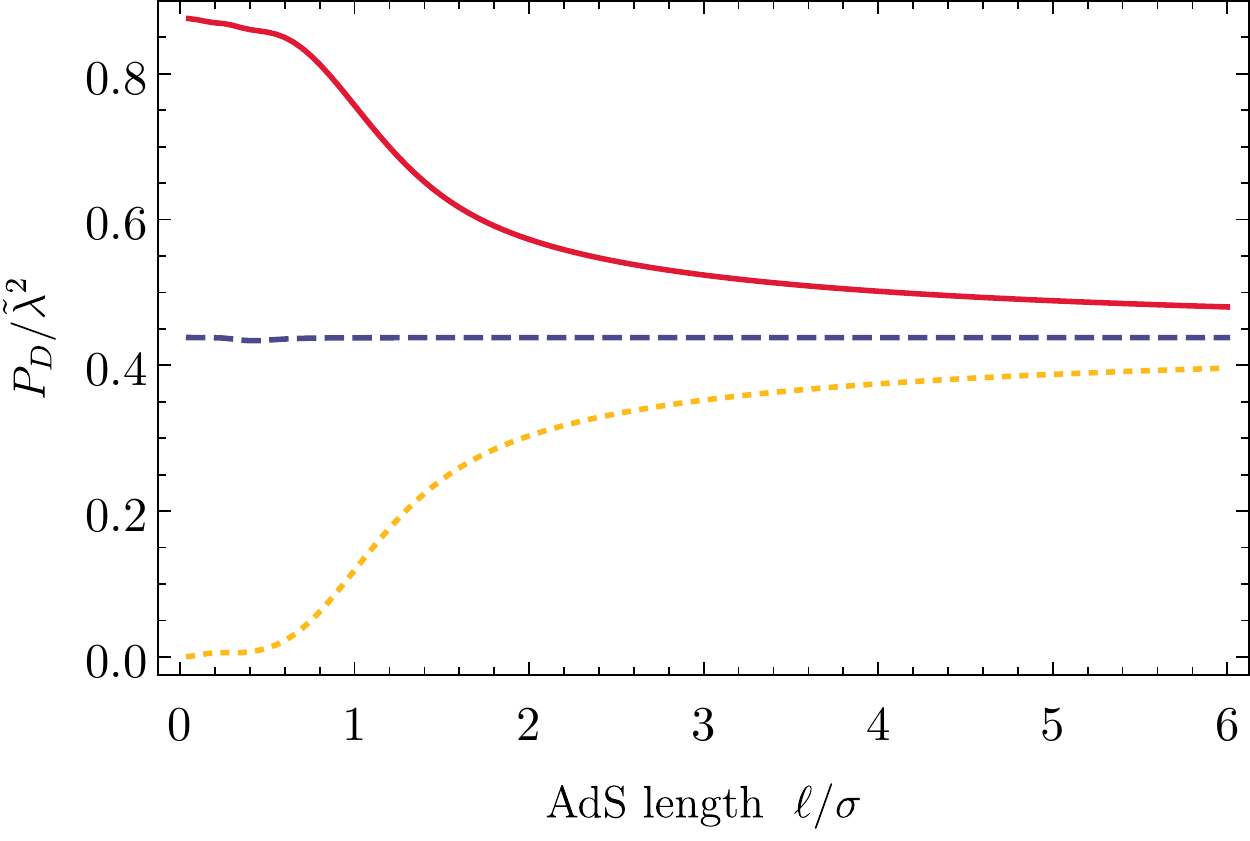}%
}\quad
\subfloat[$d(0, R_D)/\sigma=1$]{%
  \includegraphics[width=.3\linewidth]{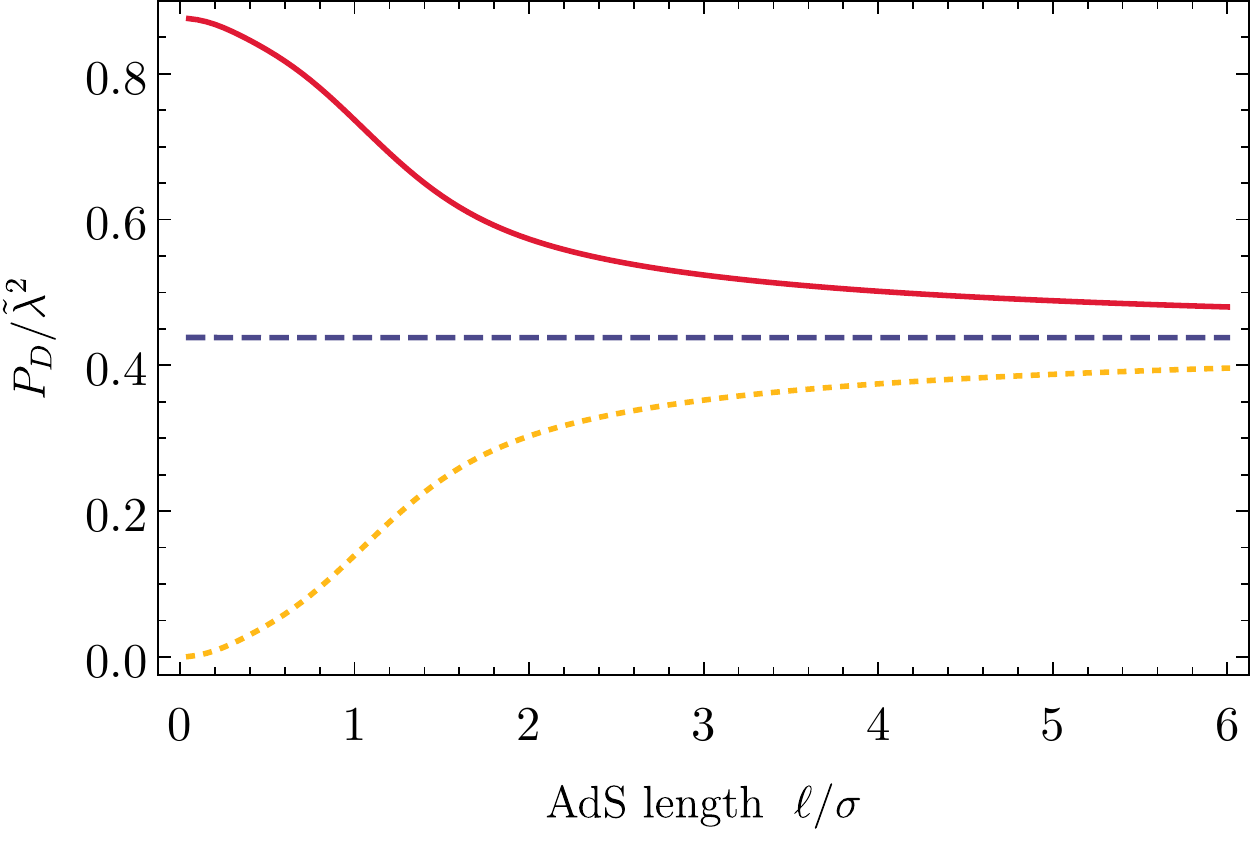}
} \\
  \includegraphics[width=.3\linewidth]{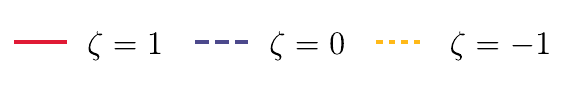}

\caption{The transition probability $P_D$ of a static detector with energy gap $\Omega\sigma = 1/100$ is plotted as a function of the AdS length $\ell/\sigma$ for each of the three boundary conditions $\zeta =\{1,0,-1\}$. Plots (a), (b), and (c) correspond to detectors located at different proper distances $d(0,R_D)/\sigma$ away from the origin. We introduce the dimensionless coupling strength $\tilde{\lambda}=\lambda\sqrt{\sigma}$ for convenience.}
\label{prob1}
\end{figure*}
%----------------------------------------------------------------------

%========================================
%========================================
\subsection{The transition probability}

We now evaluate the transition probability of a static detector, given in Eq.~\eqref{PAPB}, for a wide range of scenarios. We begin by noting that the transition probability in Eq.~\eqref{PAPB} can be evaluated perturbatively in $\sigma/\ell$. We find that  the leading contributions take a very simple form
\begin{align}
\frac{P_D}{\sigma \lambda^2} &= \frac{\sqrt{\pi}}{4}\left(1 - \erf(\sigma\Omega) \right) - \zeta \frac{e^{-\sigma^2\Omega^2}}{4} \left(\frac{\sigma}{\ell} \right) \nonumber\\
&- \frac{ \sigma \Omega e^{-\sigma^2\Omega^2}}{24} \left(\frac{\sigma}{\ell} \right)^2 - \zeta  \frac{ e^{-\sigma^2\Omega^2} \left(1 - 2 \sigma^2 \Omega^2 \right) }{16 } \left( \frac{\sigma}{\ell} \right)^3
	\nonumber\\
&+ \frac{ \sigma \Omega e^{-\sigma^2\Omega^2}}{2880}  \left[120  \frac{ d(0,R_D)^2}{\sigma^2} + 14 \sigma^2 \Omega^2 -21 \right] \left( \frac{\sigma}{\ell} \right)^4 	\nonumber\\
&+ {\cal O} \left( \left(\frac{\sigma}{\ell}\right)^{5} \right).
\label{pertP}
\end{align}

%----------------------------------------------------------------------
\begin{figure*}
\subfloat[$\ell/\sigma=1/2$ and $d(0,R_D)/\sigma = 0$]{%
  \includegraphics[width=.3\linewidth]{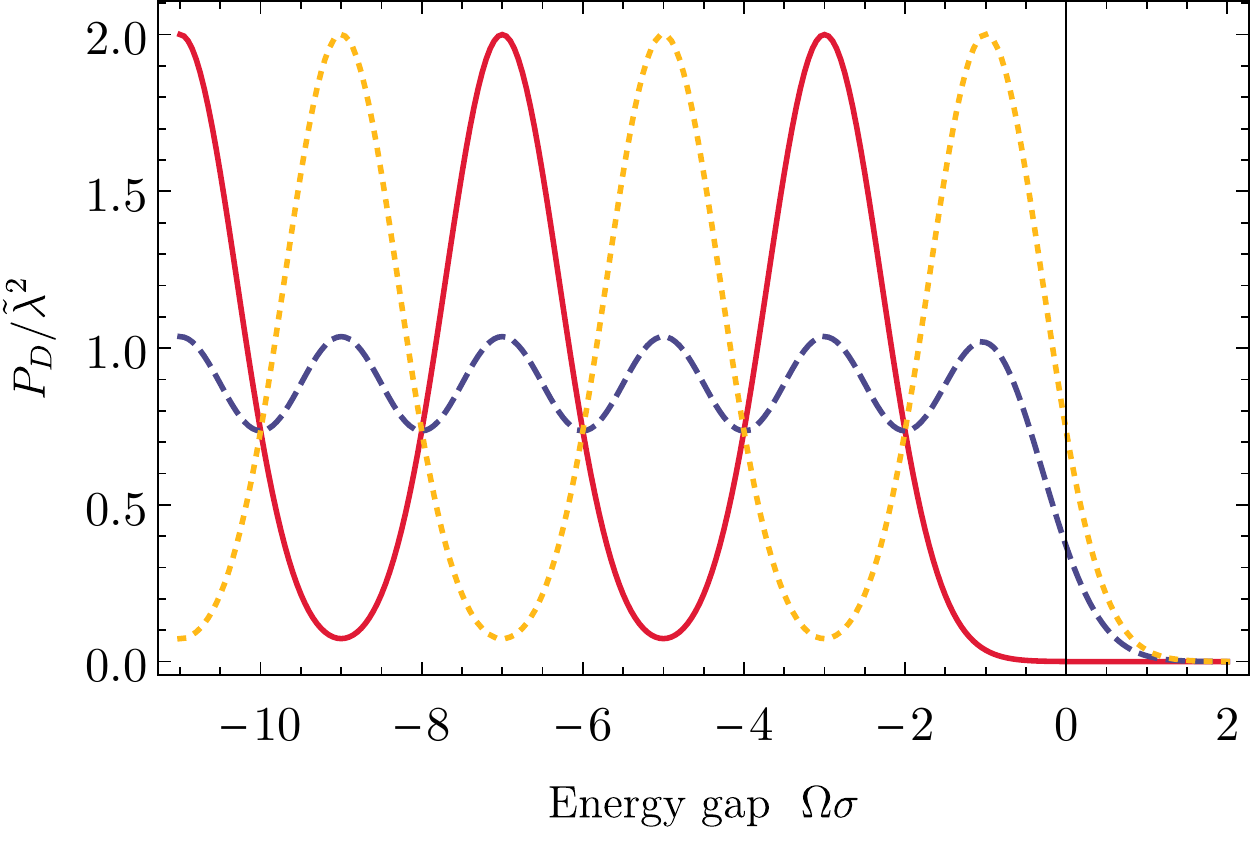}%
}\quad
\subfloat[$\ell/\sigma=1$ and $d(0,R_D)/\sigma = 0$]{%
  \includegraphics[width=.3\linewidth]{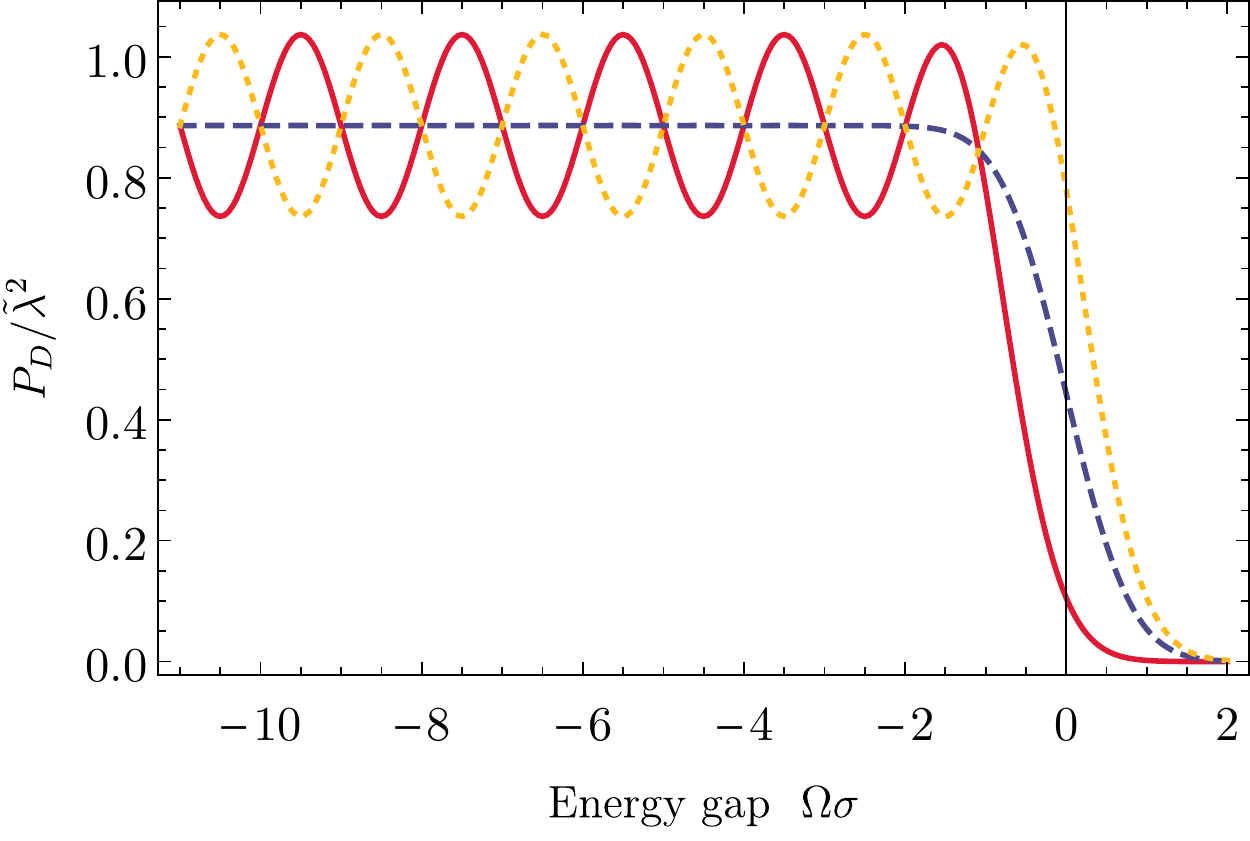}%
}\quad
 \subfloat[$\ell/\sigma=10$ and $d(0,R_D)/\sigma = 0$]{%
  \includegraphics[width=.3\linewidth]{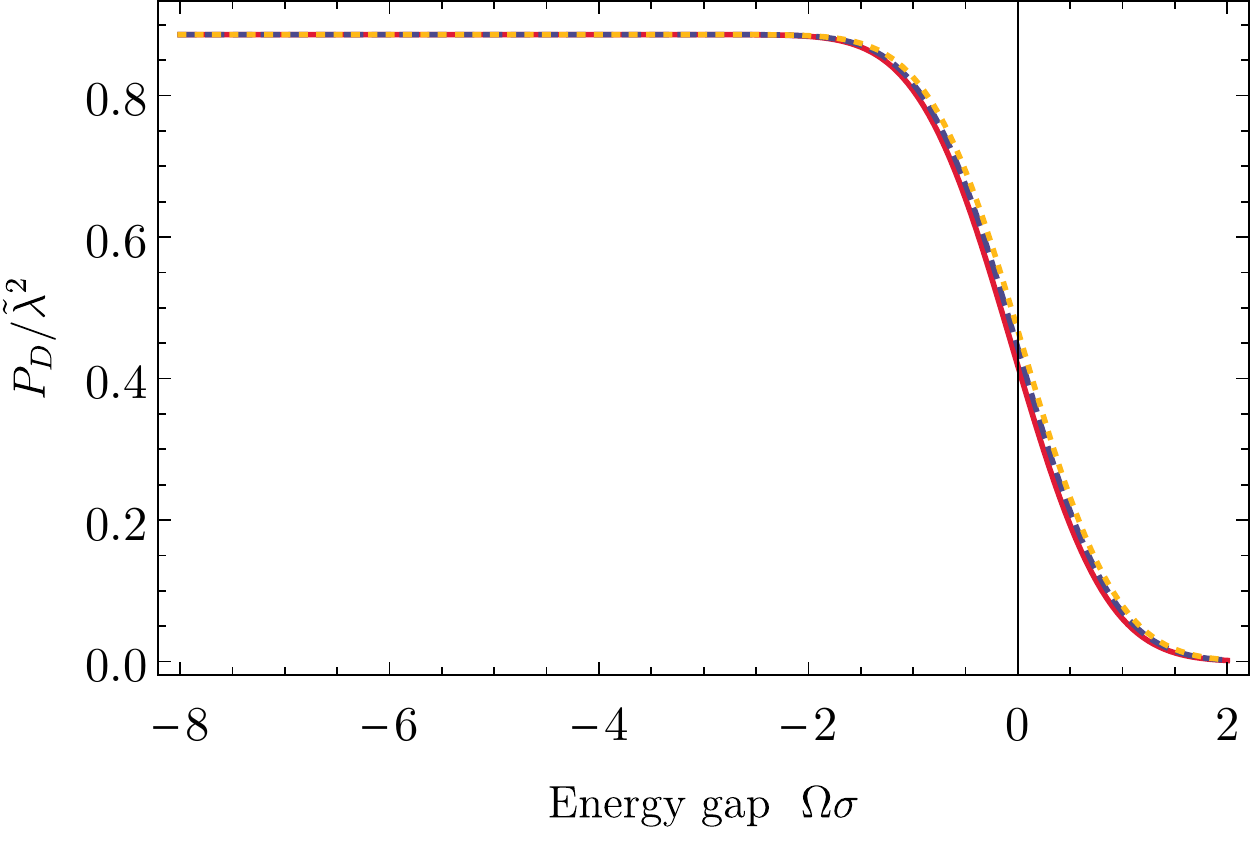}%
}
\\
\subfloat[$\ell/\sigma=1/2$ and $d(0,R_D)/\sigma = 1$]{%
  \includegraphics[width=.3\linewidth]{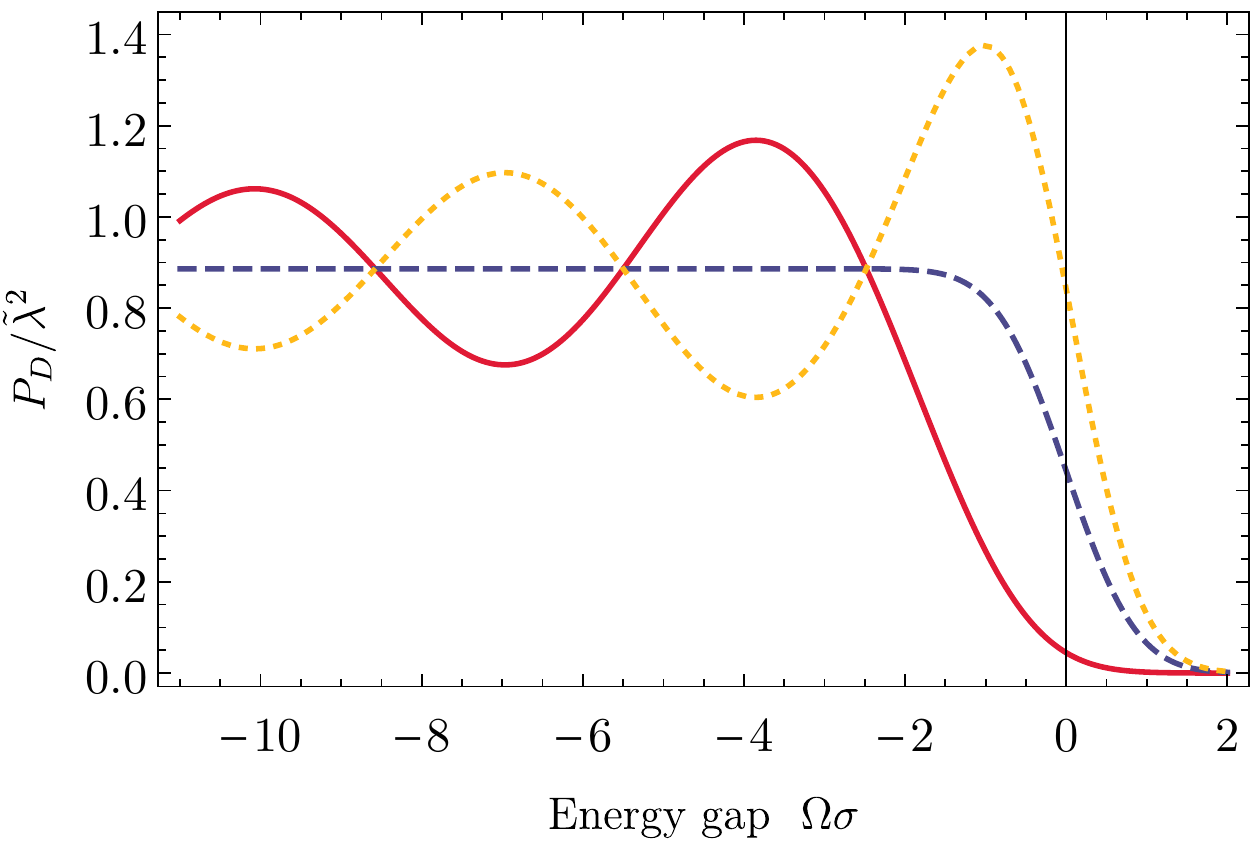}%
}\quad
\subfloat[$\ell/\sigma=1$ and $d(0,R_D)/\sigma = 1$]{%
  \includegraphics[width=.3\linewidth]{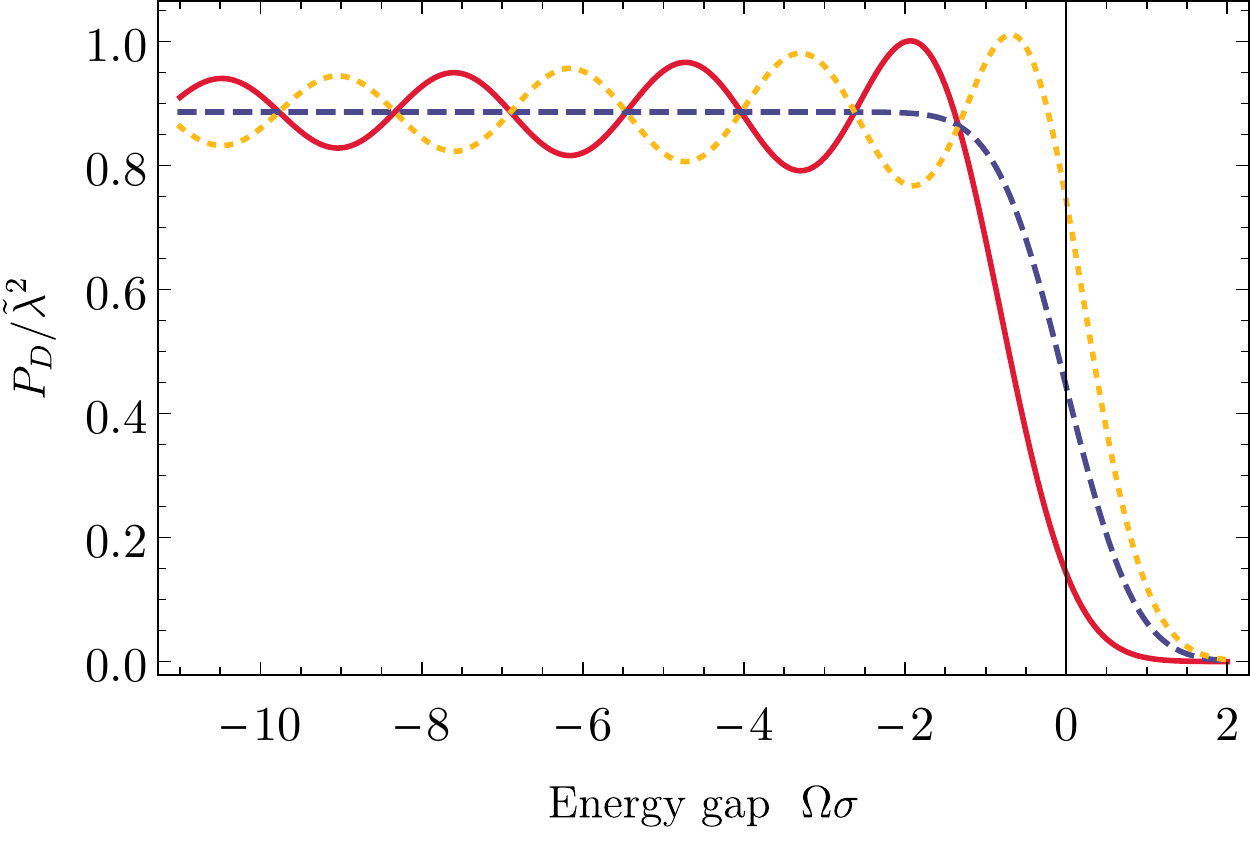}%
}\quad
 \subfloat[$\ell/\sigma=10$ and $d(0,R_D)/\sigma = 1$]{%
  \includegraphics[width=.3\linewidth]{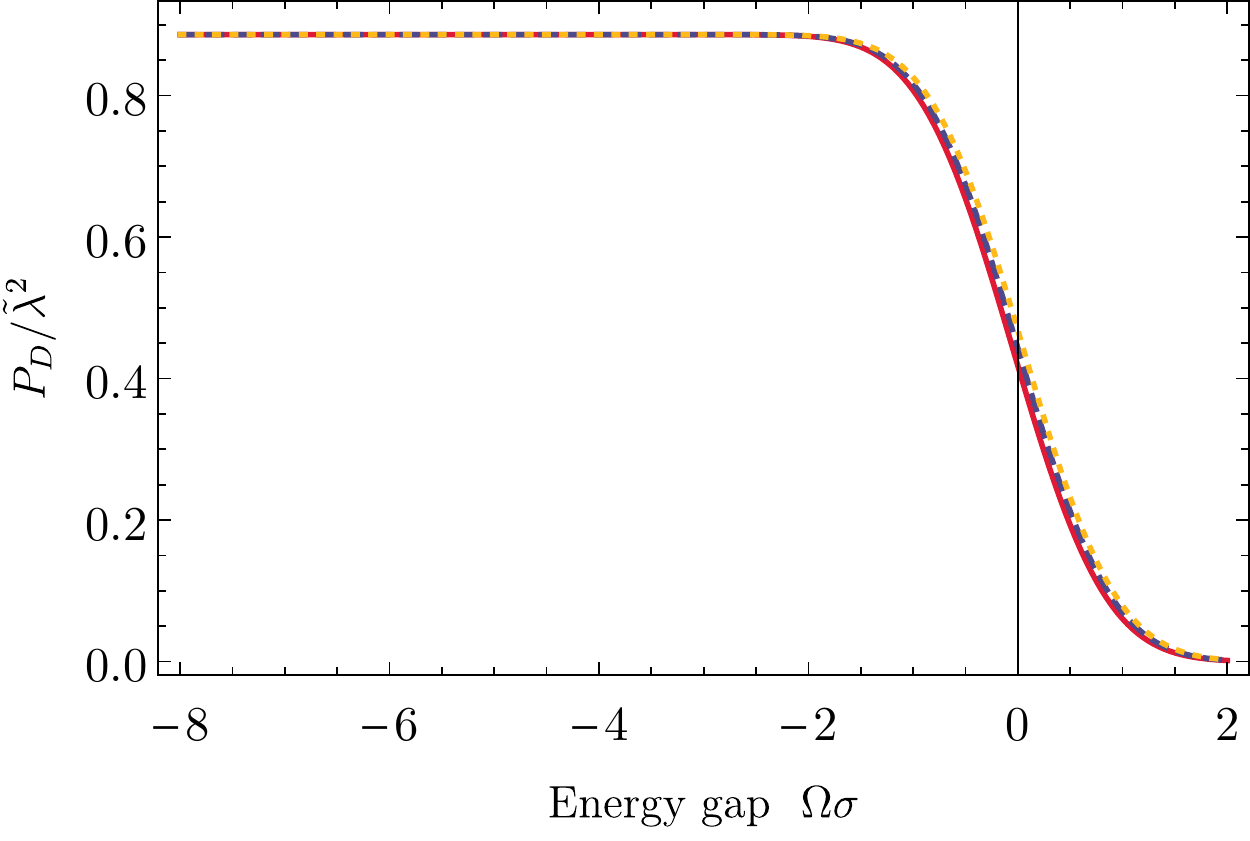}%
}\\
  \includegraphics[width=.3\linewidth]{legend.pdf}% \
\caption{The transition probability $P_D$ of a static detector is plotted as a function of its energy gap $\Omega \sigma$ for all  boundary conditions $\zeta =\{1,0,-1\}$ and various values of the AdS length $\ell/\sigma$. In plots (a), (b), and (c) the detector is located at the origin, while in plots (d), (e), and (f) the detector is located a proper distance $d(0,R_D)/\sigma = 1$ away from the origin. Negative energy gaps $\Omega \sigma <0$ correspond to a detector that is prepared in its excited state $\ket{1}_D$ prior to interacting with the field.}
\label{PAPB-omega}
\end{figure*}
%----------------------------------------------------------------------

The first term appearing above is simply the transition probability in (2+1)-dimensional Minkowski space (flat space).  The proper distance $d( 0, R_D)$ first enters at order $(\sigma/\ell)^{4}$ and its appearance can be understood as the manifestation of redshift (time dilation) due to the detector being on a static trajectory.

 The perturbative expression in Eq.~\eqref{pertP} is particularly useful for understanding how the transition probability in AdS$_3$ deviates from the flat space result.  We see that the leading order correction to the flat space result  is proportional to $\zeta$. This means that the most important factor in the deviation from flat spacetime is due to the boundary conditions satisfied by the field. A small amount of negative curvature will result in the detector clicking more (less) than it would in flat spacetime if the field satisfies Neumann (Dirichlet) boundary conditions. If the field satisfies transparent boundary conditions then there is no correction to the flat space result at order $\sigma/\ell$.

%  If the field satisfies transparent boundary conditions, then the transition probability will decrease relative to the flat space value, provided the detector is initially in its ground state.

 The next order correction is also quite interesting. It is independent of the boundary conditions satisfied by the field and so  can be thought of as a ``universal'' contribution due purely to negative curvature. Since this term in the perturbative expansion is always negative for detectors starting in their ground state, we can interpret this as the statement that negative curvature tends to decrease the transition probability of a detector. Conversely, since this correction will always be positive for detectors that are initially excited ($\Omega_D<0$), this means excited detectors in negatively curved spaces are more likely to relax to their ground state than they would be in flat spacetime.

\begin{figure}[t]
\subfloat[$\ell/\sigma=1/2$]{%
  \includegraphics[width=.9\linewidth]{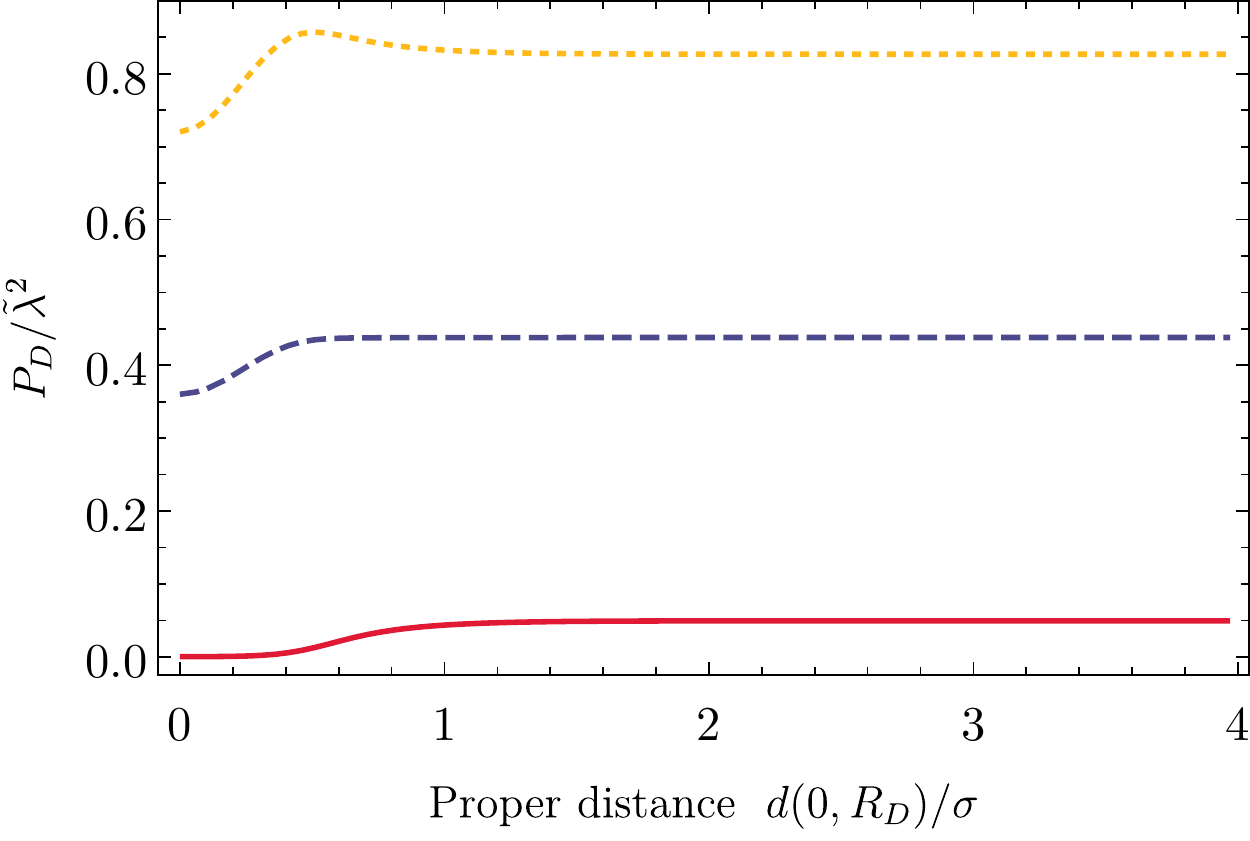}%
}\qquad
\\
 \subfloat[$\ell/\sigma=1$]{%
  \includegraphics[width=.9\linewidth]{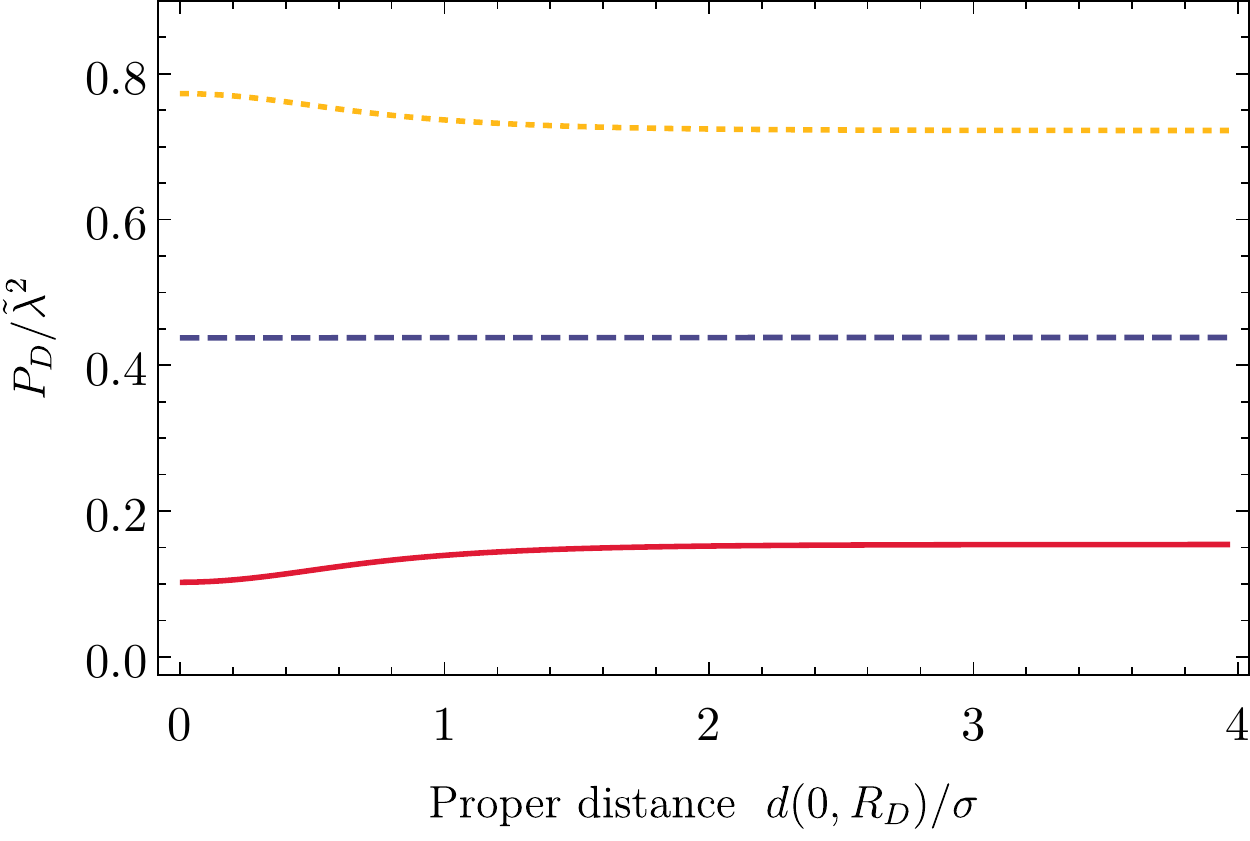}%
} \\
  \includegraphics[width=.65\linewidth]{legend.pdf}% \
\caption{The transition probability $P_D$ of a static detector with energy gap $\Omega\sigma=1/100$ is plotted as function of its proper distance from the origin for an AdS length of (a) $\ell/\sigma=1/2$ and \mbox{(b) $\ell/\sigma=1$ for all  boundary conditions $\zeta =\{1,0,-1\}$. }}
\label{PBposition-L}
\end{figure}

The perturbative expansion, while insightful, cannot be used when $\sigma/\ell$ is large, and there are many interesting phenomena in this regime. To explore this regime we must resort to a full numerical evaluation of the integral in Eq.~\eqref{PAPB} which is used in producing Figs.~\ref{prob1}-\ref{PBposition-L}.

As seen upon comparison of the plots in Fig.~\ref{prob1}, for all boundary conditions, the transition probability $P_D$ is only sensitive to the proper distance $d(0,R_D)$ the detector is from the origin when the switching parameter $\sigma$ is greater than or comparable to the AdS length $\ell$. As $\ell/\sigma$ grows, regardless of the boundary conditions, $P_D$ approaches its corresponding value in Minkowski space, indicated by the dashed line. Furthermore, for all boundary conditions, the transition probability of a detector located at the origin vanishes as $\ell/\sigma \to 0$, whereas for a detector positioned some proper distance away from the origin, the transition probability remains almost the same for any $\ell$. The most notable property of a detector located at the origin is that for a field satisfying Dirichlet boundary conditions ($\zeta=1$), its transition probability reaches a maximum near $\ell/\sigma \approx 0.7$.

In Fig.~\ref{PAPB-omega} the transition probability is plotted as a function of the detector's energy gap $\Omega \sigma$ for fixed AdS length $\ell/\sigma=1/2$, where $\Omega < 0$ corresponds to the  detector initially prepared in its excited state.  We observe that for $\Omega > 0$, the transition probability decays to zero exponentially with increasing $\Omega\sigma$ without oscillation, regardless of the boundary condition. Moreover, the decay is much faster for the Dirichlet condition ($\zeta=1$) than for the other boundary conditions.  However for $\Omega < 0$, the transition probability oscillates as a function of $\Omega \sigma$, this oscillation being suppressed by both the transparent boundary condition and for detectors located at large distances $d(0,R_D)$ from the origin. This feature
is evidently dependent on dimensionality,  and \textit{not} on  the general structure of AdS spacetime, since
the analogous graph in $(3+1)$-dimensions exhibits a transition rate (modulo oscillations) that is roughly proportional to $|\Omega|$ for  large negative energy gap \cite{KRE}.  It is well-known that the detector response exhibits different dependence on $\Omega$ as a function of spacetime dimension, and so this difference is not surprising. The dimension dependence arises due to the short distance behaviour of the Wightman function~\cite{Hodgkinson:2011pc} or, equivalently, due to the energy scaling of the density of states of the field quanta sensed by the detector which goes as $\Omega^{d-3}$ (see, e.g.,~\cite{densityOfStates}).

% The two dimensionalities have different state spaces, with an additional angular momentum mode being present in $(3+1)$-dimensions at any given energy, and so the density of states scales  as $O(\omega)$; however in $(2+1)$-dimensions the absence of this mode yields a density of states scaling as $O(1)$ and the detector response is flat~\cite{KRE}.

In Fig.~\ref{PBposition-L}, we illustrate the behaviour of the transition probability for a detector located at different spatial positions in AdS$_3$.  The transition probability, regardless of the boundary conditions, is hardly influenced by the change of position if $\ell/\sigma$ is not vanishingly small.  This is to be expected:  for large AdS lengths the detector is highly localized in spacetime and so any change in the AdS curvature negligibly affects the detector. It is only when the AdS length is comparable to the width of the switching function that we see a dependence of transition probability of detector $B$ on its proper distance from the origin.

%----------------------------------------------------------------------
%----------------------------------------------------------------------

%========================================
%========================================
\subsection{Entanglement harvesting with static detectors }

We now consider harvesting entanglement with detector $A$ at the origin and detector $B$ placed
at various  fixed proper   distances  from $A$, and for now we assume no relative time delay in the detectors' switching functions $(t_0=0)$. The transition probabilities $P_A$ and $P_B$ and matrix element $X$ can be obtained numerically using Eqs.~\eqref{PAPB} and \eqref{X-equation}, after which the concurrence given in Eq.~(\ref{concurrence-eq}) can be easily evaluated as a measure of the resulting entanglement between the detectors. We find that the effects of curvature, spatial separation, and detector energy gap are notably more dramatic for entanglement harvesting as compared to the transition probability, which we shall now demonstrate.

%----------------------------------------------------------------------
\begin{figure}[t]
\includegraphics[width=.9\linewidth]{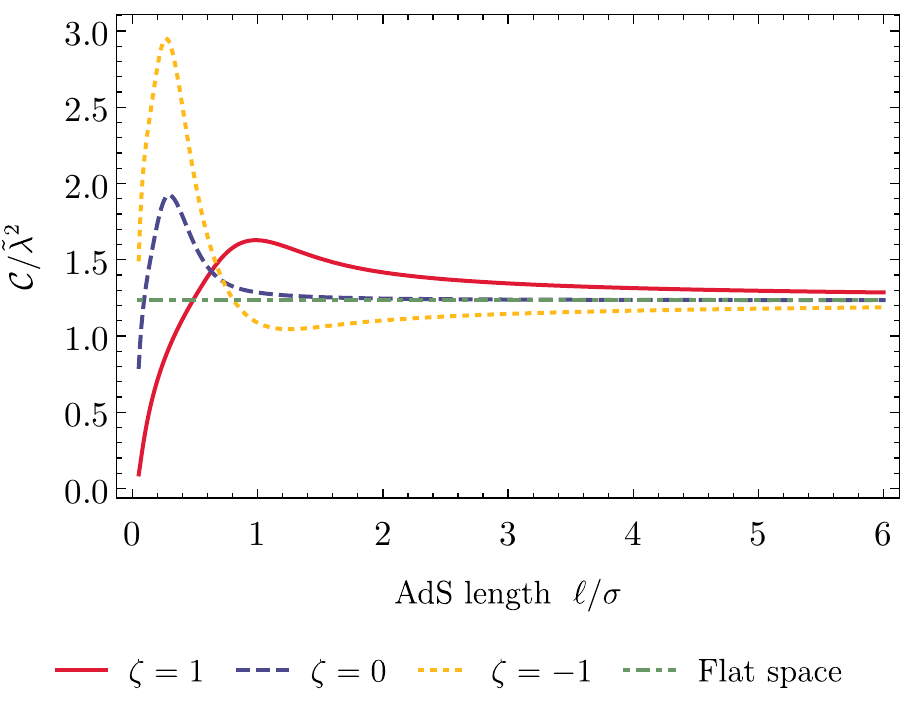}
\caption{The concurrence $\mathcal{C}/\tilde{\lambda}^2$ associated with the state $\rho_{AB}$ describing two static detectors is plotted as a function of the AdS length $\ell/\sigma$ for all boundary conditions $\zeta =\{1,0,-1\}$. The proper separation of the detectors is $d(R_A,R_B)/\sigma=1/10$, their energy gap is $\Omega\sigma=1/100$, and detector $A$ is located at the origin.}\label{Negativity-ell}
%\end{figure}
%%----------------------------------------------------------------------
%
%%----------------------------------------------------------------------
%\begin{figure}[h!]
\ \\
\includegraphics[width=.9\linewidth]{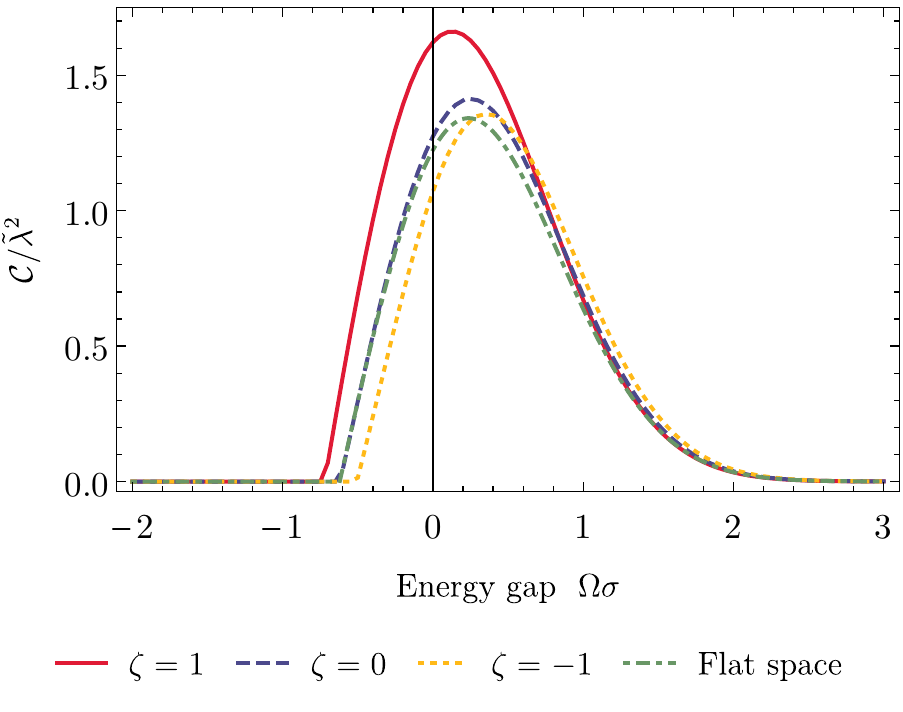}
\caption{The concurrence $\mathcal{C}/\tilde{\lambda}^2$ associated with the state $\rho_{AB}$ describing two static detectors is plotted as a function of the detectors' energy gap  $\Omega \sigma$ for all boundary conditions $\zeta =\{1,0,-1\}$. The AdS length is chosen to be $\ell/\sigma=1$, the detectors are separated by a proper distance of $d(R_A,R_B)/\sigma=1/10$, and detector $A$ is located at the origin.}
\label{Negativity-omega}
\end{figure}
%----------------------------------------------------------------------

%----------------------------------------------------------------------
\begin{figure}[]
  \subfloat[$d(R_A,R_B)/\sigma=1/10$]{%
    \includegraphics[width=.9\linewidth]{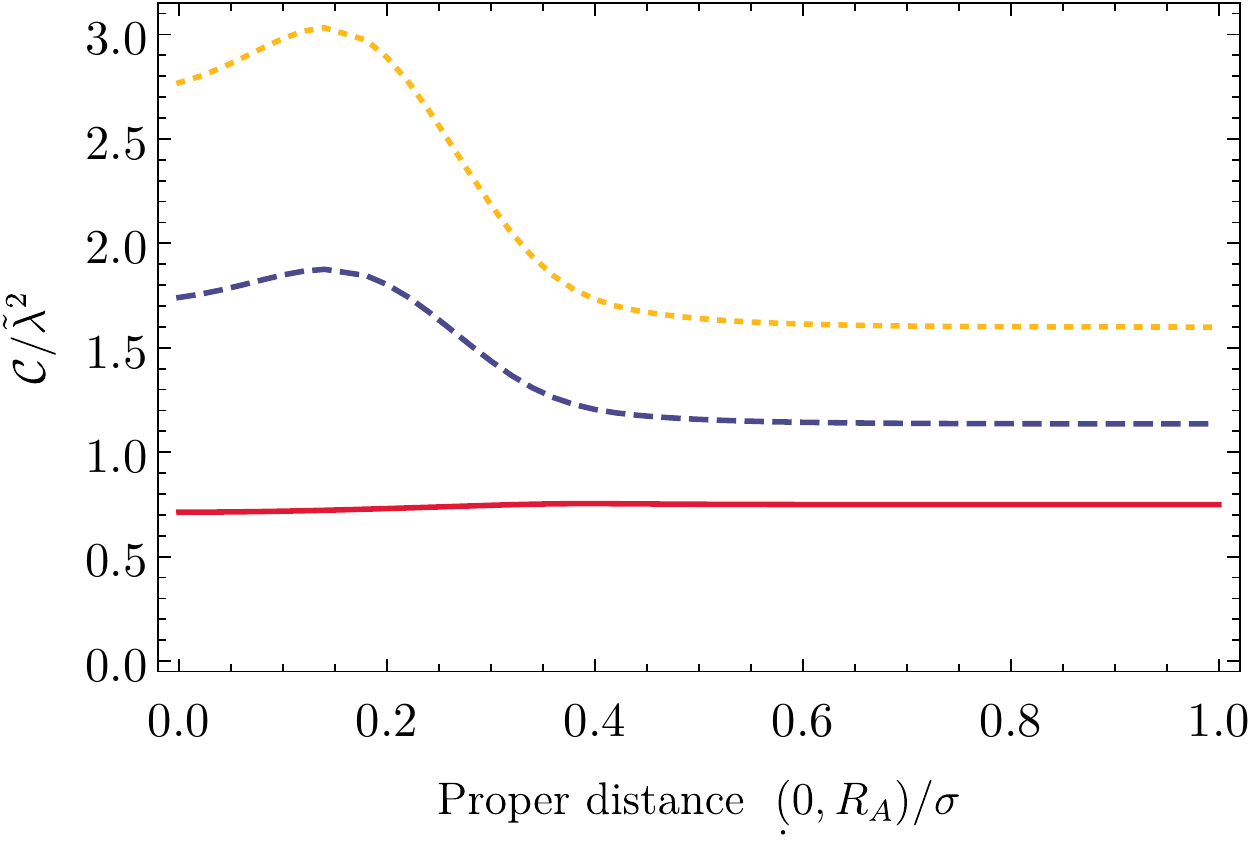}%
  } \\
  \subfloat[$d(R_A,R_B)/\sigma=1/2$]{%
    \includegraphics[width=.9\linewidth]{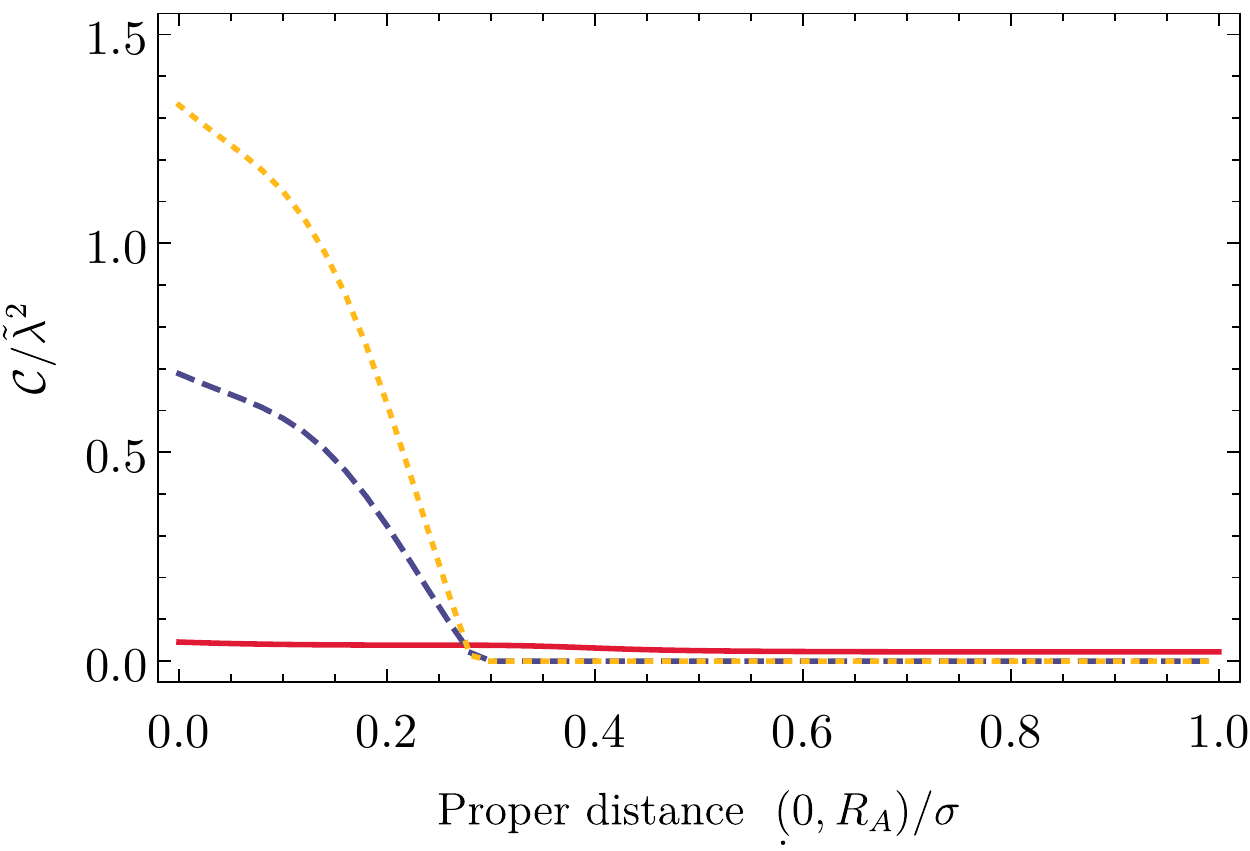}%
  }\\
  \subfloat[$d(R_A,R_B)/\sigma=1$]{%
    \includegraphics[width=.9\linewidth]{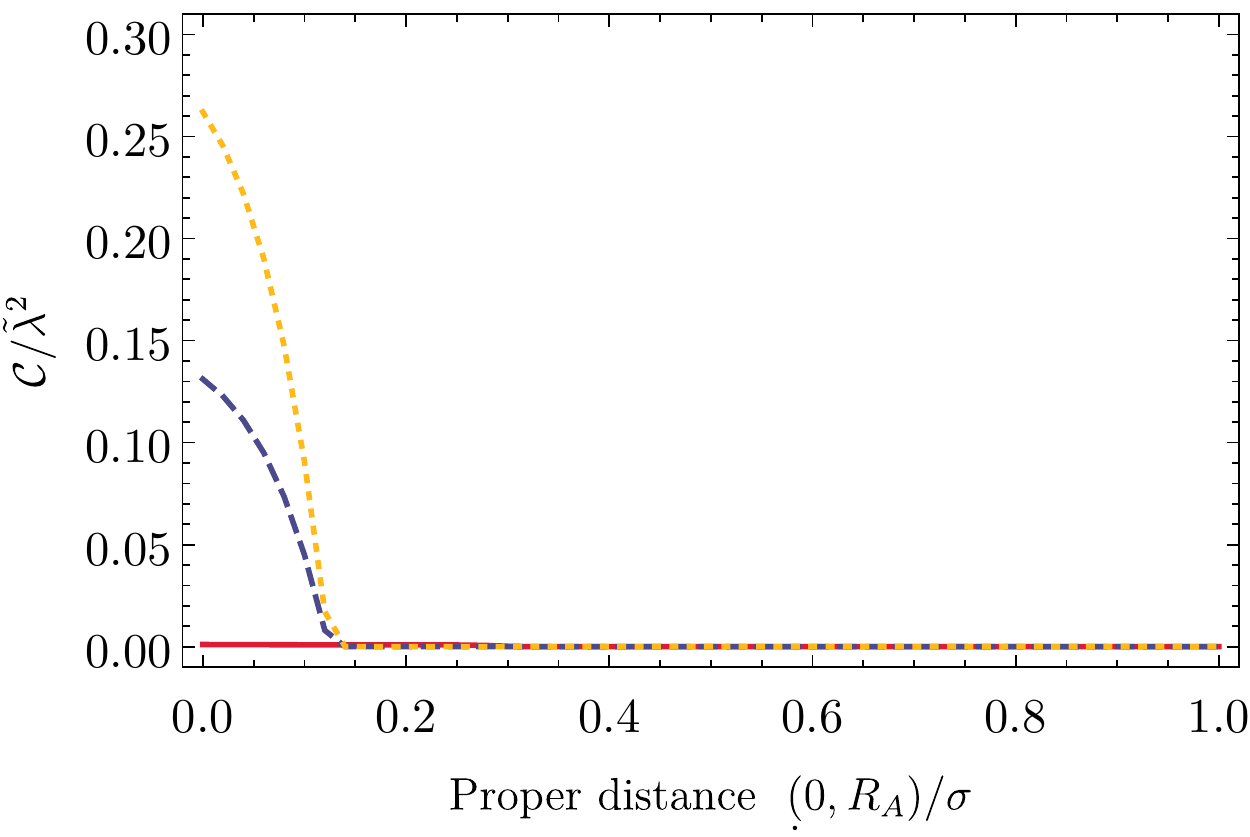}%
  }\\
  \includegraphics[width=.65\linewidth]{legend.pdf}
  \caption{%
    The concurrence $\mathcal{C}/\tilde{\lambda}^2$ associated with the state $\rho_{AB}$ describing two static detectors is plotted as a function of the proper distance $d(0,R_A)$ detector $A$ is from the origin, for different proper separations $d(R_A,R_B)$ between the detectors; all boundary conditions $\zeta =\{1,0,-1\}$ are shown. The AdS length is chosen to be $\ell/\sigma =5$ and the energy gap of the detectors is $\Omega\sigma=1/100$.
    }
  \label{fig:StatCvsPropRA}
\end{figure}
%----------------------------------------------------------------------

%----------------------------------------------------------------------
\begin{figure*}[t]
\subfloat[$\zeta =-1$ and $\ell/\sigma = 1/2$]{%
  \includegraphics[width=.3\linewidth]{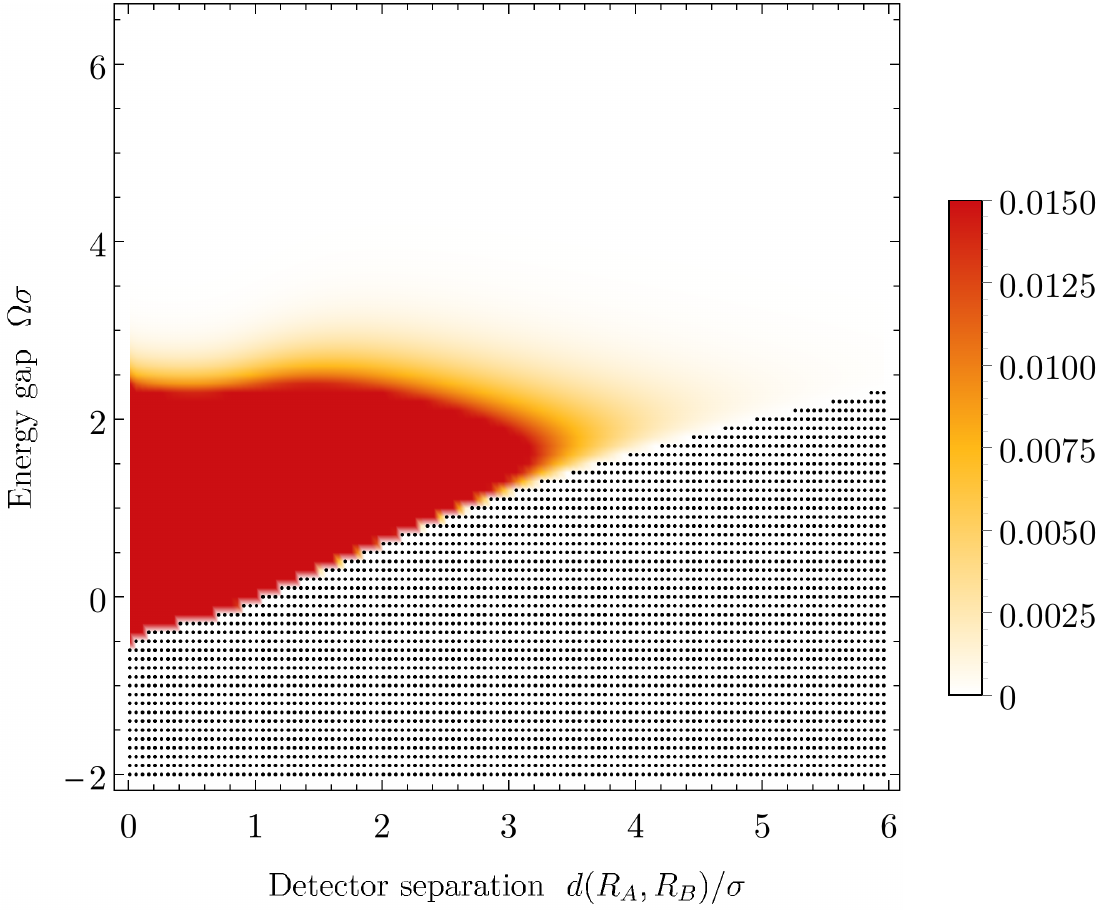}%
}%
\quad
\subfloat[$\zeta =-1$ and $\ell/\sigma = 5/2$]{%
  \includegraphics[width=.3\linewidth]{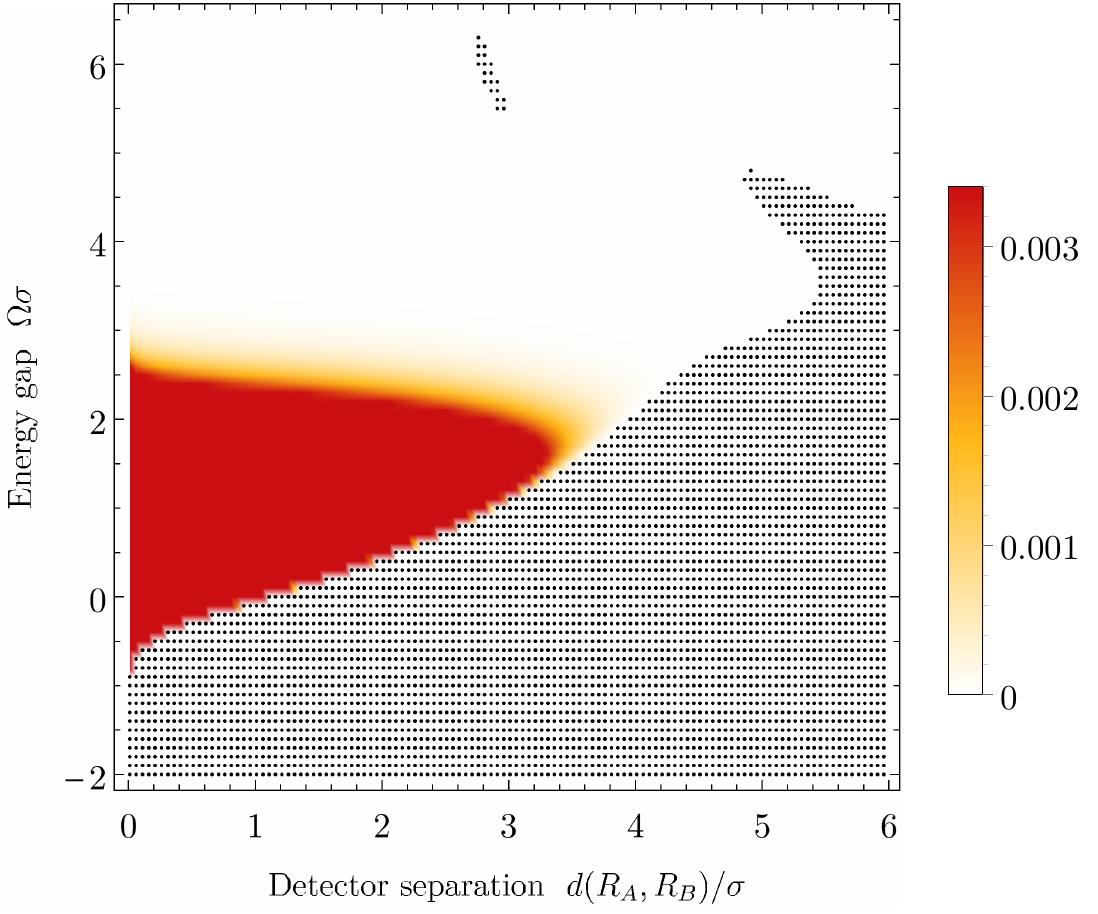}%
}%
\quad
\subfloat[$\zeta =-1$ and $\ell/\sigma = 20$]{%
  \includegraphics[width=.3\linewidth]{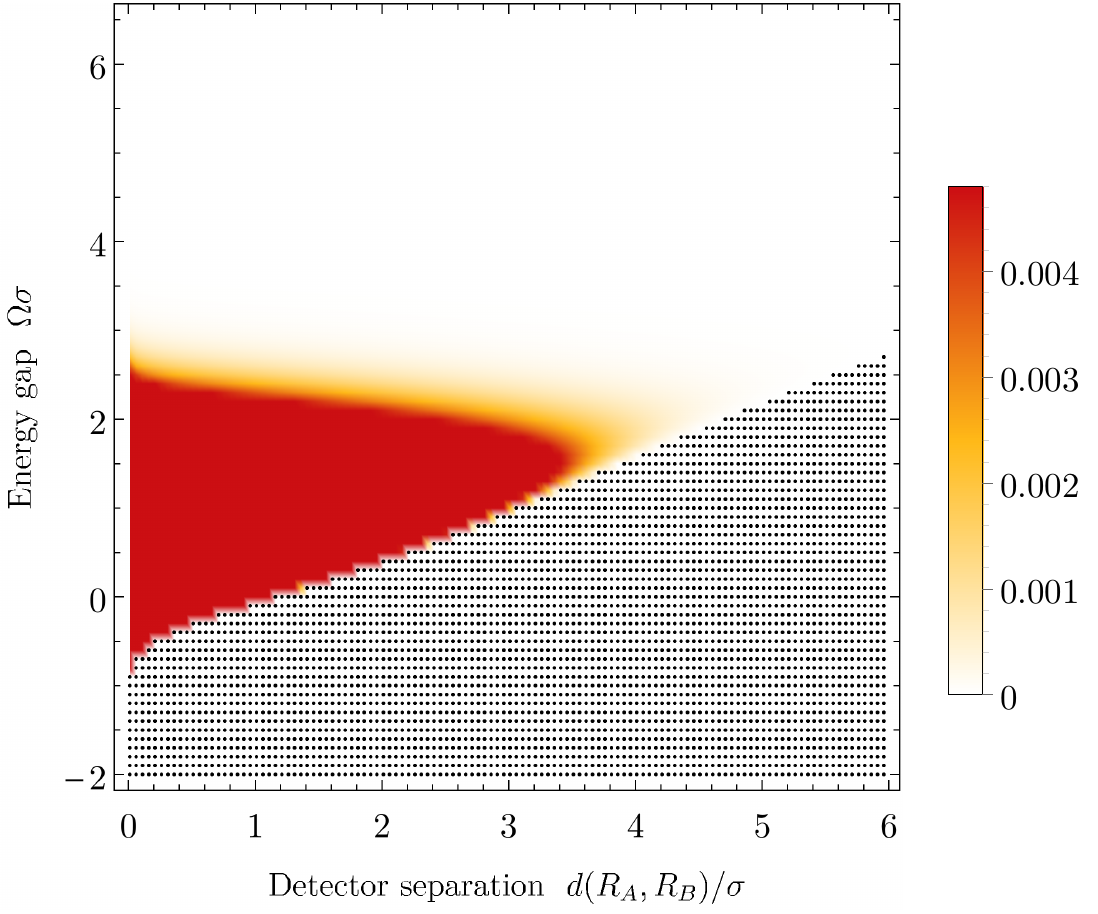}%
}%
\\ %NEWLINE
\subfloat[$\zeta = 0$ and $\ell/\sigma = 1/2$]{%
  \includegraphics[width=.3\linewidth]{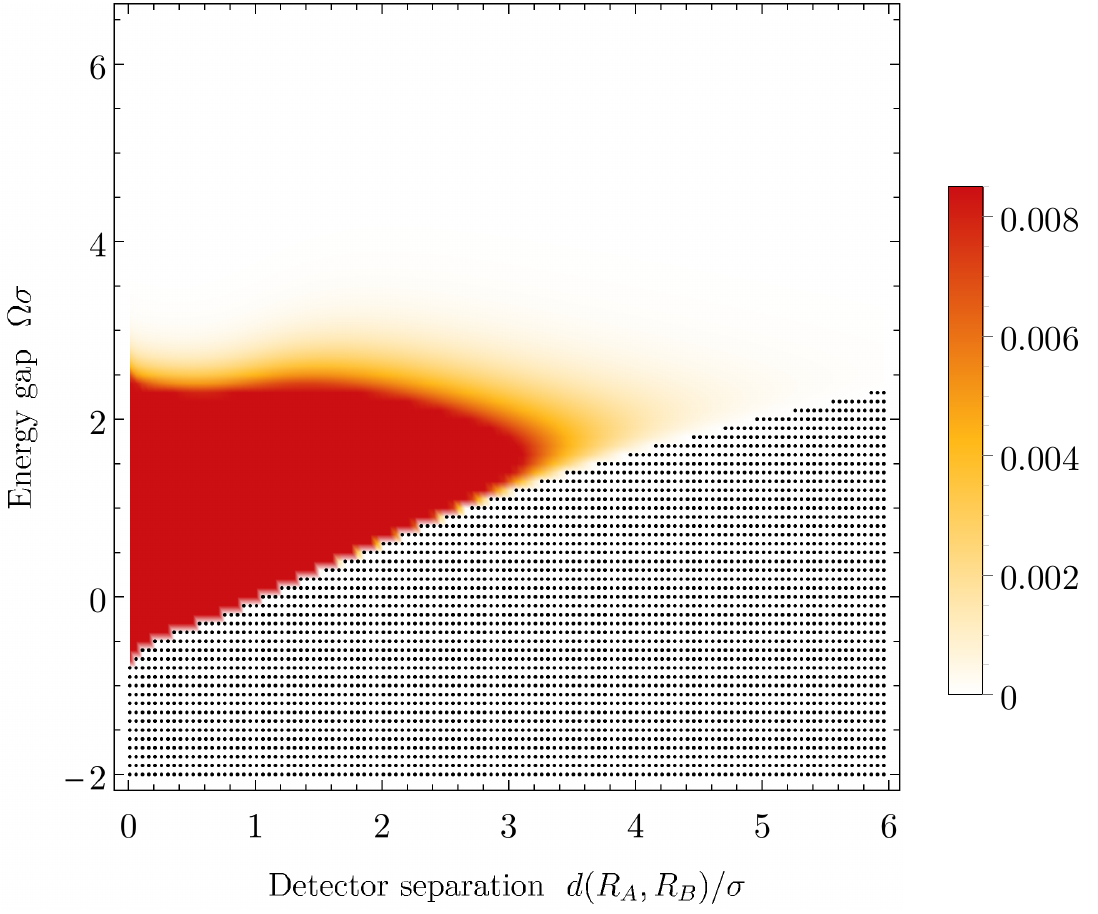}%
}%
\quad
\subfloat[$\zeta = 0$ and $\ell/\sigma = 5/2$]{%
  \includegraphics[width=.3\linewidth]{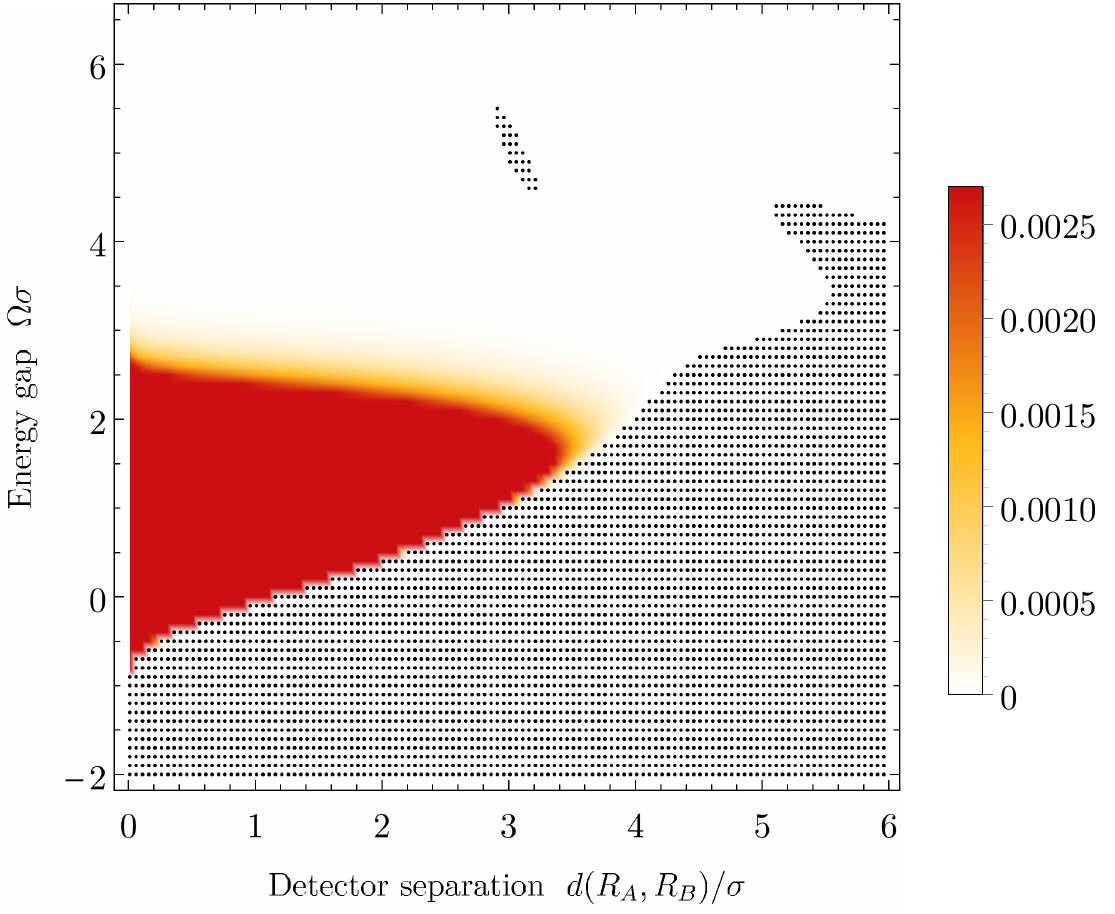}%
}%
\quad
\subfloat[$\zeta = 0$ and $\ell/\sigma = 20$]{%
  \includegraphics[width=.3\linewidth]{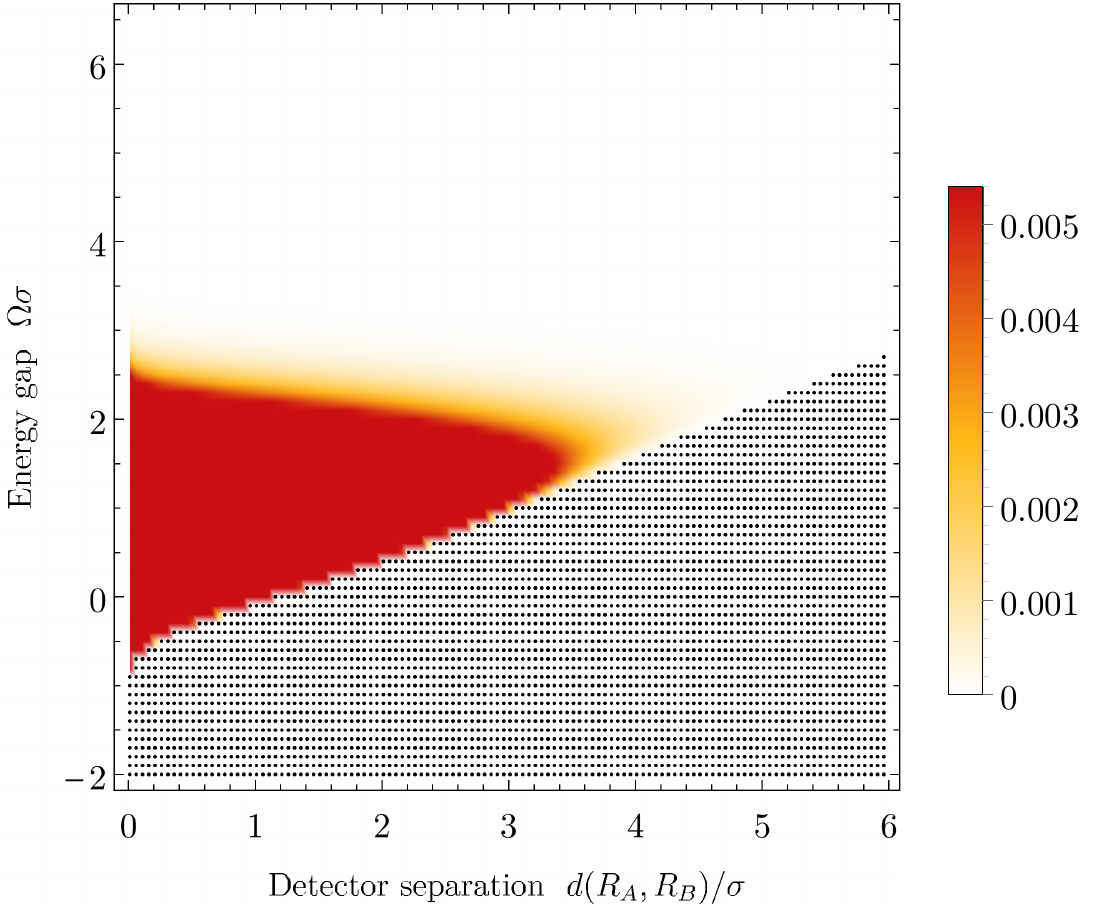}%
}%
\\ %NEWLINE
\subfloat[$\zeta = 1$ and $\ell/\sigma = 1/2$ ]{%
  \includegraphics[width=.3\linewidth]{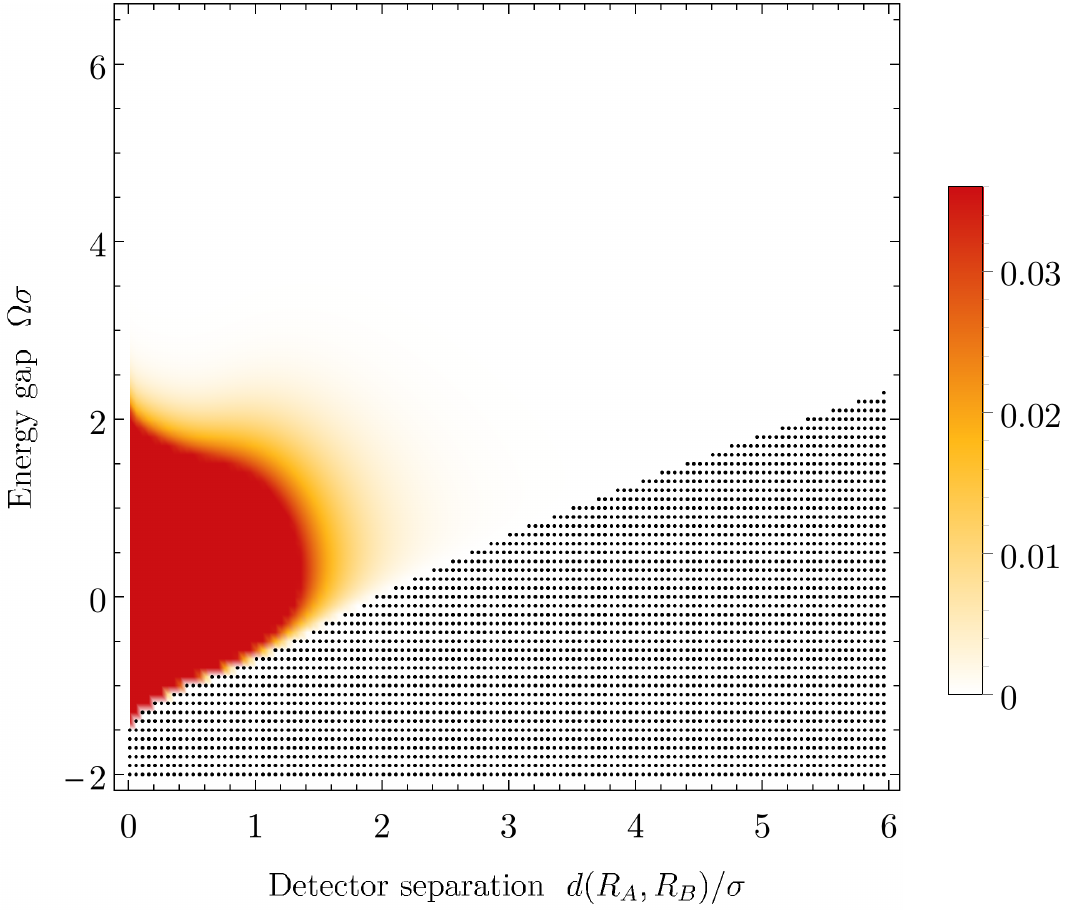}%
}%
\quad
\subfloat[$\zeta = 1$ and $\ell/\sigma = 5/2$]{%
  \includegraphics[width=.3\linewidth]{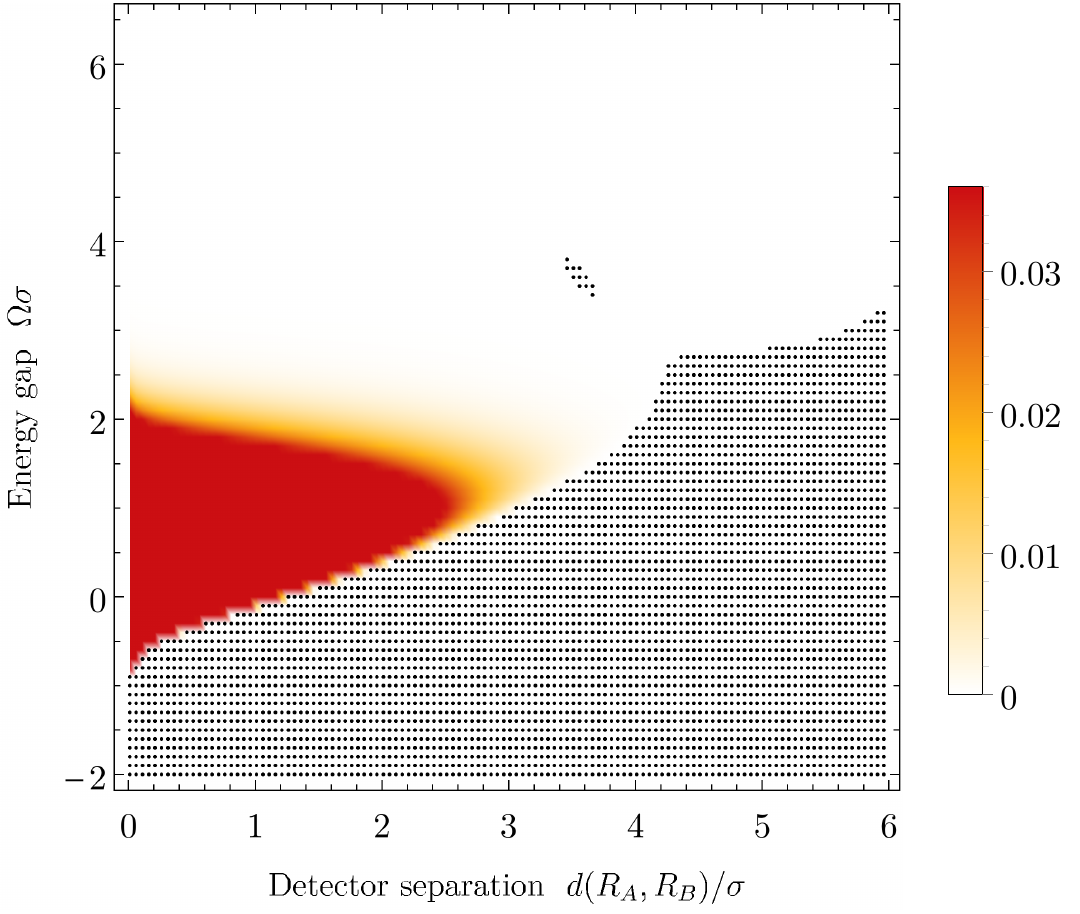}%
}%
\quad
\subfloat[$\zeta = 1$ and $\ell/\sigma = 20$]{%
  \includegraphics[width=.3\linewidth]{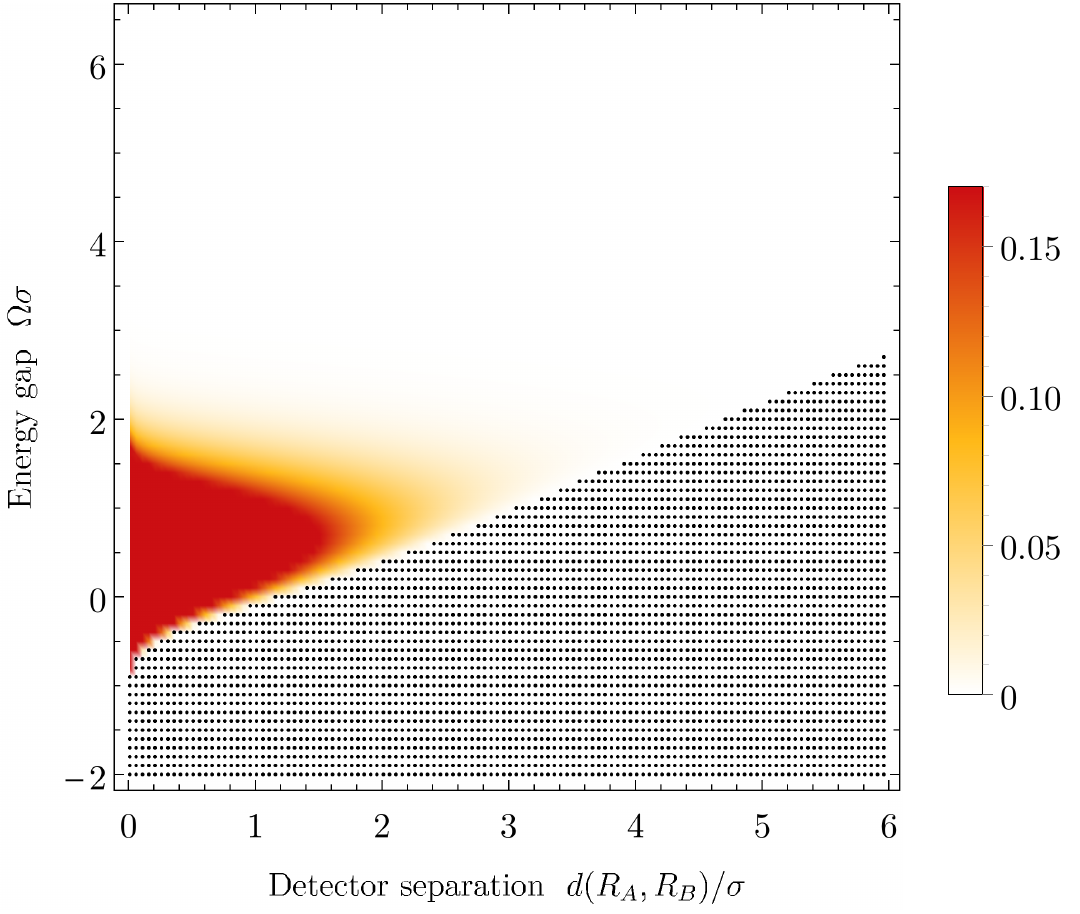}%
}%

\caption{The concurrence $\mathcal{C}/\tilde{\lambda}^2$ associated with the state $\rho_{AB}$ describing two static detectors is plotted as a function of their proper separation $d(R_A, R_B)/\sigma$ and energy gap $\Omega \sigma$ for all boundary conditions $\zeta = \{-1,0,1\}$ and different values of the AdS length $\ell/\sigma$. Detector $A$ is located at the origin. The area filled with black dots represents the region where the concurrence vanishes and thus no entanglement harvesting is possible.
}
\label{densityplot1}
\end{figure*}
%----------------------------------------------------------------------

We first consider how the concurrence depends on the AdS length $\ell/\sigma$. From Fig.~\ref{Negativity-ell} we see that the concurrence vanishes for all boundary conditions as $\ell/\sigma\to 0$. The concurrence attains its maximum value in the interval $0.5\lesssim \ell/\sigma \lesssim1$, and this maximum is largest if the field satisfies Dirichlet boundary conditions ($\zeta=1$).  We also note that as $\ell$ grows, the concurrence asymptotes to the flat space value for all  boundary conditions, as expected.

 Similar peaking behaviour occurs as the energy gap is changed for fixed $\ell$.
 In Fig.~\ref{Negativity-omega}  we observe a peak in concurrence  for positive $\Omega\sigma$ for each boundary condition, the peak again being largest  for the Dirichlet boundary condition for the value of $\ell/\sigma$ considered.
 %{\bf [Based on fig. 4 and 5 it looks like a different boundary condition in each case has the highest peak.]}
For initially excited detectors ($\Omega < 0$) the decrease in concurrence with increasing $|\Omega|$  is much more rapid than for detectors initially in their ground states. Hence, similar to what we found for $(2+1)$-dimensional flat space (see Appendix~\ref{EH in flat spacetime}), it is  easier to perform entanglement harvesting   using  detectors in their  ground-state instead of their excited-state.

%\tcb{\bf [Comparison to flat space case for Fig.~(5), analogous to what was done for Fig(4), should be added to the text.]}

%Hence, similar to what we found for $(2+1)$-dimensional flat space (see Appendix B), it is  easier to perform entanglement harvesting   using  detectors in their  ground-state instead of their excited-state. \tcb{\bf [Comparison to flat space case for Fig.~(5), analogous to what was done for Fig(4), should be added to the text.]}

Next, we analyze the dependence of the concurrence on the proper distance $d(0,R_A)$ detector $A$ is from the origin for a fixed proper separation $d(R_A,R_B)$ of the detectors. Figure~\ref{fig:StatCvsPropRA} shows that the large $d(0,R_A)$ behaviour changes depending on the proper separation of the detectors.  When the detectors are close together, the concurrence approaches a constant value, but for larger separations, the concurrence rapidly falls to zero with increasing distance from the origin.  We also note that the concurrence is maximum when the detectors are close to the origin.  In this region they will experience a smaller acceleration and a lower transition probability, which likely leads to the larger value of the concurrence as seen from Eq.~\eqref{concurrence-eq}.

Finally, we consider the behaviour of the concurrence for different proper separations $d(R_A,R_B)$ between the two detectors. From the plots shown in Fig.~\ref{fig:StatCvsPropRA} we see that that the larger the value of $d(R_A,R_B)$, the less the concurrence, commensurate with analogous results in flat spacetime and consistent with ones expectations from the fall off of the Wightman function in Eq.~\eqref{wightmanf} as the spatial distance between $x$ and $x'$ grows. For Dirichlet boundary conditions ($\zeta=1$) the decay rate is slowest and the complete elimination of entanglement (sudden death) occurs at the largest proper separation, with the opposite true for Neumann boundary conditions ($\zeta=-1$).

Clearly the concurrence exhibits interesting behaviour as a function of AdS length $\ell$, and the detectors' energy gap $\Omega$ and proper separation $d(R_A,R_B)$. To illustrate this more fully, we now provide density plots of the concurrence as a function of the detectors' energy gap $\Omega$ and proper separation for different boundary conditions in Fig.~\ref{densityplot1}.  The shaded areas in each plot are regions in the parameter space where no entanglement harvesting is possible. We do not plot values of $\ell/\sigma > 20$ as we have found no appreciable quantitive changes for these values of $\ell/\sigma$.

We see from Fig.~\ref{densityplot1} several common features for all boundary conditions over the range of $\ell/\sigma$.  Regions of large concurrence are always in the lower-left corner (smaller energy gap and smaller detector separation), and regions of zero concurrence are in the lower-right corner (smaller energy gap and larger separation). We note also that entanglement harvesting is generally possible at any detector separation for sufficiently large energy gap, albeit in minuscule amounts compared to smaller energy gaps and detector separation. We also
note that the boundary of entanglement in the $\{d(R_A,R_B)/\sigma,\Omega\sigma\}$ parameter space is approximately a straight line for
large enough $\ell$, a feature observed in $(3+1)$-dimensions~\cite{KRE}, and also in flat space in all spatial dimensionalities $D \leq 3$, noted in~\cite{Pozas-Kerstjens:2015}, and which we have observed in $(2+1)$-dimensions (though we have not displayed the graph).

Distinct features arise when we consider the boundary of the zero-entanglement region (shaded region in Fig.~\ref{densityplot1}). The intersection point of the boundary of this region with a horizontal line at $\Omega\sigma= 0$ slowly moves to smaller values of $d(R_A,R_B)/\sigma$ as $\ell/\sigma$ gets larger. More interestingly is the behaviour of this boundary as a function of $\ell/\sigma$.  We see for intermediate values of $\ell/\sigma$ that the shape of this region changes significantly for all boundary conditions at large detector separations.

\begin{figure}[t]
\subfloat[]{%
  \includegraphics[width=.9\linewidth]{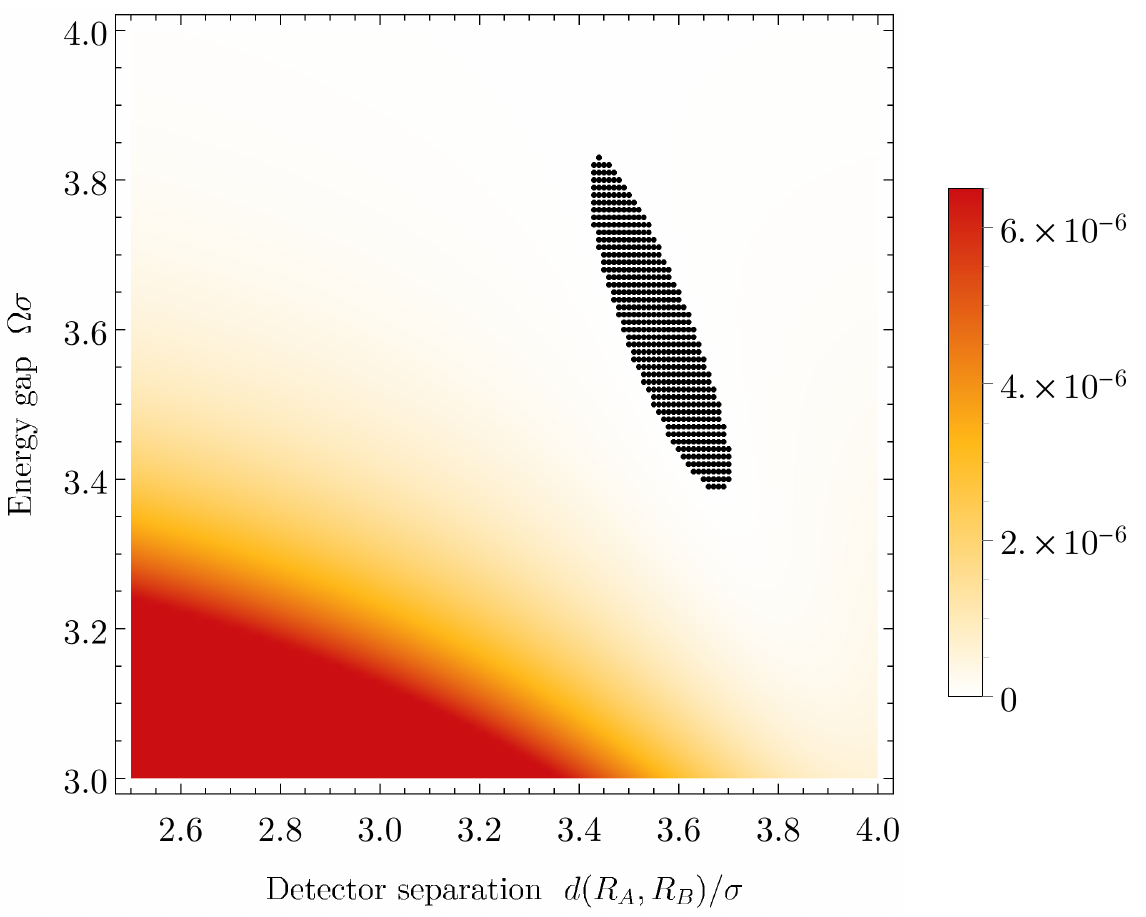}%
}%
 \\
\subfloat[]{%
  \includegraphics[width=.9\linewidth]{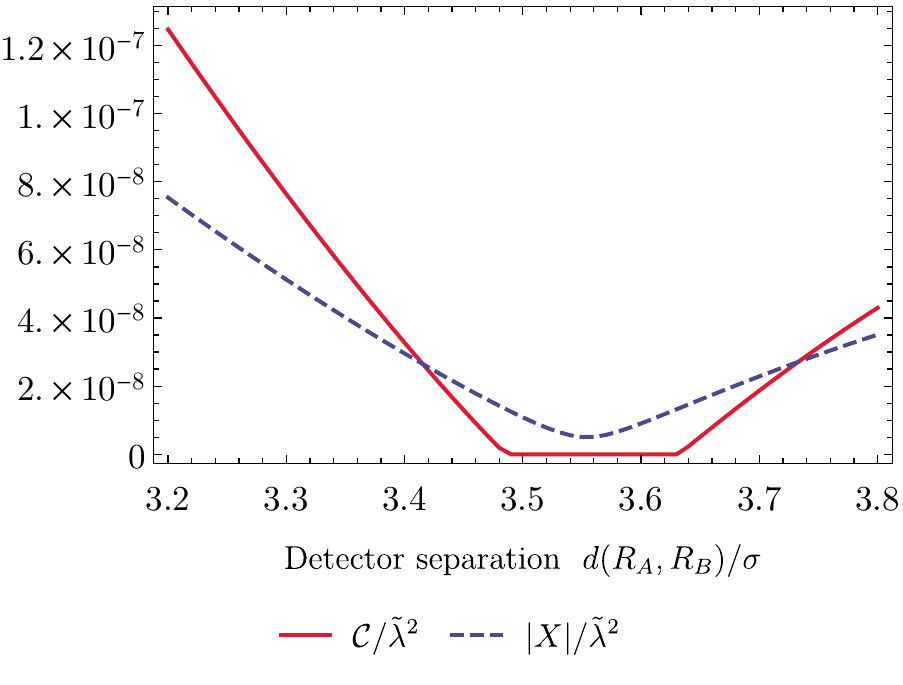}%
}
\caption{ For Dirichlet boundary conditions ($\zeta=1$) and an AdS length of $\ell/\sigma=5/2$, plot (a) depicts a close-up of the separability island in the parameter space spanned by proper detector separation $d(R_A,R_B)/\sigma$ (with $A$ at the origin) and energy gap $\Omega\sigma$ in which no entanglement harvesting is possible; and in (b) the concurrence $\mathcal{C}/\tilde{\lambda}^2$ and the absolute value of the matrix element $X$ are plotted as a function of the detectors proper separation $d(R_A,R_B)/\sigma$ for detectors with an energy gap of $\Omega\sigma=3.6$.  It is easy to see that the vanishing region of concurrence is located about the local minimum
of  $|X|$.  }
\label{islandplot}
\end{figure}

Most intriguingly is the emergence of an ``island" of no entanglement harvesting for intermediate values of $d(R_A,R_B)/\sigma$ and positive values of the detectors energy gap $\Omega \sigma$ for all boundary conditions (see plots (b), (e), and (h) in Fig.~\ref{densityplot1}); we will refer to this region as a separability island. In order to see this separability island clearly, we provide a close-up view of this region for Dirichlet boundary conditions ($\zeta=1$) in Fig.~\ref{islandplot}; other boundary conditions yield qualitatively similar plots. We see that the island is in the region $3.4 \lesssim \Omega\sigma \lesssim 3.85$.
%{The plot of concurrence indicates a small  oscillation with   increasing $d(R_A,R_B)/\sigma$ in this region.}  \tcb{\bf [Presumably the plot being referred to here is Fig.~8 (b), in which I don't seen any oscillation.]}

This behaviour is due to the matrix element $X$. Over this region, as $d(R_A,R_B)$ increases,  $X$ decreases to a minimum and then begins to increase again, while $P_B$ remains nearly constant throughout. Considering the expression for $X$, and recalling that $\Delta_T=0$ in Eq.~\eqref{X-equation}, specifically the term appearing in the numerator $e^{-a_Xy^2} \cos \left( \beta_X u\right)$, and the cosine factor $\cos\beta_X$ in Eq.~\eqref{Kxeq}, we can deduce that when $|\beta_X|\gg{a_X}$ and the corresponding $X$ must oscillate. Otherwise, were $|\beta_X|\ll{a_X},$ then the Gaussian factor $e^{-a_Xy^2}$ would play a dominant role in the integration, suppressing the resulting decrease/increase behaviour.

For larger AdS length, $\ell/\sigma\gg5/2$, $\gamma_B$ approaches $\gamma_A=1$ and in  general  $|\beta_X|\ll{a_X}$, which results in the oscillatory behaviour of $X$ being suppressed. Of course, $\gamma_B$ can increase for detectors with very large proper detector separation $d(R_A,R_B)$ so that  $|\beta_X|\gg{a_X}$; then the factor $K_X\propto\gamma_B^{-1/2}e^{-\Omega^2\sigma^2/2} $  will result in $|X|$ being extremely small, and  tiny fluctuations of $|X|$ will have no impact on the concurrence.

%----------------------------------------------------------------------
%----------------------------------------------------------------------

Conversely, when the AdS length is very small, \mbox{$\ell/\sigma\ll1$},  it is easy to deduce that $\gamma_B\propto{e}^{d(R_A,R_B)/\ell}\gg \gamma_A=1$  and $|\beta_X|\gg{a_X}$ because  $|\beta_X|\propto\Omega\ell$ and ${a_X}\propto\ell^2/(2\sigma^2)$. Therefore, the result of the integration of $X$ can oscillate as a function of the position of detector $B$. However, the factor $K_X$ then becomes approximately
\begin{align}
K_X\propto \exp{\Big[-\frac{d(R_A,R_B)/\sigma}{2\ell/\sigma}-\frac{\Omega^2\sigma^2}{2}\Big]}
\end{align}
where we have used Eq.~\eqref{properdistance} in arriving at the above expression. Despite oscillations in $X$, the above behaviour of $K_X$ highly suppresses their intensity, and so such behaviour is negligible. Similar conclusions can also be obtained for transparent ($\zeta = 0$) and Neumann ($\zeta = -1$) boundary conditions.

Summarizing, the shape of the separability island in the $\{ d(R_A,R_B)/\sigma,\Omega\sigma \}$ parameter space  depends sensitively on the   AdS length $\ell$, and vanishes for values of  $|\ell/\sigma - 2.5| \gtrsim 1 $.  Outside this region entanglement harvesting is possible because of the influence of $K_X$ and $a_X$ on the matrix element $X$.
We also note that a similar island has been observed in the same general region of parameter space in
$(3+1)$-dimensions~\cite{KRE}.  It is clear that this phenomenon merits further study.
%\onecolumngrid

\begin{figure*}[t]
\subfloat[$\zeta=-1$ and $\ell\sigma=1$]{%
  \includegraphics[width=.3\linewidth]{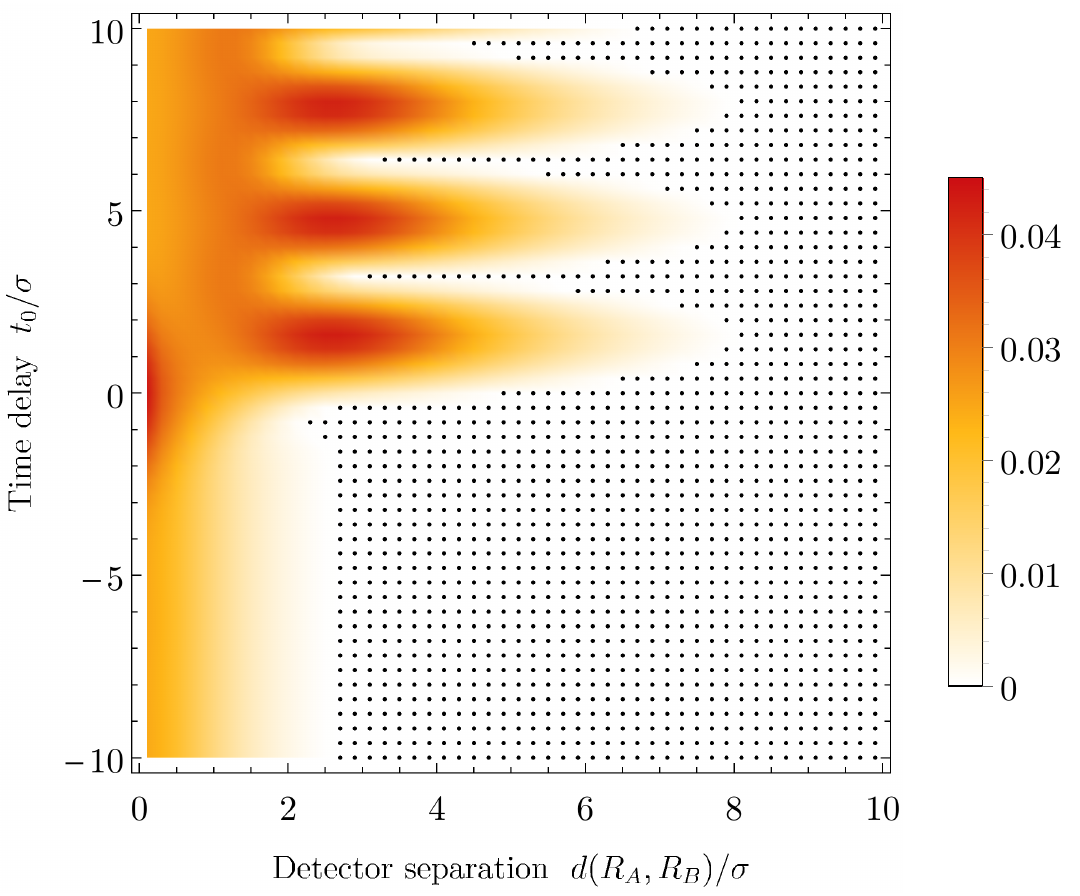}%
}%
\quad
\subfloat[$\zeta=-1$ and $\ell\sigma=5$]{%
  \includegraphics[width=.3\linewidth]{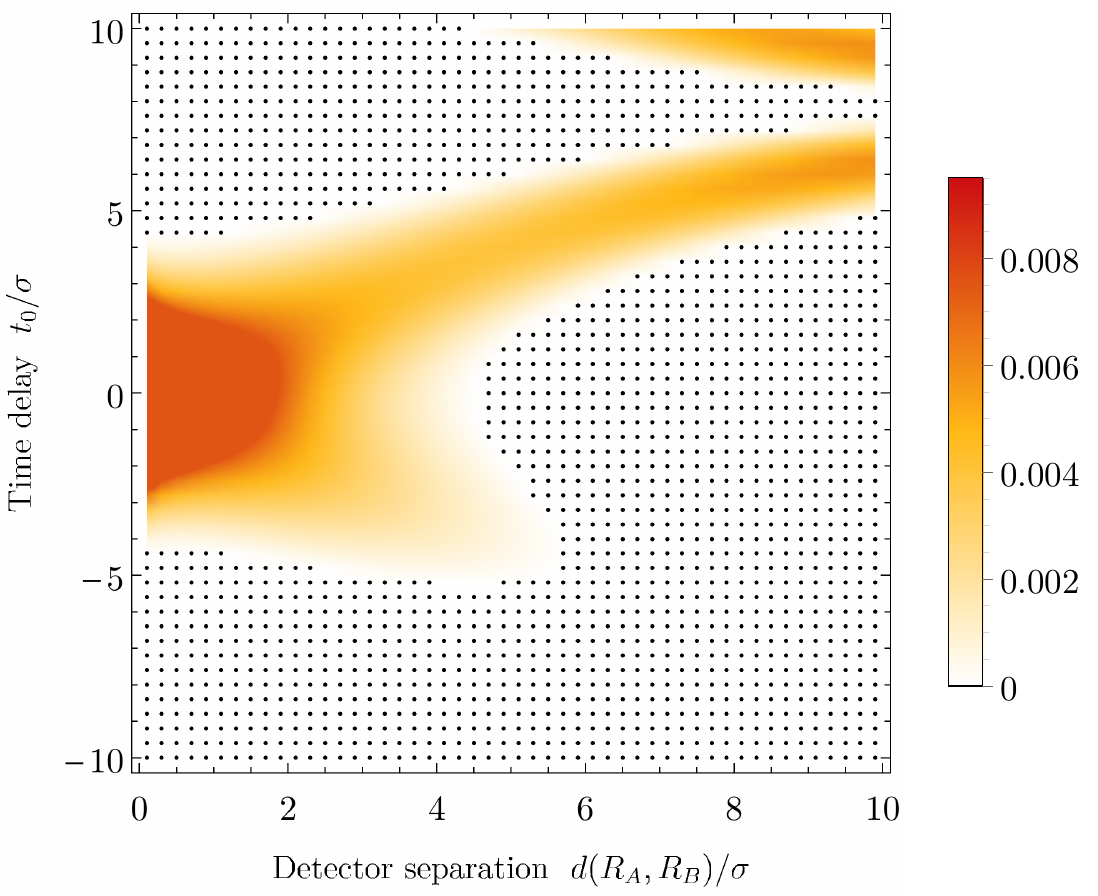}%
}%
\quad
\subfloat[$\zeta=-1$ and $\ell\sigma=20$]{%
  \includegraphics[width=.3\linewidth]{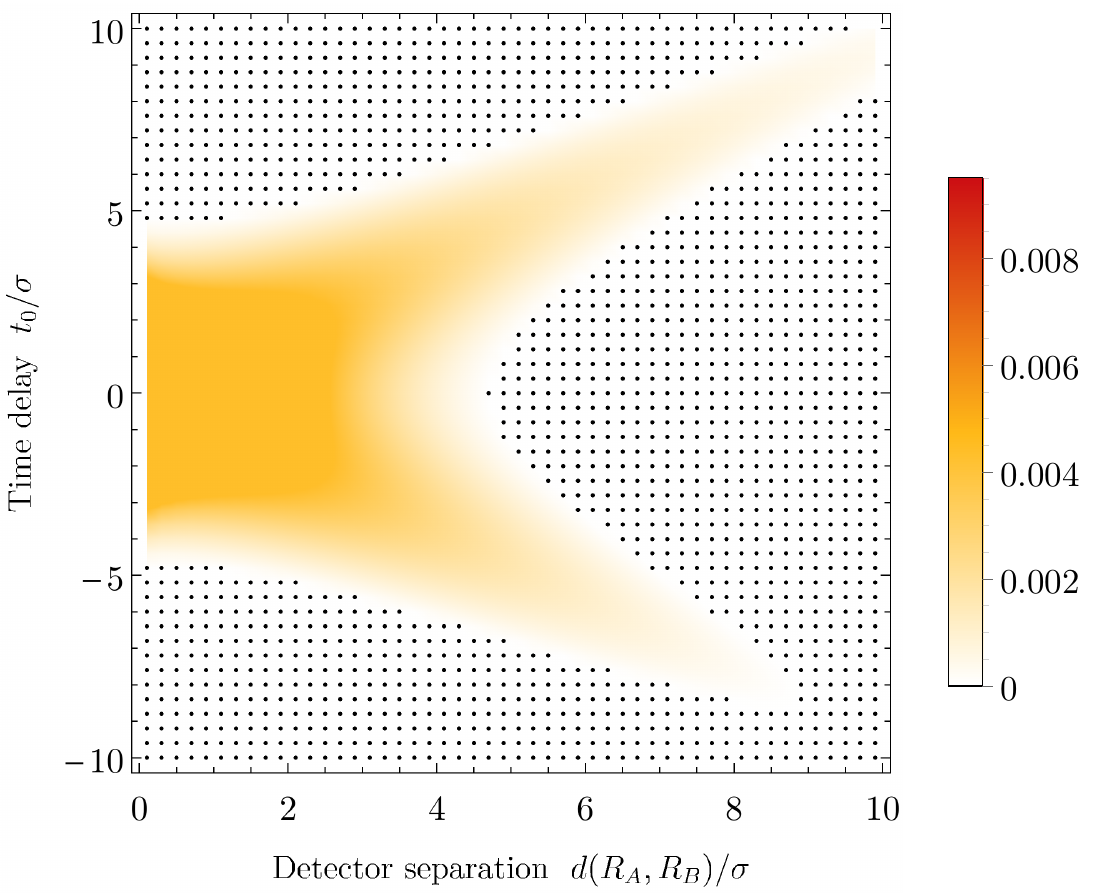}%
}%
\\ %NEWLINE
\subfloat[$\zeta = 0$ and $\ell/\sigma = 1$]{%
  \includegraphics[width=.3\linewidth]{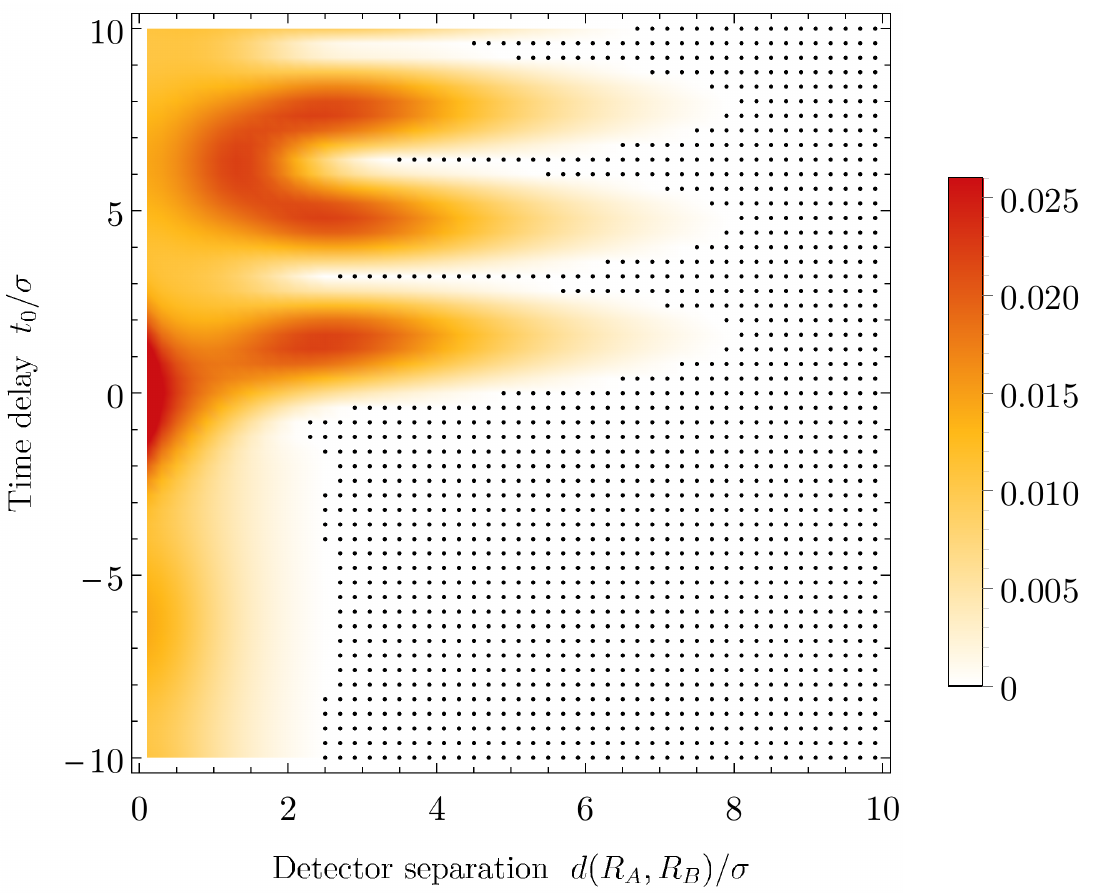}%
}%
\quad
\subfloat[$\zeta = 0$ and $\ell/\sigma = 5$]{%
  \includegraphics[width=.3\linewidth]{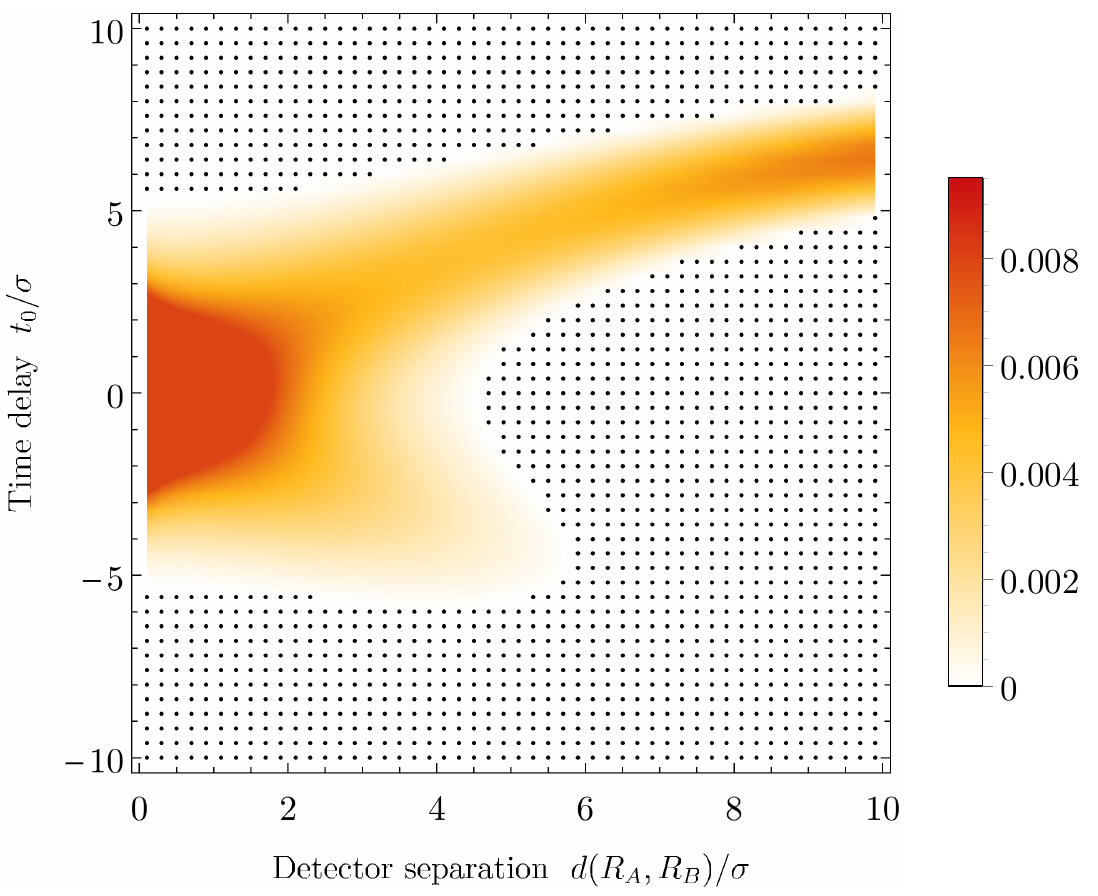}%
}%
\quad
\subfloat[$\zeta = 0$ and $\ell/\sigma = 20$]{%
  \includegraphics[width=.3\linewidth]{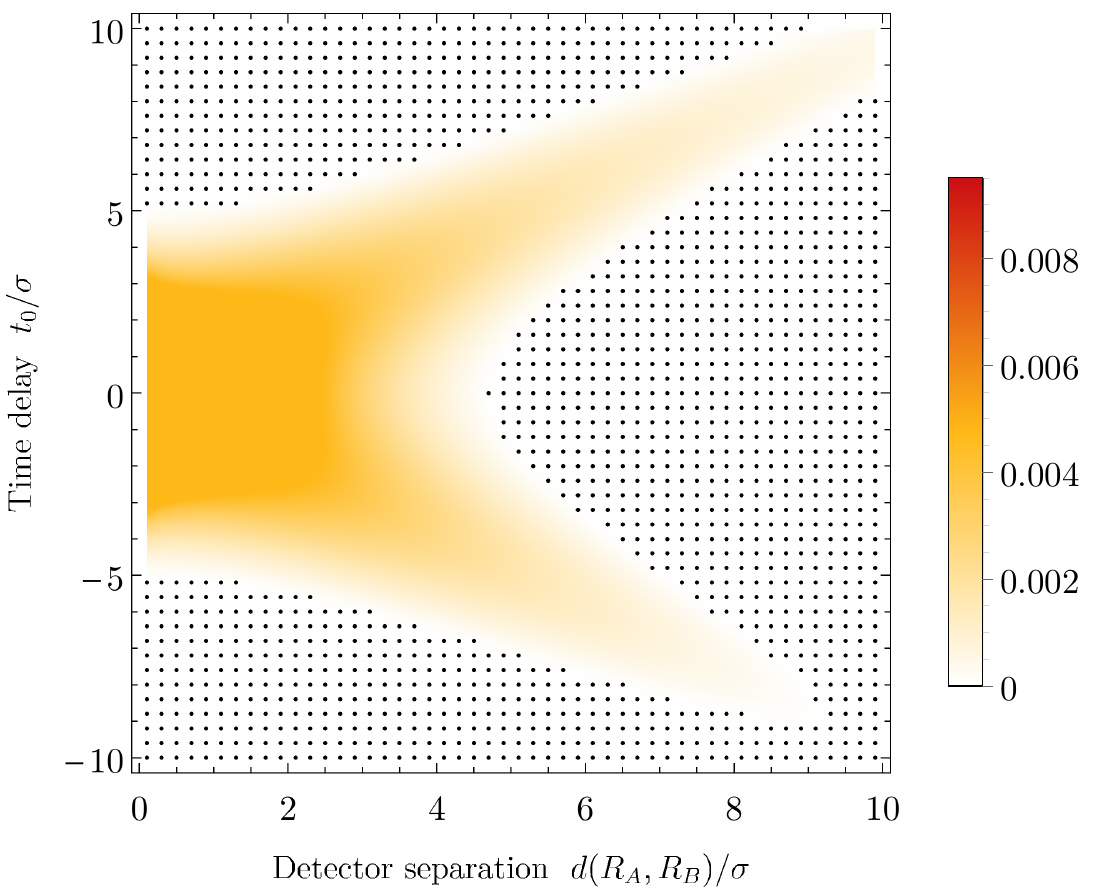}%
}%
\\ %NEWLINE
\subfloat[$\zeta = 1$ and $\ell/\sigma = 1$ ]{%
  \includegraphics[width=.3\linewidth]{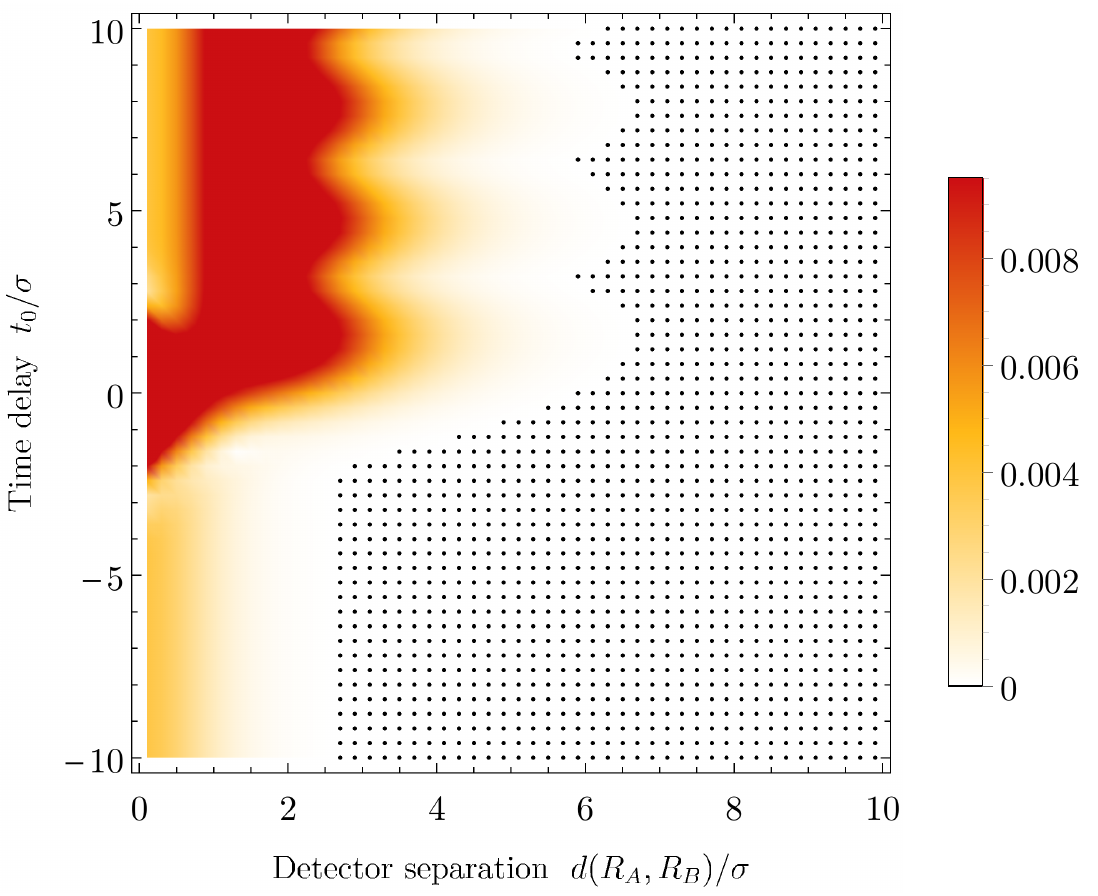}%
}%
\quad
\subfloat[$\zeta = 1$ and $\ell/\sigma = 5$]{%
  \includegraphics[width=.3\linewidth]{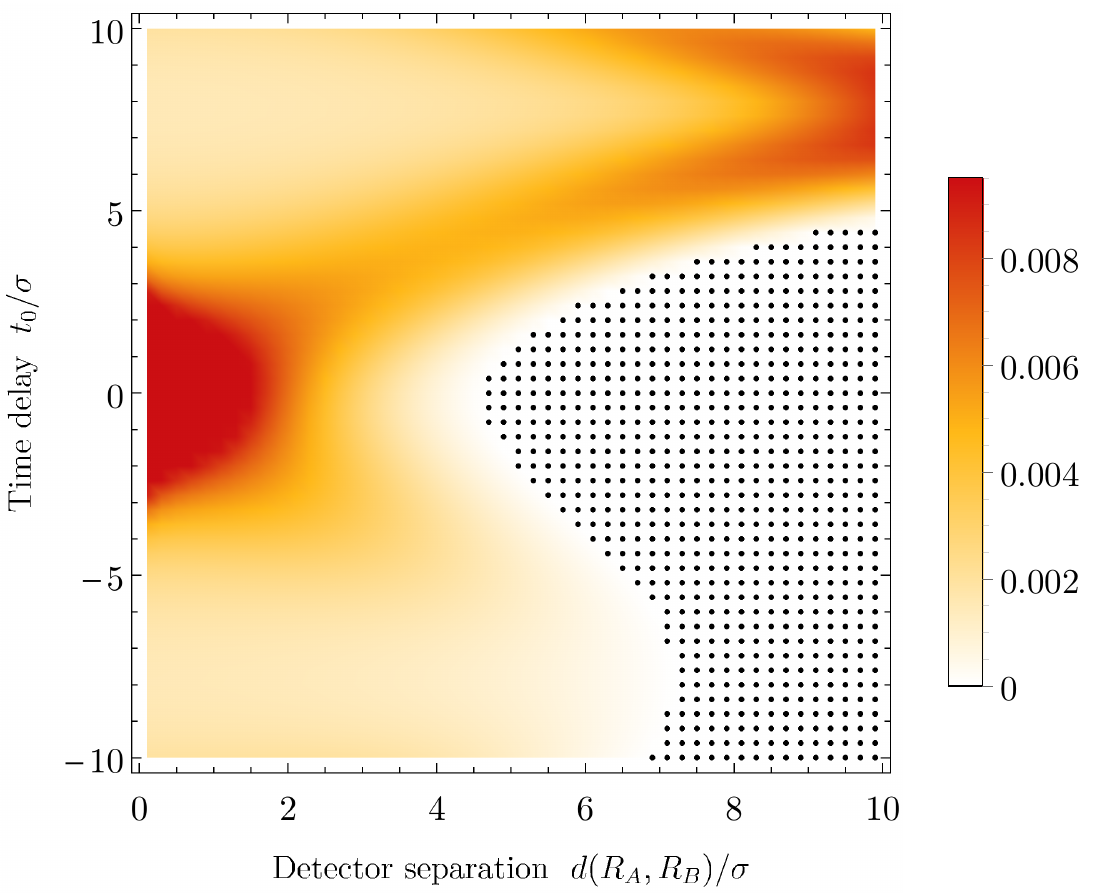}%
}%
\quad
\subfloat[$\zeta = 1$ and $\ell/\sigma = 20$]{%
  \includegraphics[width=.3\linewidth]{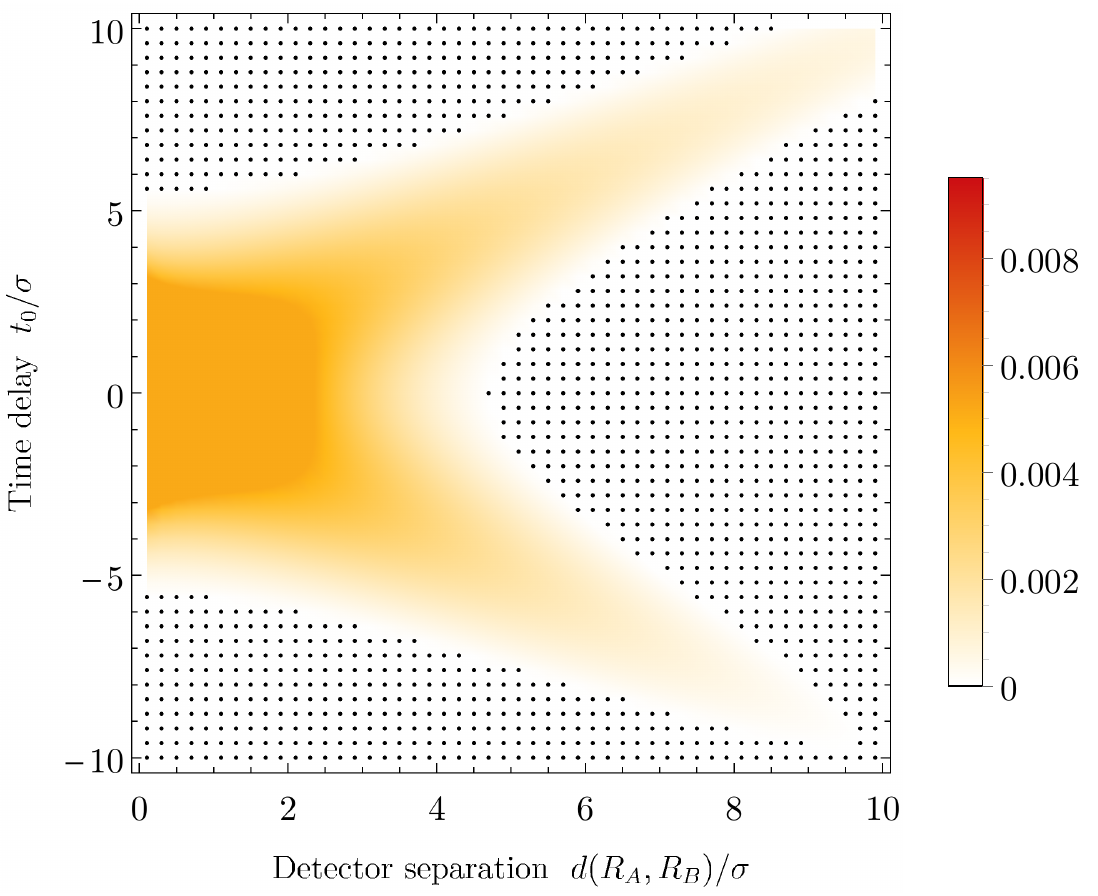}%
}%

\caption{%
  The concurrence, $\mathcal{C}/\tilde{\lambda}^2$ associated with the state $\rho_{AB}$ describing two static detectors is plotted as a function of their proper separation $d(R_A,R_B)/\sigma$ and the relative time delay in their switching functions $t_0/\sigma$ for all boundary conditions $\zeta={-1,0,1}$ and different value of the AdS length $\ell/\sigma$.  A negative $t_0$ means that detector $B$ switches before detector $A$.  Detector A is located at the origin, and the energy gap of the detectors is $\Omega\sigma=2$. The area filled with black dots represents the region where the concurrence vanishes and thus no entanglement harvesting is possible.
}
\label{fig:TimeDelay}
\end{figure*}

%========================================
\subsection{Entanglement harvesting with a time delay}
\label{sub-TimeDelay}

We now consider the case when the switching functions of the two static detectors are offset by some $t_0\ne0$ in the coordinate time $t$.  It is clear from the definition of $X$ in Eq.~\eqref{X-equation} that the concurrence will not be symmetric under the transformation $t_0\to-t_0$; in other words, the amount of entanglement harvested will depend on whether detector $A$ or $B$ interacts with the field first.  Indeed, this can be seen in Fig.~\ref{fig:TimeDelay} where we consider the dependence of concurrence on detector separation and switching delay when the detectors' gaps are equal to $\Omega \sigma = 2$.  The concurrence is non-zero for greater detector separation when detector $A$ switches first.  This effect is most noticeable in the case of Dirichlet $(\zeta=1)$ boundary conditions.

At moderate AdS length ($\ell/\sigma = 5$) we note the formation of  two ``peninsulas'' in the $\{ d(R_A,R_B)/\sigma, t_0/\sigma) \}$ parameter space, which is largest around the region of small time delay and detector separation.  When the AdS length is large $(\ell/\sigma=20)$, the peninsulas are nearly symmetric about $t_0=0$, look very similar for all three boundary conditions, and approach the flat space behaviour.  In the limit of large $\ell$, $\beta_X$ vanishes as $\ell^{-1}$ so that $(\Delta_T+i\beta_X)\to\Delta_T$. Thus in the limit $\ell \to \infty$, $X$ will be even in $t_0$, recovering the Minkowski result.
%\sout{In the large $\ell$ limit $(\Delta_T+i\beta_X)\to\Delta_T$ and $X$ will be approximately even in $t_0$.}

\begin{figure}[t]
\centering{
\includegraphics[width=.9\linewidth]{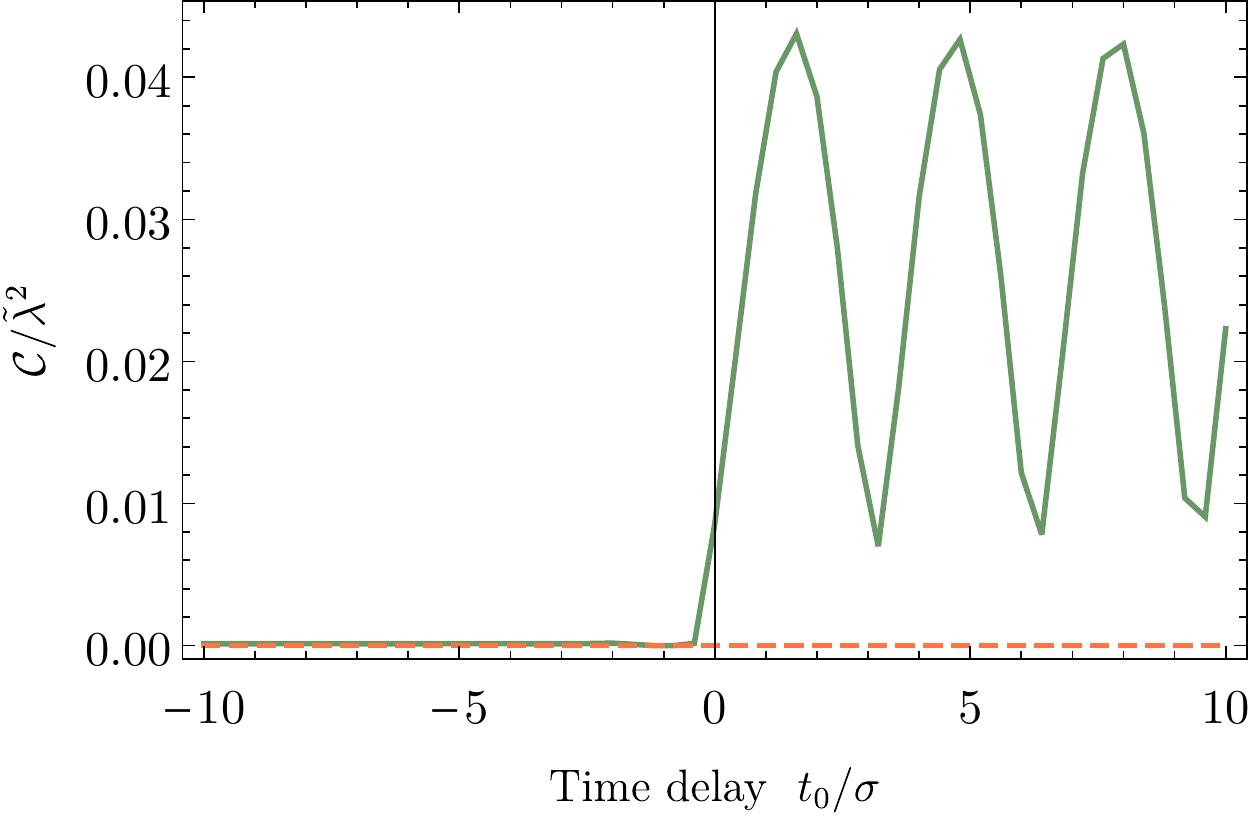}
\\
\includegraphics[width=.45\linewidth]{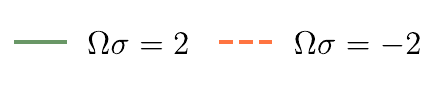}
\caption{%
The concurrence, $\mathcal{C}/\tilde{\lambda}$ associated with the state $\rho_{AB}$ of the two detectors as a function of the the relative time delay in their switching functions $t_0/\sigma$ for an AdS length of $\ell/\sigma=1$, a detector separation of $d(R_A,R_B)=5\sigma/2$, Neumann boundary conditions $(\zeta=-1)$ and an energy gap of $\Omega\sigma=2$ and $\Omega\sigma=-2$.
%\tcp{\textbf{[This figure is not strictly necessary.  It can be removed to save space.]}} \tcb{\bf [I have no strong feelings. I will at least leave it for the others to see]}
}
\label{fig:DelayNegOm}
}
\end{figure}

For small AdS length and small detector separation, the concurrence is also largest around the region of small time delay, but for moderate detector separation, the concurrence is much larger when detector $A$ interacts with the field first.  Again, this effect is seen for all three boundary conditions, but is most exaggerated for Dirichlet $(\zeta=1)$ boundary conditions.  Additionally, in this regime, and when $t_0>0$, we find oscillations
dependent on time delay
in the concurrence that occur for all three boundary conditions, although they are much more pronounced for Neumann $(\zeta=-1)$ and transparent $(\zeta=0)$ boundary conditions where the concurrence goes to zero in the troughs for some parameters.  In the case of Neumann    boundary conditions $(\zeta=-1)$, these oscillations can be seen more clearly in Fig.~\ref{fig:DelayNegOm}.
%----------------------------------------------------------------------
%----------------------------------------------------------------------

 It can be seen from the definition of $X$ in Eq.~\eqref{X-equation} that it will be invariant under the transformation $t_0\to-t_0$ and $\Omega\to-\Omega$, however, as depicted in figure \ref{fig:DelayNegOm}, the concurrence is not.  As illustrated in Fig.~\ref{PAPB-omega}, the transition probability of a detector initialized in its excited state ($\Omega<0$) is much higher than when the detector is initialized in its ground state, meaning in the former case, the term $\sqrt{P_A P_B}$ dominates over $|X|$ in Eq.~\eqref{concurrence-eq} and the concurrence is zero.

%----------------------------------------------------------------------
%----------------------------------------------------------------------

%----------------------------------------------------------------------
\begin{figure*}[t!]
\centering{
\subfloat[$\zeta=-1$]{\label{aa1}
\includegraphics[width=0.3\textwidth]{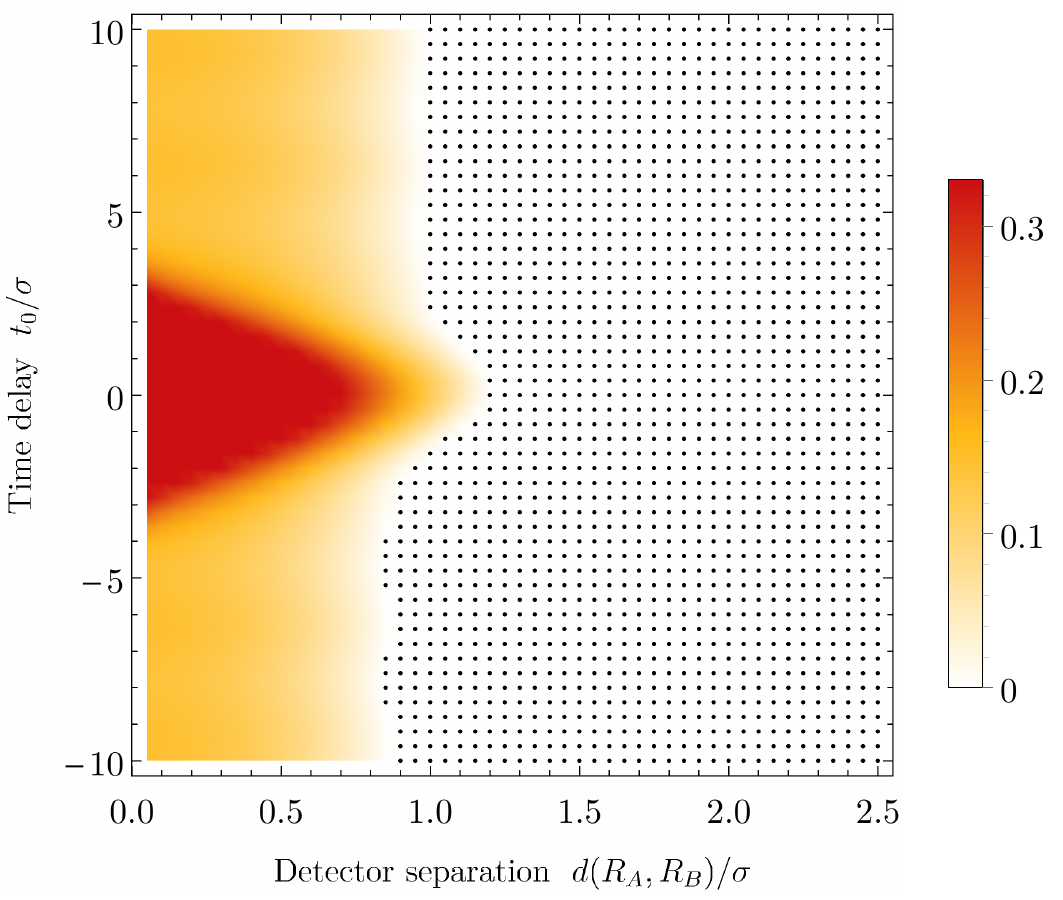}}
\quad
\subfloat[$\zeta=0$]{\label{aa2}
\includegraphics[width=0.3\textwidth]{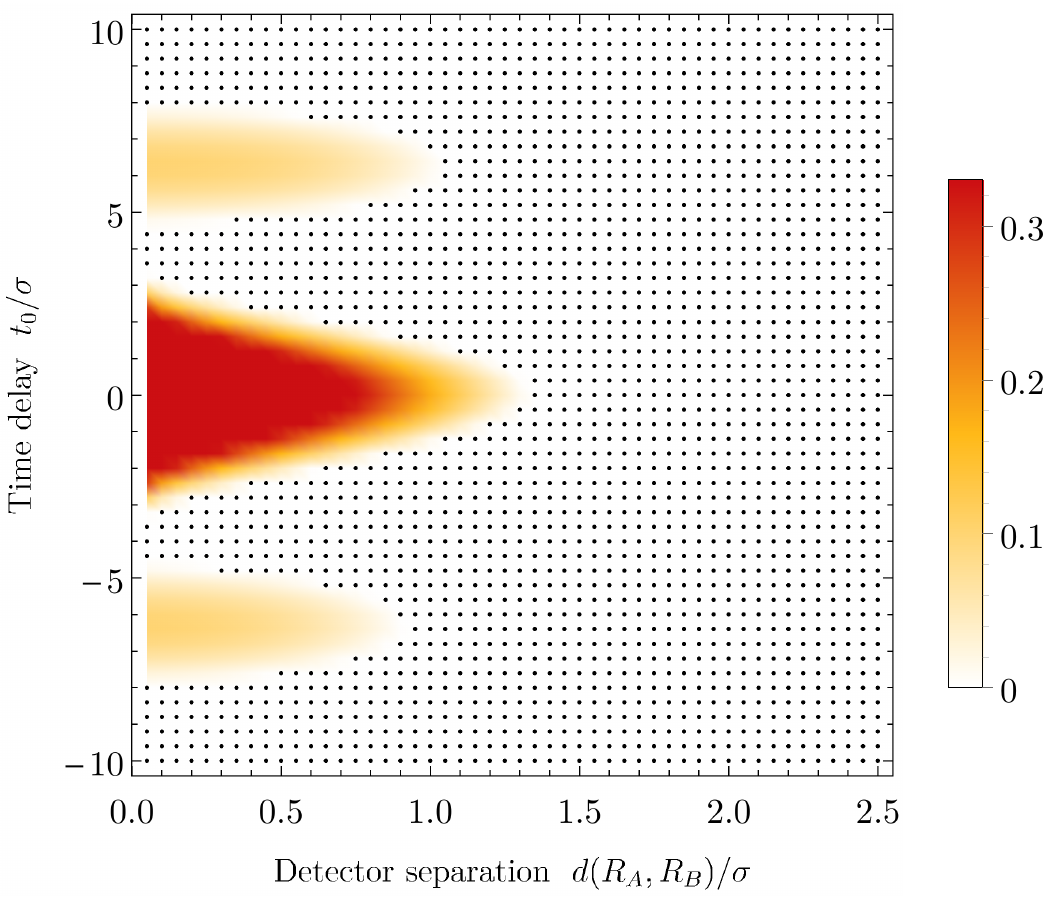}}
\quad
\subfloat[$\zeta=1$]{\label{aa3}
\includegraphics[width=0.3\textwidth]{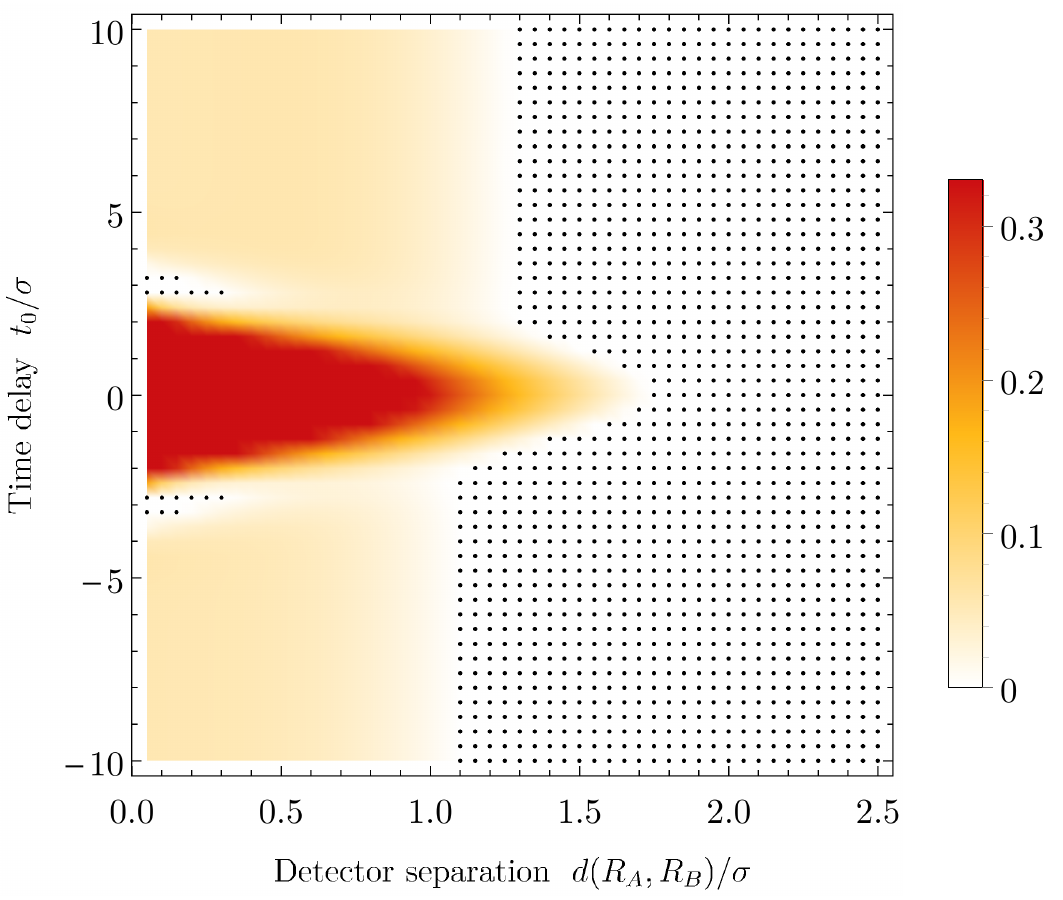}}
}
\caption{%
The concurrence, $\mathcal{C}/\tilde{\lambda}^2$ associated with the state $\rho_{AB}$ describing two static detectors is plotted as a function of their proper separation $d(R_A,R_B)/\sigma$ and the relative time delay in their switching functions $t_0/\sigma$ for all boundary conditions $\zeta={-1,0,1}$ and an AdS length of $\ell/\sigma=1$.  A negative $t_0$ means that detector $B$ switches before detector $A$.  Detector A is located at the origin, and the energy gap of the detectors is $\Omega\sigma=1/10$. The area filled with black dots represents the region where the concurrence vanishes and thus no entanglement harvesting is possible.
%\tco{\bf [ I think the horizontal axis should only be plotted from 0 to 2.5.]}
}
\label{fig:DelaySmallOm}
\end{figure*}
%----------------------------------------------------------------------

Finally, in Fig.~\ref{fig:DelaySmallOm} we consider small energy gaps, and find that the concurrence is nearly symmetric about $t_0=0$ for all three boundary conditions.  Here, $\beta_X$ is small so $X$ is approximately even in $t_0$ (the large $y$ behaviour of the integrand in Eq.~\eqref{X-equation} is exponentially suppressed by the Gaussian factor).  Unlike in Fig.~\ref{fig:TimeDelay}, there are no time delay dependent oscillations, nor is entanglement harvesting possible for moderate or large detector separation regardless of which detector interacts with the field first.  However, the concurrence is maximum around the region of small time delay for all boundary conditions.  Outside of this region, there are notable differences in behaviour for each boundary condition.  For Neumann boundary conditions $(\zeta =-1)$ there is little variation in the concurrence as $|t_0|$ increases. However, for Dirichlet boundary conditions $(\zeta=1)$ there are two regions around $|t_0|/\sigma=3$ where no entanglement harvesting is possible when the detectors are very close together, but possible again when their separation increases.  In the case of transparent boundary conditions $(\zeta=0)$, entanglement harvesting is only possible for three small regions of the parameter space.

 Let us emphasize that the time delay asymmetries observed in the harvested entanglement do not, strictly speaking, reflect something specific to $(2+1)$-AdS spacetime. Instead these asymmetries provide another example of the effects of a relative red shift (or relative time dilation) on the ability of atoms to harvest entanglement from the quantum vacuum. Note that the asymmetries can all be traced back to the non-vanishing of $\beta_X$, which is proportional to $\gamma_A - \gamma_B$. If one traces the origin of the $\beta_X$ term, it arises from the term $\exp\left(-i\Omega_A \tau_A - i\Omega_B \tau_B\right)$ in the expression for $X$~\eqref{defX} which does not explicitly depend on any properties of the Wightman function. As a result, we expect that similar asymmetries should arise whenever two detectors have a relative red-shift factor when compared to the frame in which the time delay is measured. Characterizing the differences in these asymmetries between equivalent detector setups in different spacetimes could then provide a probe of the underlying spacetime geometry, topology, etc.

When a  time delay is present, the switching functions are modified, yielding a non-zero $\Delta_T$.  This is symmetric in $t_0$, but the presence of a differential redshift breaks this symmetry.  If $\gamma_A < \gamma_B$
and $t_0 >0$, then entanglement can be harvested.  However if $t_0 < 0$,
a simple transformation of the integrand in Eq.~\eqref{X-equation} converts the problem into one with $\Omega < 0$\,---\,the problem effectively becomes that of harvesting entanglement from initially excited detectors, for which the $\sqrt{P_AP_B}$ term appearing in Eq.~\eqref{concurrence-eq} is large and the harvested entanglement suppressed. For the case at hand $\gamma_A=1  < \gamma_B$ and so entanglement harvesting is suppressed for $t_0<0$, the effect diminishing as the AdS length increases, causing  $\gamma_B\to 1$. It also diminishes for small gap (Fig.~\ref{fig:DelaySmallOm}), which likewise suppresses the asymmetry.

This asymmetry has also been observed in AdS$_4$ \cite{KRE}, though the effect is considerably less pronounced.  We attribute this to Hugyens' principle being operative in this case, which has the effect of suppressing harvested entanglement within the light cones of the detectors.

%========================================
%========================================
\section{Detectors on circular geodesics}
\label{Detectors on circular geodesics}

\begin{figure*}[]
\vspace{0pt}
\centering{
\subfloat[$d(R_A,R_B)=\sigma/10$]{\label{a1}
\includegraphics[width=0.3\textwidth]{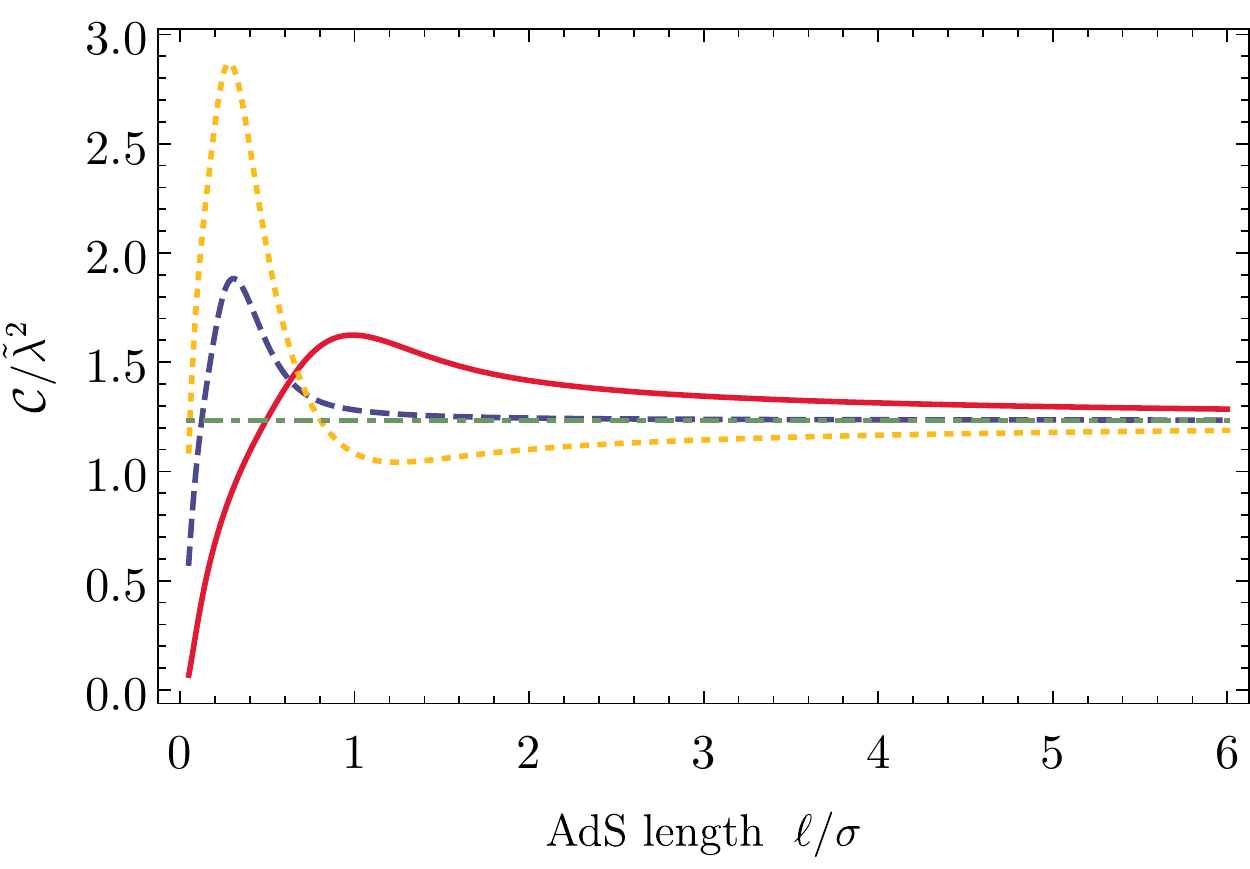}}
\quad
\subfloat[$d(R_A,R_B)=\sigma$]{\label{a2}
\includegraphics[width=0.3\textwidth]{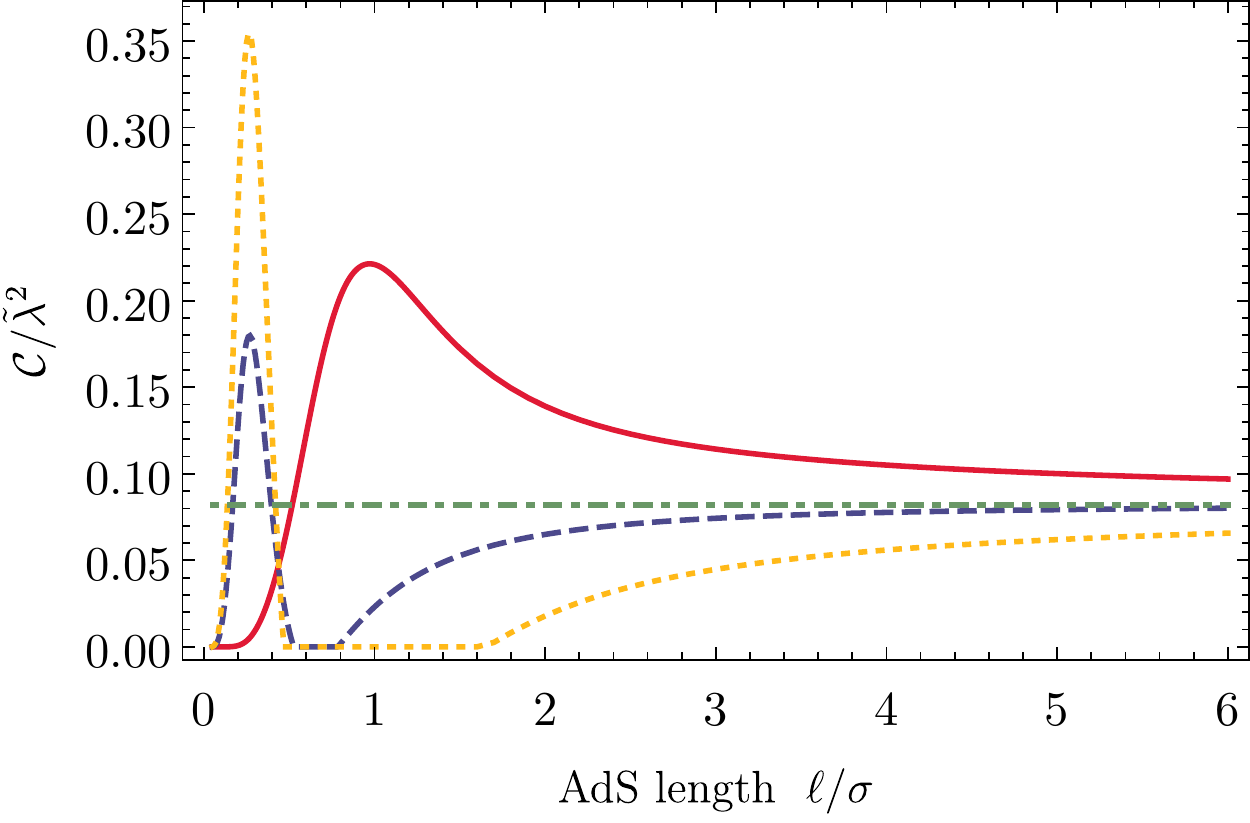}}
\quad
\subfloat[$d(R_A,R_B)=2\sigma$]{\label{a2}
\includegraphics[width=0.3\textwidth]{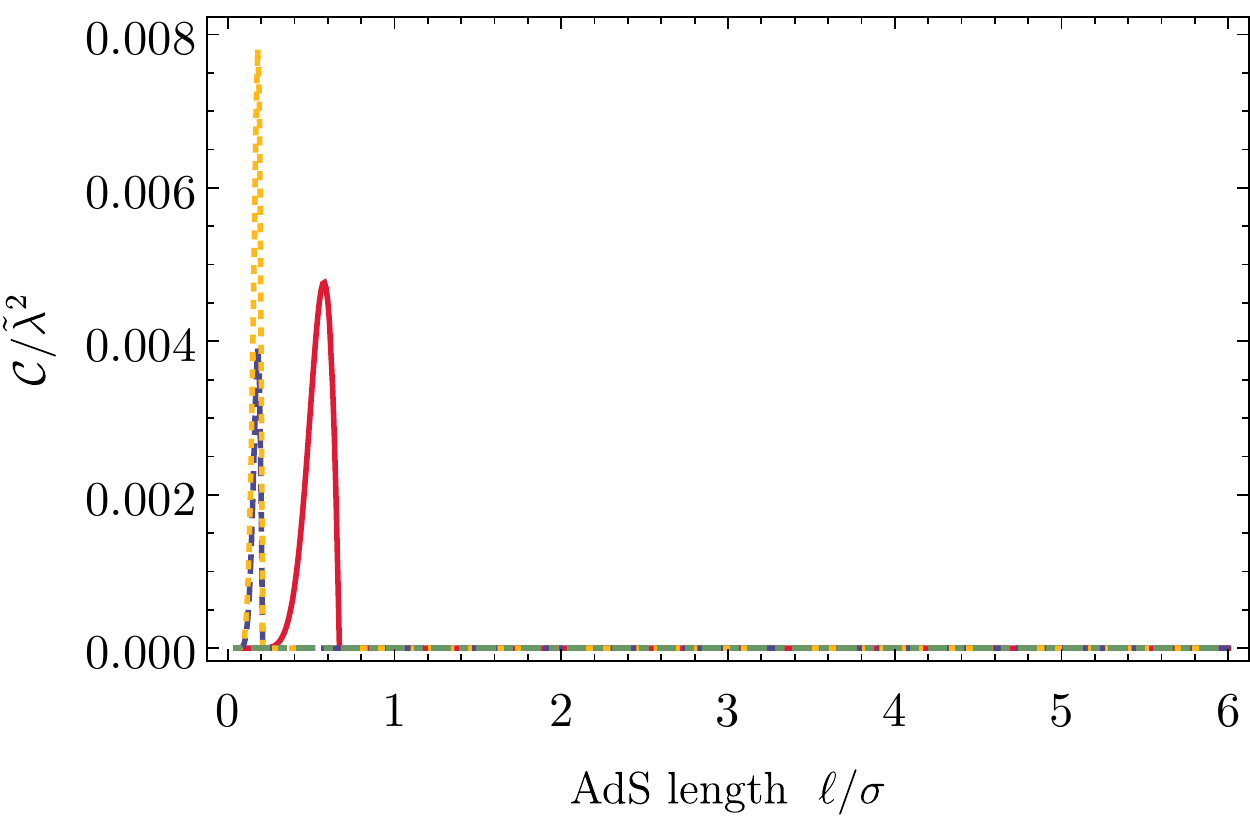}}
}
 \includegraphics[width=.4\linewidth]{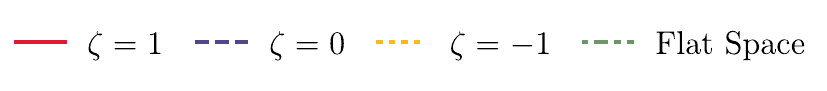}
\caption{The concurrence $\mathcal{C}/\tilde{\lambda}^2$ associated with the state $\rho_{AB}$ describing detector $A$ located at the origin and detector $B$ on a circular geodesic orbit around the origin is plotted as a function of the AdS length $\ell/\sigma$ for different values of the detectors proper separation $d(R_A,R_B)/\sigma$. The energy gap of the detectors is $\Omega\sigma=1/100$.  }
\label{Cvsell-Om}
\end{figure*}
%\twocolumngrid

%========================================

From the analysis presented so far we have seen that redshift effects, manifested through $\gamma_A$ and $\gamma_B$, can significantly affect entanglement harvesting when considering two static detectors in AdS$_3$. While these effects have led to a rich structure, it is certainly worth understanding what happens in the case where there is no relative redshift between the detectors.  In this section we consider such a detector configuration, with detector $A$ and detector $B$ in circular geodesic motion about the origin. The trajectories of such detectors are
\begin{align}
\label{traj}
x_D(\tau_D) &\ce \left\{ t =  \tau_D,  \  r = R_D, \  \phi= \tau_D/ \ell\right\},
\end{align}
where the angular velocity of the detectors in the coordinate frame $(t,\,r,\,\phi)$ is $1/\ell$. As can be seen from the detector trajectories, the proper times of each detector are equal and coincide with the coordinate time $t$. As a consequence the energy gaps of both detectors are equal in the coordinate frame. Additionally, we take the two detectors to have identical switching functions given by Eq.~\eqref{eq:TimeDelay}.

Since the detectors have the same proper time, their transition probabilities are equal, $P_A= P_B =\tilde{P}_D$, and given by
\begin{align}
\tilde{P}_D &= \frac{\lambda^2 \sigma }{4\sqrt{\pi}}  \left( - \PV \int_{0}^{\infty} dy \,  \frac{e^{- \tilde{a} y^2} \sin \left(\ell \Omega {y} \right)}{\sin \left(y/2 \right)} \right. \nn \\
&\quad \left. + \pi \sum_{n \in \mathbb{Z} }^{}(-1)^{n} \cos \left(2n\pi \ell \Omega \right )e^{-4n^2\pi^2 \tilde{a}} \right. \nn \\
&\quad  \left.-\zeta \left[  \PV \int_{0}^{\infty} dy \, \frac{e^{- \tilde{a} y^2} \cos \left(\ell \Omega  y \right)}{\cos \left(y/2 \right)} \right. \right. \nn \\
&\quad \left. \left. {-\pi\sum_{n \in \mathbb{Z}}^{}  (-1)^{n}}\sin \left [  (2n+1) \pi \ell \Omega \right] e^{-\tilde{a} \pi^2(2n+1)^2} \right] \right),
\label{transitionCircular}
\end{align}
where $\tilde{a}\ce \ell^2/4\sigma^2$; we will used a tilde over quantities to remind us that detectors are moving along a circular geodesic centred at the origin. Note that transition probability of the detectors in this case is identical to that of a static detector located at the origin.

%----------------------------------------------------------------------
\begin{figure*}[t!]
\centering{
\subfloat[$\ell/\sigma=1/2$]{\label{aa1}
\includegraphics[width=0.3\textwidth]{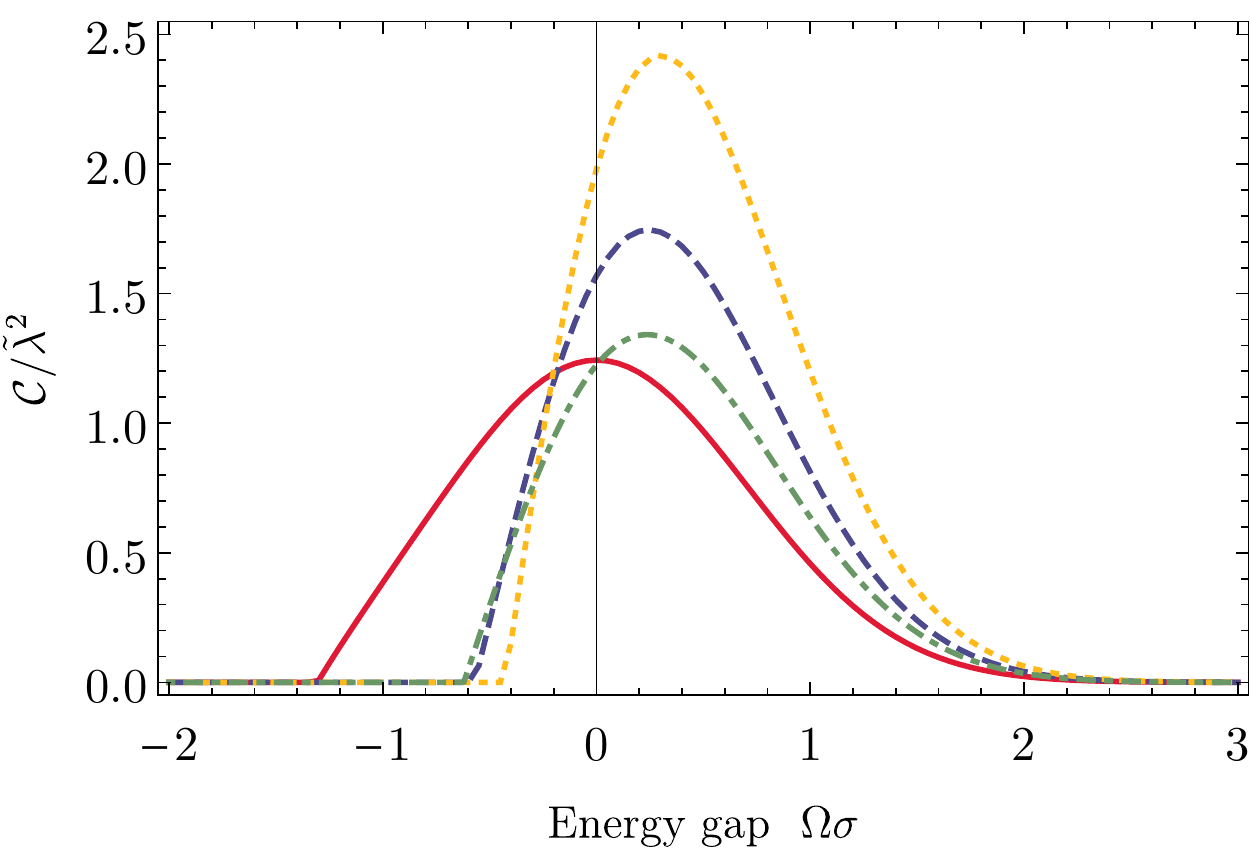}}
\quad
\subfloat[$\ell/\sigma=1$]{\label{aa2}
\includegraphics[width=0.3\textwidth]{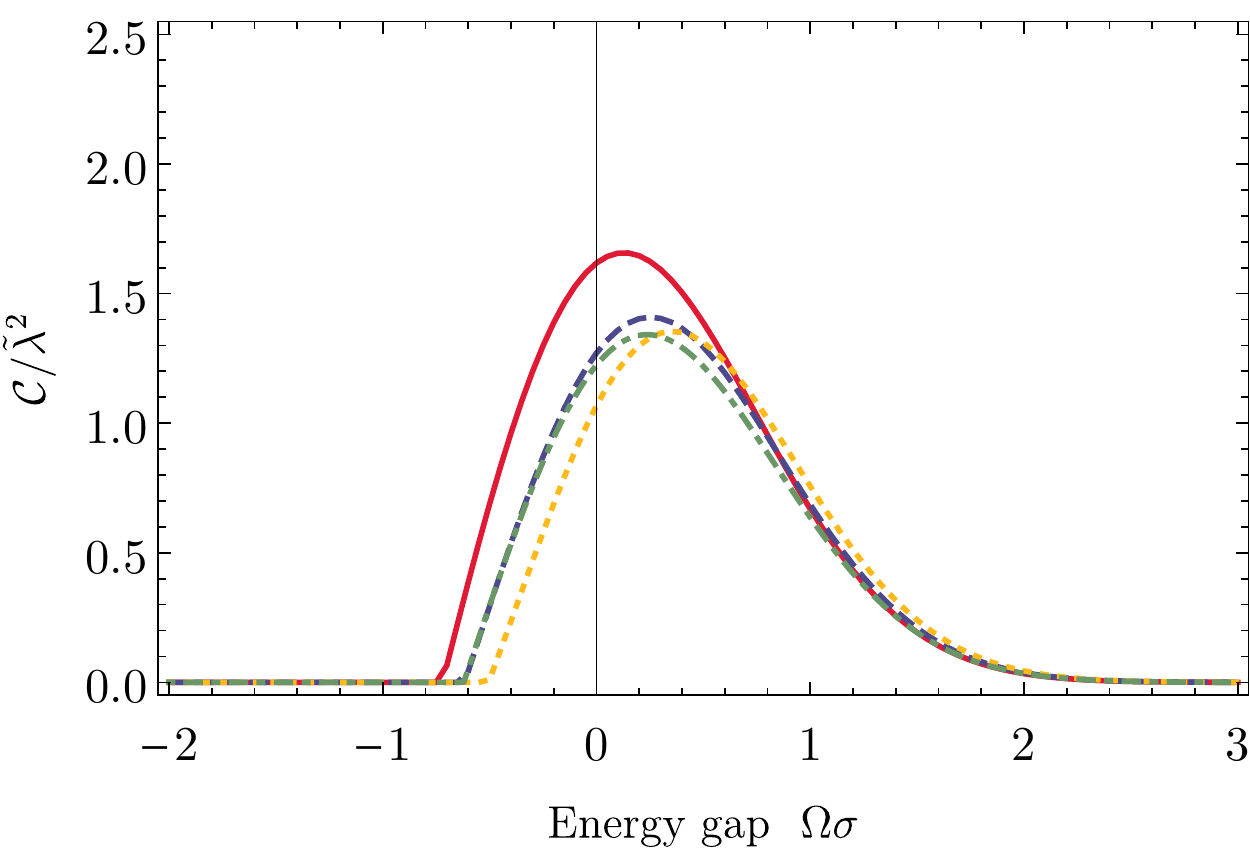}}
\quad
\subfloat[$\ell/\sigma=3$]{\label{aa3}
\includegraphics[width=0.3\textwidth]{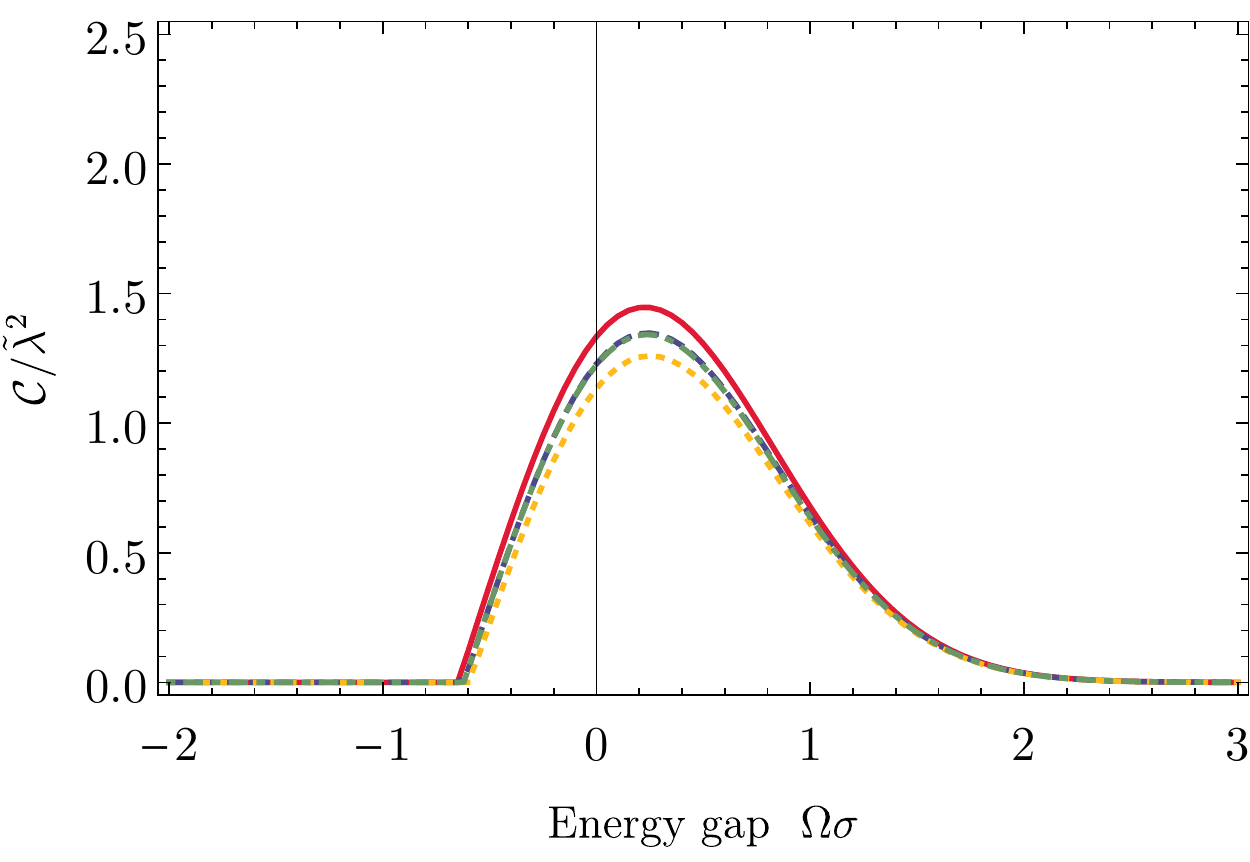}}
}
 \includegraphics[width=.45\linewidth]{legendFlat.pdf}
\caption{The concurrence $\mathcal{C}/\tilde{\lambda}^2$ associated with the state $\rho_{AB}$ describing detector $A$ located at the origin and detector $B$ on a circular geodesic orbit around the origin is plotted as a function of their energy gap $\Omega \sigma$ for different values of the AdS length $\ell/\sigma$. The proper separation of the detectors is $d(R_A,R_B)/\sigma=1/10$.}
\label{CvsOm-ell}
\end{figure*}
%----------------------------------------------------------------------

The matrix element $\tilde{X}$ appearing in the final joint state of these two detectors is
\begin{align}
\tilde{X}&=-\frac{\lambda^2 \sigma}{2 \sqrt{2 \pi }}
\tilde{K}_X \sum_{n \geq 0} (-1)^n \left(   \int_{\tilde\theta_{X_n}^-}^{\tilde\theta_{X_{n+1}}^-} dy \,  \frac{e^{-\tilde{a} y^2} \cosh(\tilde\Delta_Ty)}{\sqrt{\cos y -\tilde\alpha_X}} \right. \nn \\
&\quad \left.  -\zeta  \int_{\tilde\theta_{X_n}^+}^{\tilde\theta_{X_{n+1}}^+} dy \,  \frac{e^{-\tilde{a} y^2}\cosh(\tilde\Delta_Ty)}{\sqrt{\cos y + \tilde\alpha_X}}  \right), \label{Xcircular}
\end{align}
where we have defined
\begin{align}
\tilde{K}_X &\ce \sqrt{\tilde\alpha_X}\ \exp\left(-\sigma^2\Omega^2-\frac{t_0^2}{4\sigma^2}\right) \nn\\
\tilde\Delta_T &\ce \frac{\ell t_0}{2\sigma^2} \label{eq:tildealphaX}\\
\tilde\alpha_X &\ce \frac{\ell^2}{\sqrt{R_A^2+\ell^2}\sqrt{R_B^2+\ell^2}-R_AR_B} \nn\\
\tilde\theta_{X_n}^\pm &\ce \max\left[0,\arccos\left(\pm\tilde\alpha_X\right)+(2n-1)\pi\right]. \nn
\end{align}
A detailed derivation of Eqs.~\eqref{transitionCircular} and \eqref{Xcircular} can be found in Appendix~\ref{Derivation of PD and X}.

Both the transition probability $\tilde{P}_D$ and matrix element $\tilde{X}$ given above were evaluated numerically in Mathematica to a precision of at least 25 significant digits using the method ``DoubleExponential'' or ``DoubleExponentialOscillatory'' depending on the oscillatory nature of the integrand. Mathematica's ``PrincipalValue'' procedure was implemented to handle the principal value terms appearing in the matrix elements.

We first consider the scenario when detector $A$ is static and located at the origin,   detector $B$ orbits around it, and there is no relative time delay in the switching functions $(t_0=0)$. In Figs.~\ref{Cvsell-Om} and \ref{CvsOm-ell} we show how the concurrence depends on the energy gap of the detectors $\Omega\sigma$ and AdS length $\ell/\sigma$, respectively. In Fig.~\ref{Cvsell-Om} we see that the concurrence asymptotes to a constant value for large $\ell/\sigma$,  corresponding to the flat space result; similar behaviour was observed for static detectors in  Fig.~\ref{Negativity-ell}. This is expected because in the limit $\ell/\sigma \to \infty$, AdS$_3$ approaches Minkowski space. On the other hand, when the detector energy gap $\Omega\sigma$ is small, we observe a maximum in the concurrence for $\ell/\sigma \lesssim 1$, the precise location and magnitude of which depends on the boundary conditions satisfied by the field. In particular, the maximum is largest for Neumann boundary conditions $(\zeta = -1)$. Increasing the energy gap $\Omega \sigma$ results in the broadening of the peak and, for Dirichlet  boundary conditions $(\zeta = 1)$, its eventual disappearance.

Comparing Figs.~\ref{CvsOm-ell} and \ref{Negativity-omega}, we also see similar behaviour.
The concurrence peaks for small positive detector energy gap, $\Omega\sigma > 0$, and rapidly decays for $\Omega\sigma < 0$.   This behaviour is present regardless of the AdS length; however we find that for very small AdS length, $\ell/\sigma=1/2$, the largest peak corresponds to Neumann boundary conditions $(\zeta=-1)$, but for larger AdS lengths $\ell/\sigma\ge1$, the largest peak is for Dirichlet boundary conditions $(\zeta = 1)$.  We also find that as the AdS length becomes large, the harvested entanglement for all three boundary conditions approach the flat space results.

In addition, we find that the case where both detectors $A$ and $B$ move in circular geodesic orbits around the origin is mathematically equivalent to the case when detector $A$ is fixed at the origin and detector $B$ is in a circular geodesic orbit around the origin.  The coordinates of the detectors only appear in the term $\tilde\alpha_X$; however when the proper distance between the two detectors is kept constant, then $\tilde\alpha_X=\text{sech}\big(d(R_A,R_B)/\ell\big)$ and the concurrence is only dependent on the proper distance between the two detectors. This means that the calculation is insensitive to the proper distance of either detector from the origin, and without loss of generality we can simply consider the scenario where detector $B$ orbits detector $A$ located at the origin.

%----------------------------------------------------------------------
\begin{figure*}[]
\centering{
\subfloat[$\ell/\sigma=1/5$, \ $d(0,R_A)=0$, \ $d(R_A,R_B)/\sigma=1/10$]{\label{CS_diff_a}\includegraphics[width=0.3\textwidth]{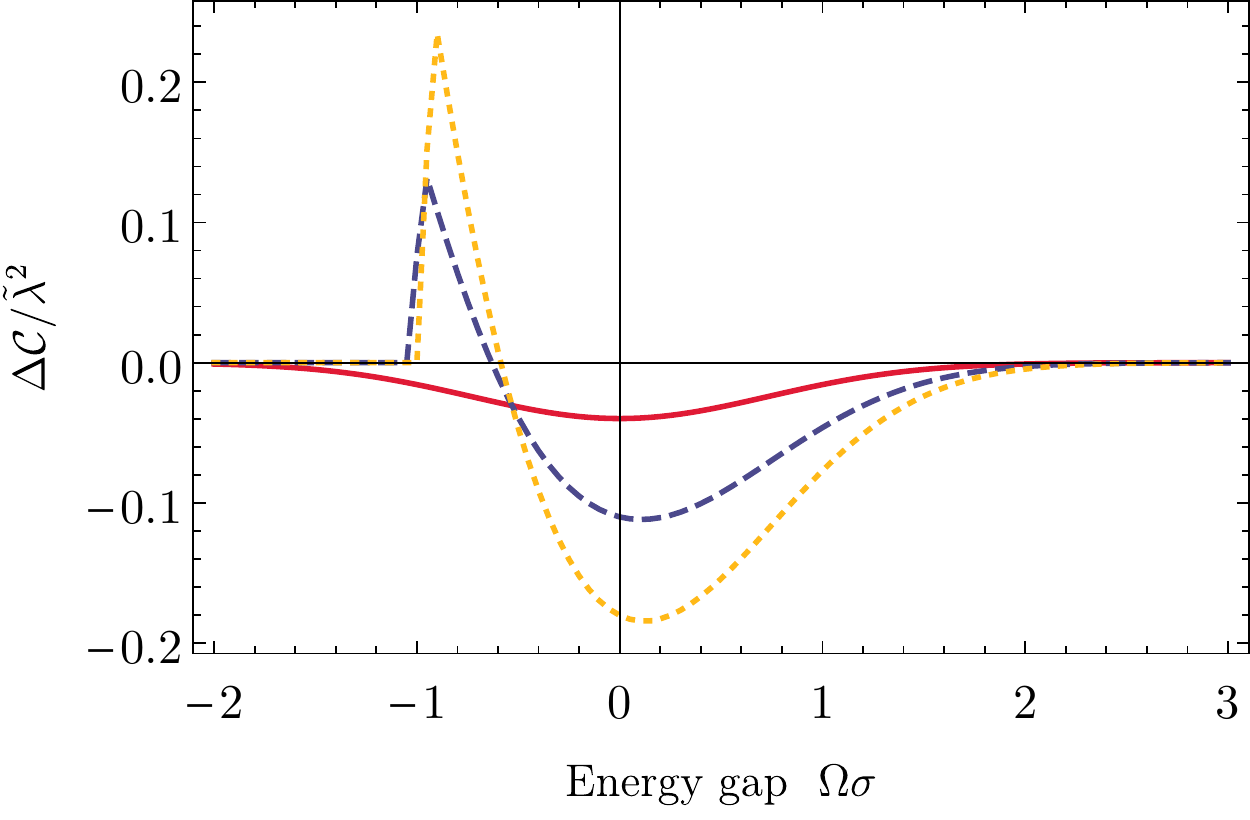}}
\qquad
\subfloat[$d(0,R_A)=0$, \ $d(R_A,R_B)/\sigma=1/10$, \ $\Omega\sigma=1/100$]{\label{CS_diff_b}\includegraphics[width=0.3\textwidth]{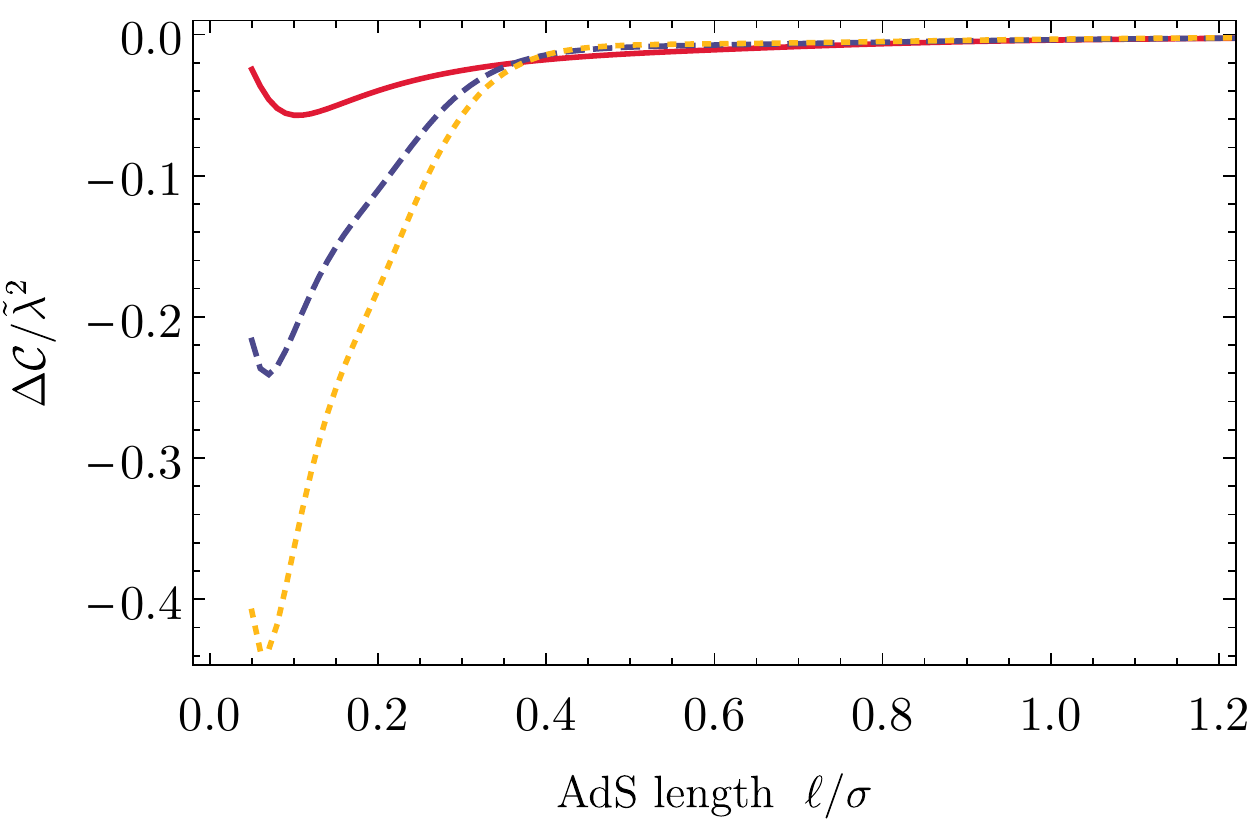}}
\qquad
\subfloat[$\ell/\sigma=1/5$, \ $d(0,R_A)=0$, \ $\Omega\sigma=1/100$]{\label{CS_diff_c}\includegraphics[width=0.3\textwidth]{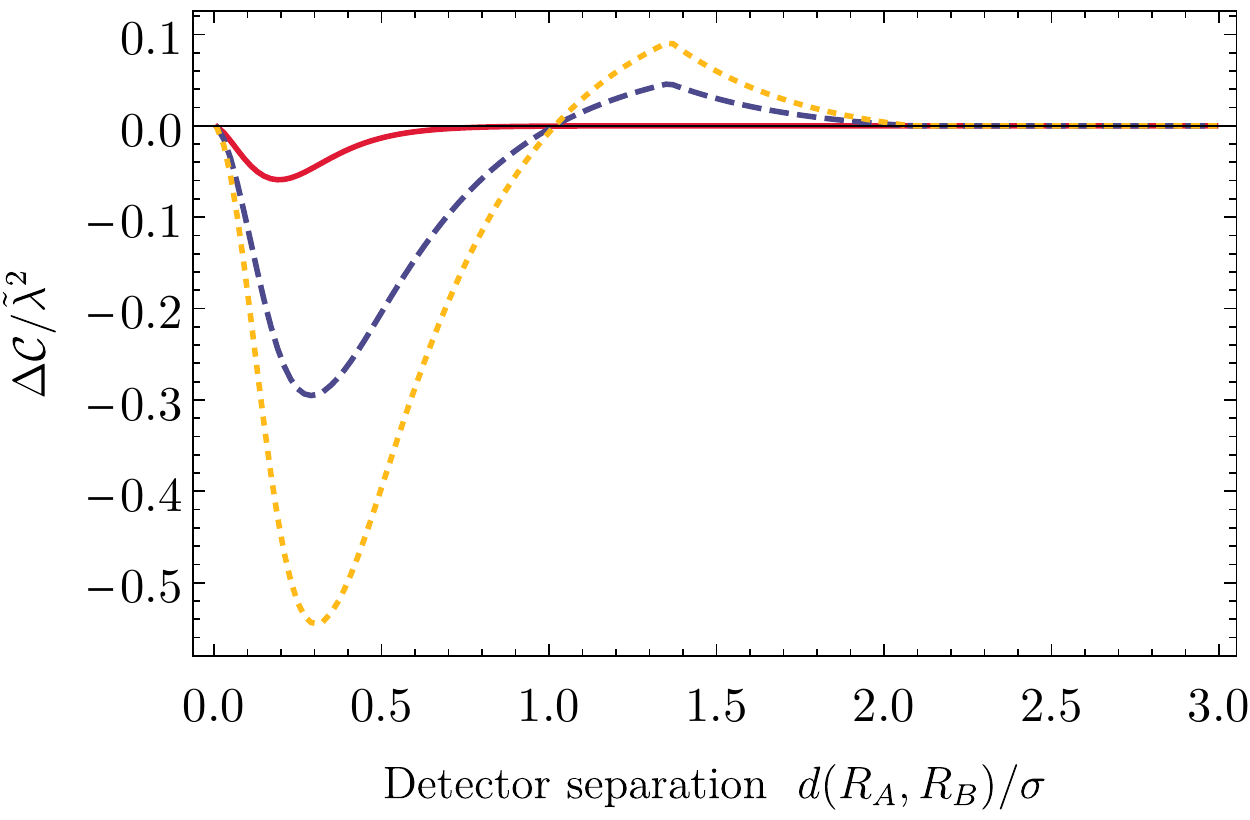}}
\\
\includegraphics[width=.3\linewidth]{legend.pdf}
}
\caption{A series of plots characterizing the difference in concurrence $\Delta\mathcal{C}\ce\mathcal{C}_{\text{circular}}-\mathcal{C}_{\text{static}}$ between static detectors and detectors on circular geodesic orbits around the origin. Plot (a) shows the difference $\Delta\mathcal{C}/\tilde{\lambda}^2$ as a function of the detectors energy gap $\Omega\sigma$ for fixed AdS length $\ell/\sigma$ and detector separation $d(R_A,R_B)$ when detector $A$ is located at the origin. Plot~(b) shows the difference $\Delta\mathcal{C}/\tilde{\lambda}^2$ as a function of AdS length $\ell/\sigma$ for fixed detector energy gap $\Omega \sigma$ and proper separation $d(R_A,R_B)$ when detector $A$ is located at the origin. Plot (c) shows the difference $\Delta\mathcal{C}/\tilde{\lambda}^2$ as a function of proper detector separation $d(R_A,R_B)$ for fixed AdS length $\ell/\sigma$ and detector energy gap $\Omega\sigma$ when detector $A$ is located at the origin.
}
\label{circ-stat}
\end{figure*}
%----------------------------------------------------------------------

Having presented results for both static detectors and detectors moving on circular geodesics, it is worth discussing the similarities and differences between the two cases. We plot the difference in the concurrence for these two cases in Fig.~\ref{circ-stat}. First, we note that in the flat space (large $\ell$) limit that the static and circular geodesic trajectories both reduce to the same flat space detector configuration. That is, taking the large $\ell$ limit in each case yields two static detectors with proper separation $d(R_A, R_B) = R_B - R_A$ in flat spacetime. Thus, we expect, and indeed observe, that any differences between the static and circular geodesic cases disappear at large AdS length $\ell/\sigma$.

At smaller AdS length, there are differences that emerge between the two cases, and we expect that these are mostly due to the lack of redshift effects for detectors moving along circular geodesics. A particularly notable difference is that, at least for small detector energy gaps, detectors on static trajectories harvest more entanglement than detectors on circular geodesic trajectories.

\begin{figure}[h]
\includegraphics[width=0.45\textwidth]{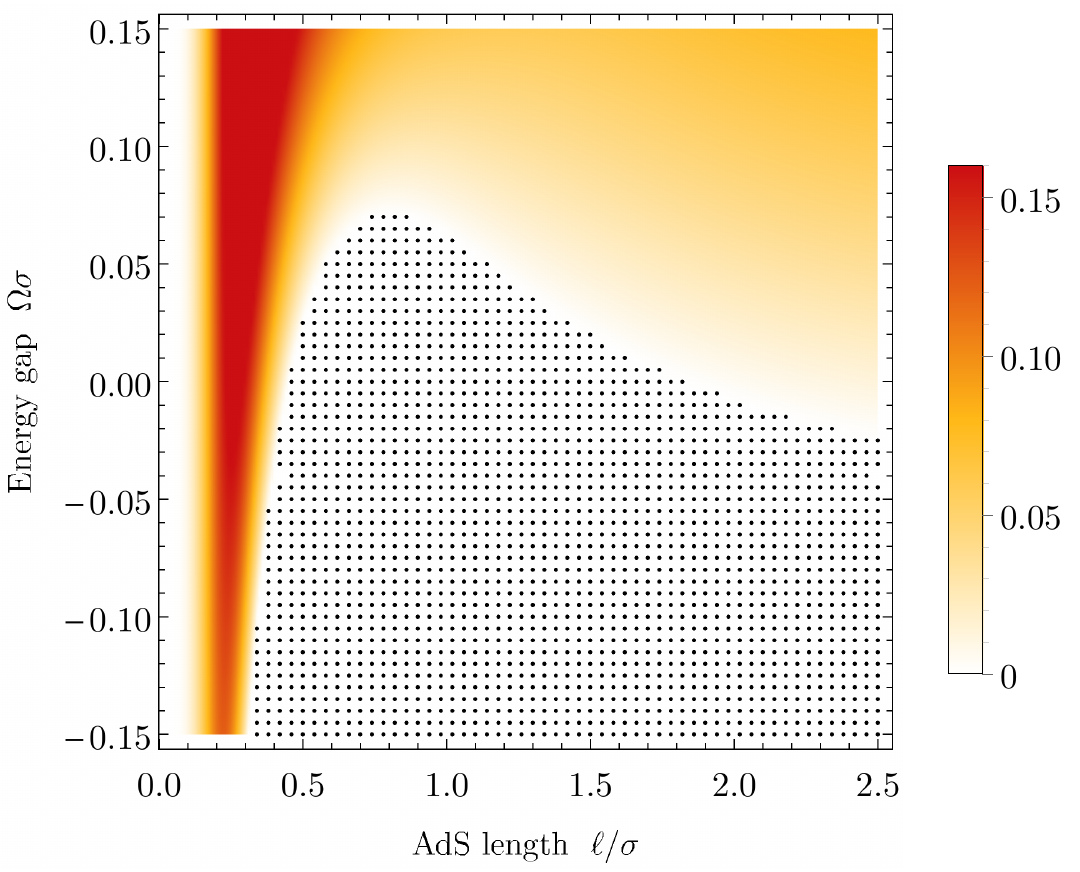}
\caption{A density plot of the concurrence $\mathcal{C}/\tilde{\lambda}^2$ as a function of the energy gap of the detectors and AdS length for the particular case of $d(R_A, R_B)/\sigma = 1$ and Neumann boundary conditions \mbox{($\zeta = -1$)}. The black dots indicate regions where the concurrence is precisely zero and no entanglement harvesting is possible.}
\label{peninsula}
\end{figure}

Another significant difference between these two cases is the lack of a separability island for detectors moving on circular geodesics, which was observed for static detectors. For static detectors and particular choices of AdS length, there was a region in the $\{d(R_A, R_B)/\sigma, \Omega \sigma\}$ parameter space (see Fig.~\ref{islandplot}) where the entanglement vanished. For detectors on circular geodesics there is no such region, suggesting that the island of no entanglement owes its existence largely to the redshift effects present in the static detector case. Instead, for detectors on circular geodesics we  observe a ``peninsula'' of large entanglement in the $\{\Omega \sigma , \ell/\sigma \}$ parameter space for small $\ell/\sigma$, most prominent
in Fig.~\ref{Cvsell-Om}(b) for Neumann boundary conditions ($\zeta = -1$) (clearly displayed in Fig.~\ref{peninsula}) but also present for all boundary conditions in Fig.~\ref{Cvsell-Om}.
We see that, for detectors with small positive energy gaps, the concurrence will vanish and then reappear as the AdS length increases.

\begin{figure*}[t!]
\subfloat[$\zeta=-1$ and $\ell\sigma=1$]{%
  \includegraphics[width=.3\linewidth]{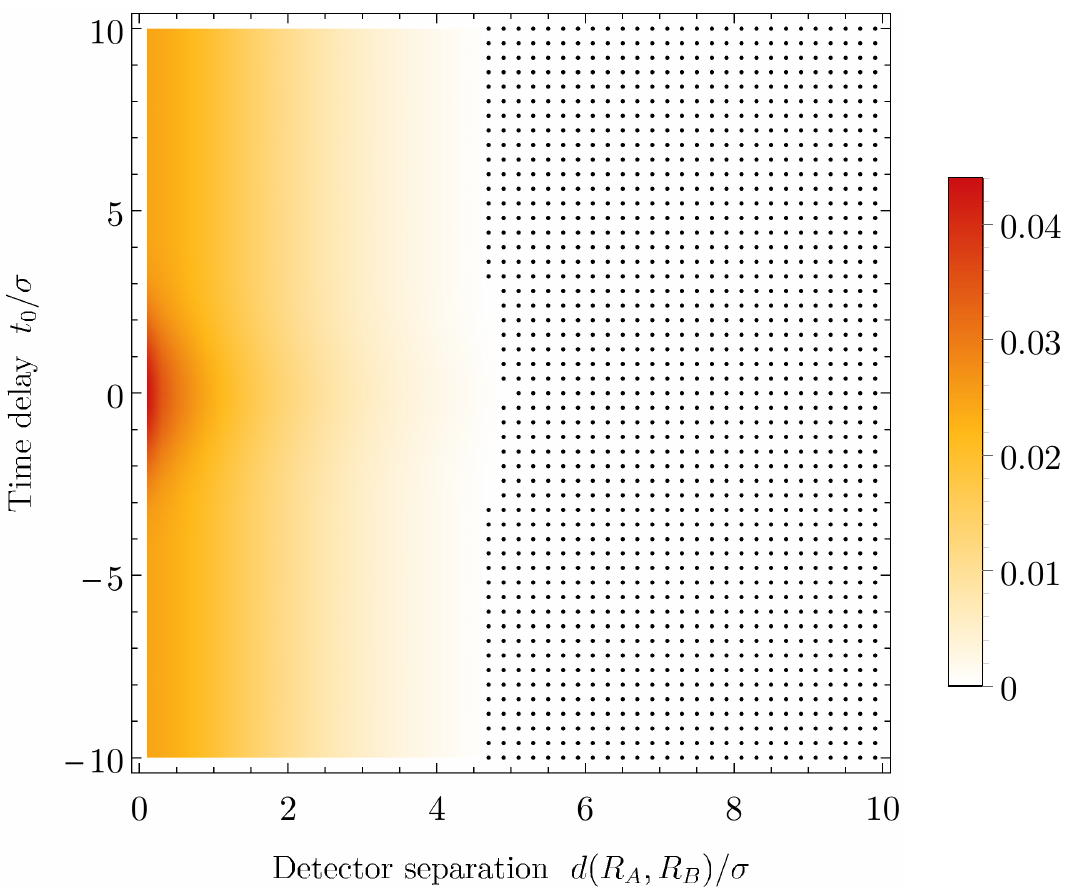}%
}%
\quad
\subfloat[$\zeta=-1$ and $\ell\sigma=5$]{%
  \includegraphics[width=.3\linewidth]{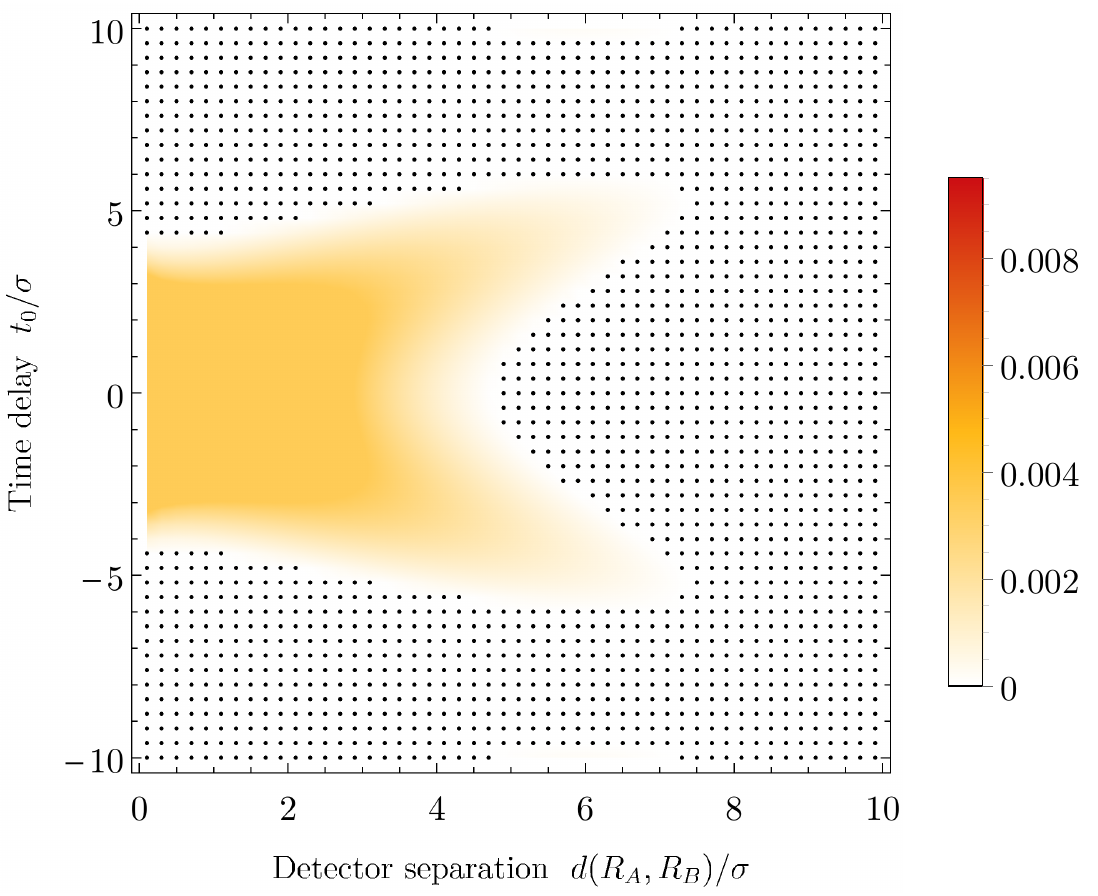}%
}%
\quad
\subfloat[$\zeta=-1$ and $\ell\sigma=20$]{%
  \includegraphics[width=.3\linewidth]{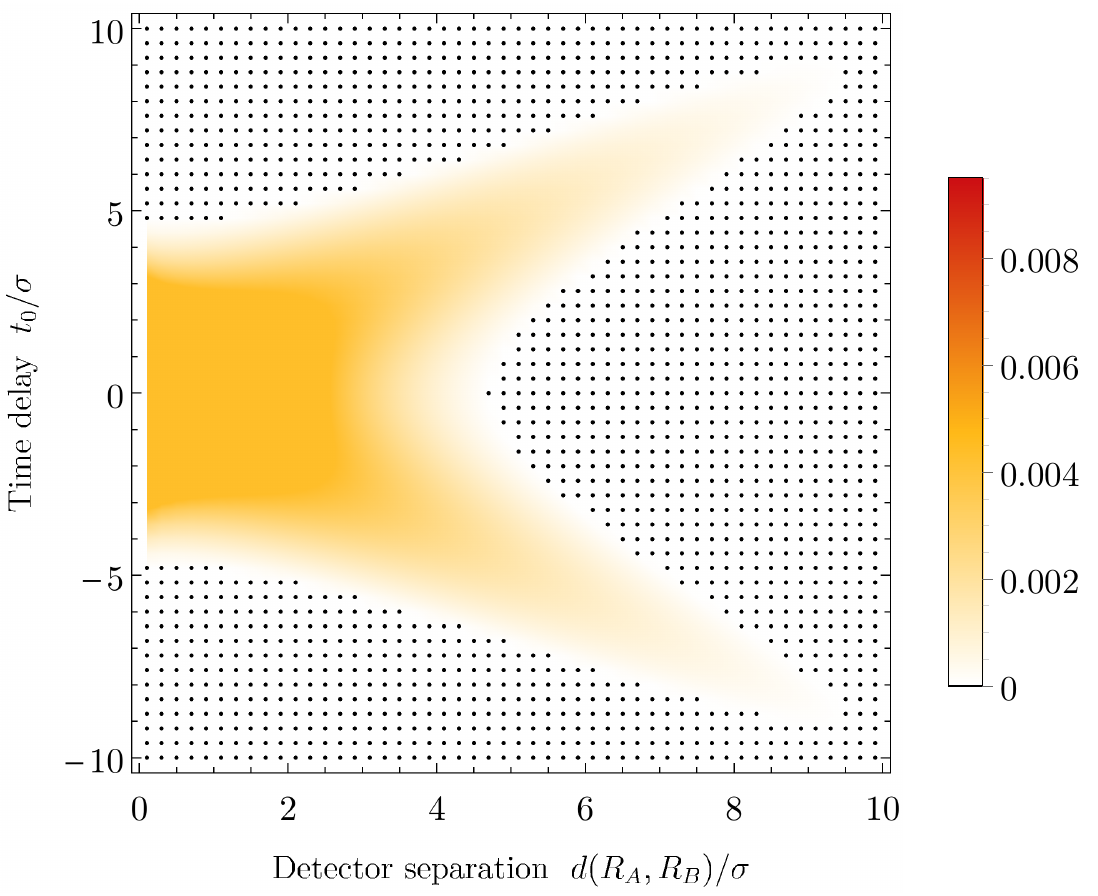}%
}%
\\ %NEWLINE
\subfloat[$\zeta = 0$ and $\ell/\sigma = 1$]{%
  \includegraphics[width=.3\linewidth]{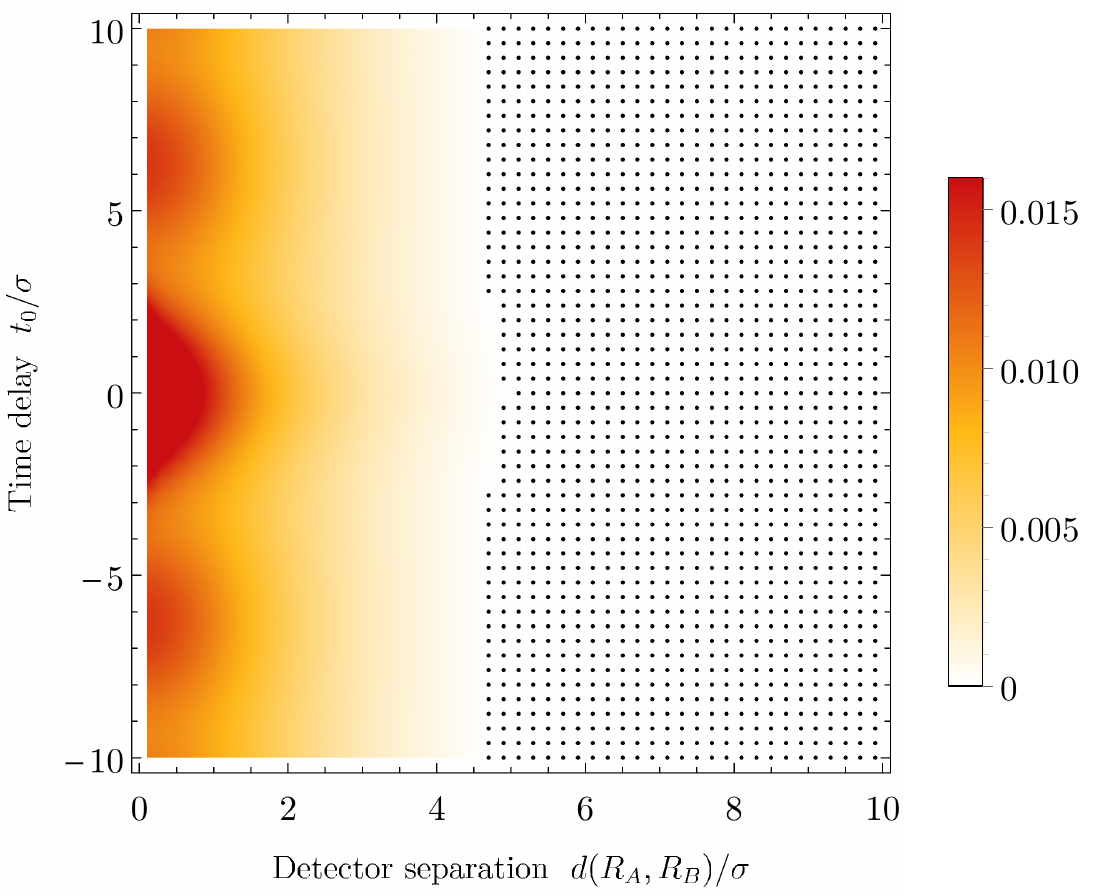}%
}%
\quad
\subfloat[$\zeta = 0$ and $\ell/\sigma = 5$]{%
  \includegraphics[width=.3\linewidth]{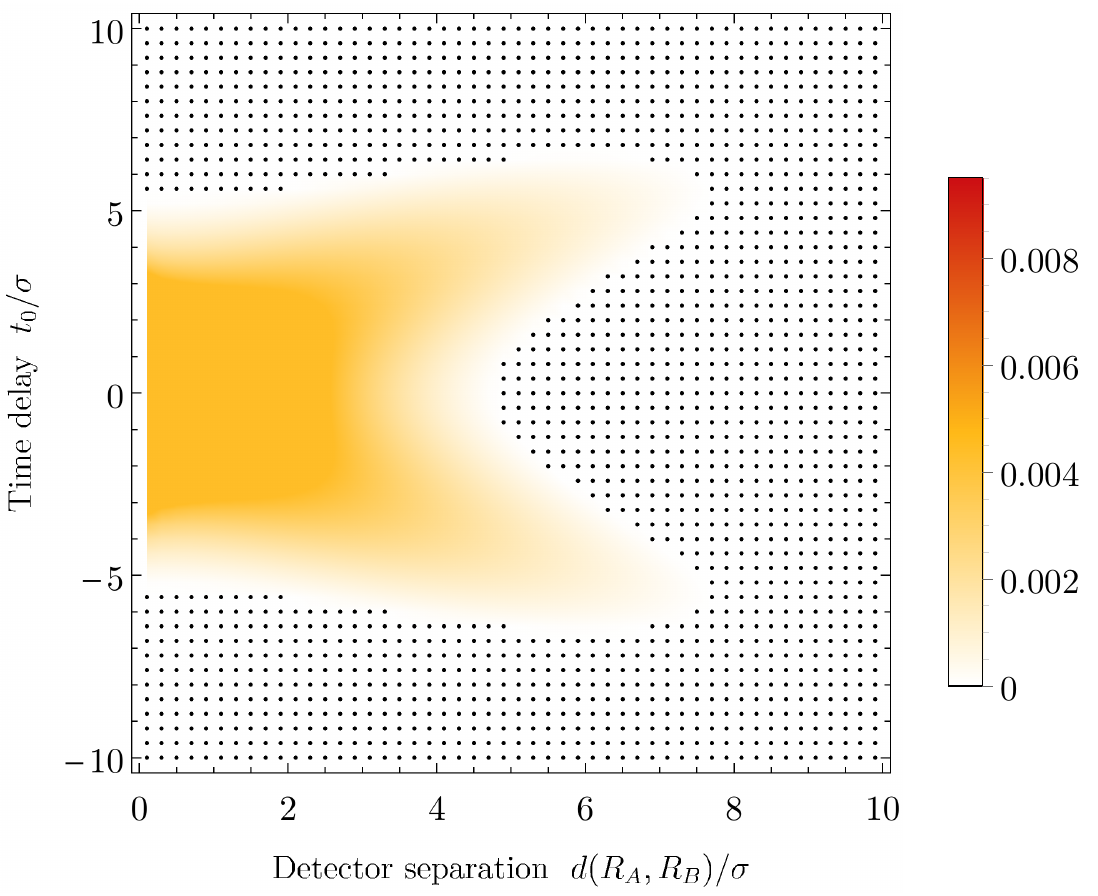}%
}%
\quad
\subfloat[$\zeta = 0$ and $\ell/\sigma = 20$]{%
  \includegraphics[width=.3\linewidth]{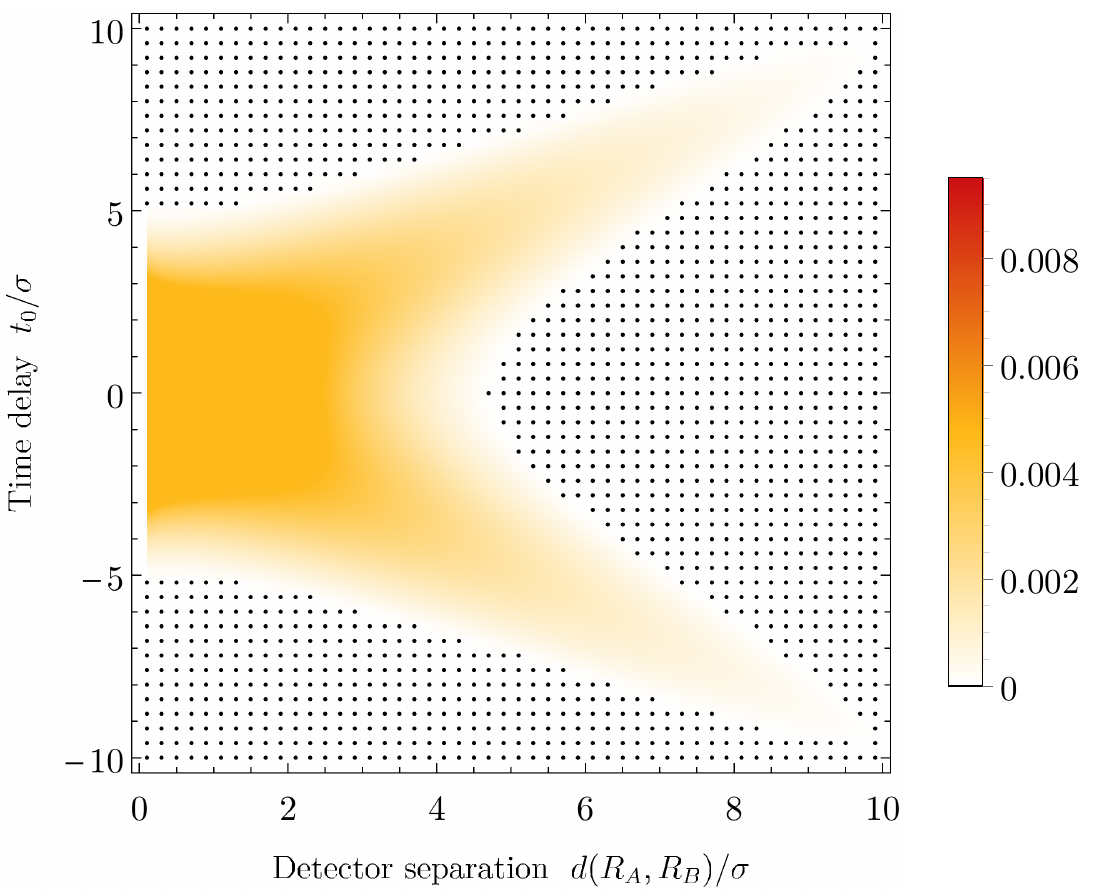}%
}%
\\ %NEWLINE
\subfloat[$\zeta = 1$ and $\ell/\sigma = 1$ ]{%
  \includegraphics[width=.3\linewidth]{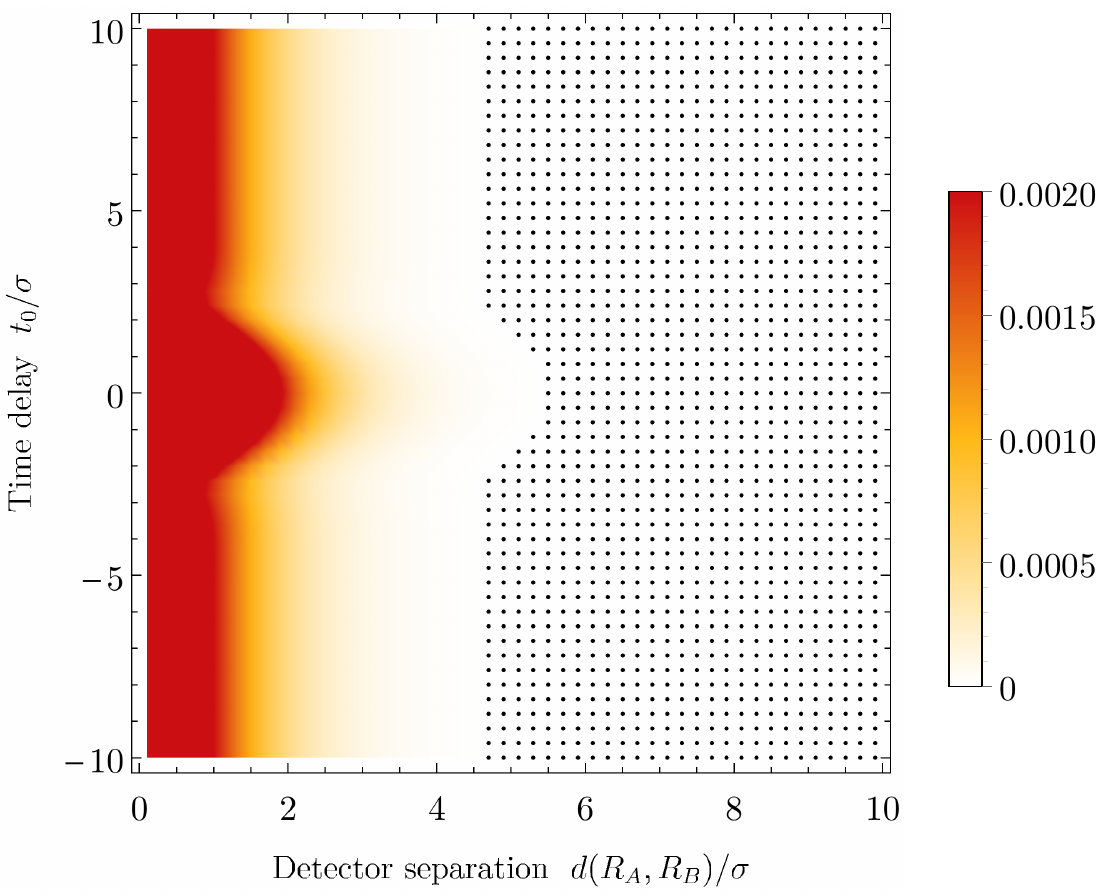}%
}%
\quad
\subfloat[$\zeta = 1$ and $\ell/\sigma = 5$]{%
  \includegraphics[width=.3\linewidth]{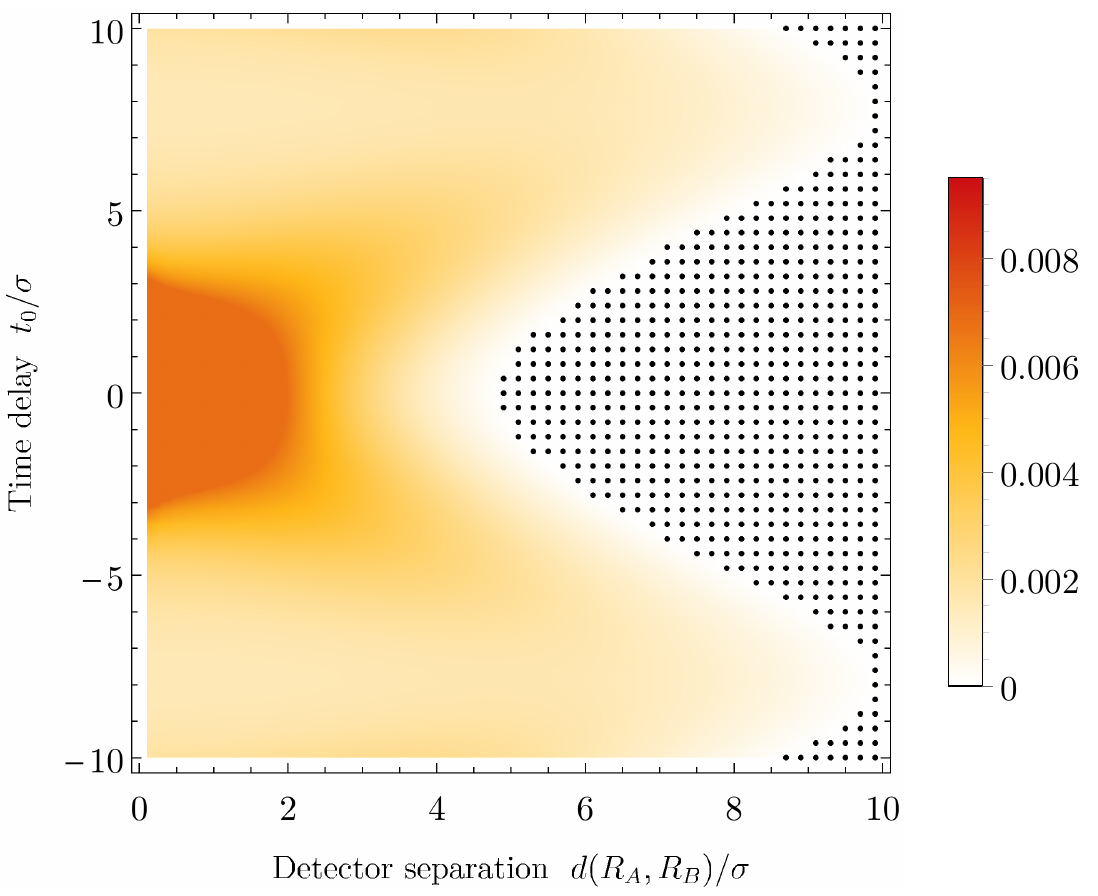}%
}%
\quad
\subfloat[$\zeta = 1$ and $\ell/\sigma = 20$]{%
  \includegraphics[width=.3\linewidth]{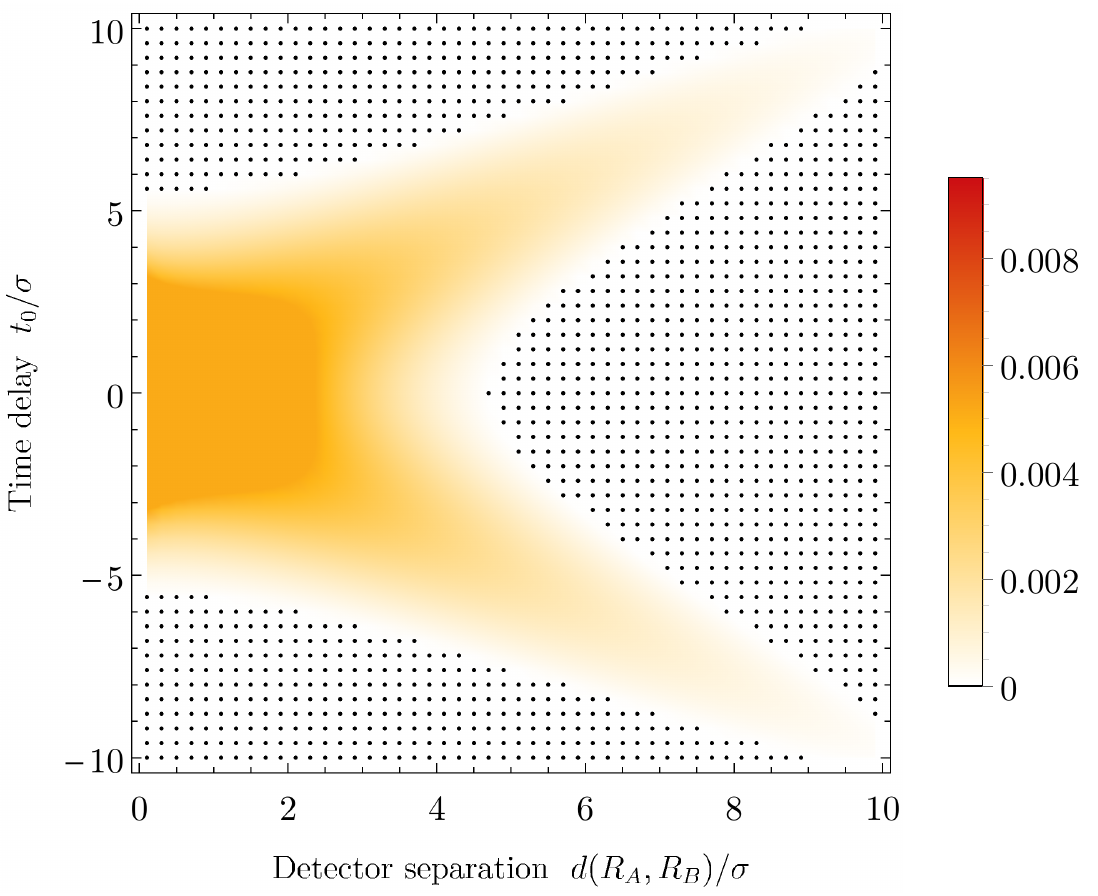}%
}%

\caption{%
The concurrence, $\mathcal{C}/\tilde{\lambda}^2$ associated with the state $\rho_{AB}$ describing two detectors on circular geodesic orbits around the origin is plotted as a function of their proper separation $d(R_A,R_B)/\sigma$ and the relative time delay in their switching functions $t_0/\sigma$ for all boundary conditions $\zeta={-1,0,1}$ and different value of the AdS length $\ell/\sigma$.  A negative $t_0$ corresponds to detector $B$ switching before detector $A$.  Detector A is located at the origin, and the energy gap of the detectors is $\Omega\sigma=2$. The area filled with black dots represents the region where the concurrence vanishes and thus no entanglement harvesting is possible.
}
\label{fig:GeoTimeDelay}
\end{figure*}

Finally, we allow the switching functions of the detectors to be offset by some $t_0\ne0$.  Unlike, in the case of static trajectories, definition of $\tilde{X}$ ensures that the concurrence is symmetric under the transformation $t_0 \to -t_0$. This is unsurprising, as the two detectors have the same proper time which is equal to the coordinate time. We depict the dependence of the concurrence on detector separation and switching  offset
 in Fig.~\ref{fig:GeoTimeDelay}, taking
 the detectors' energy gaps to be $\Omega \sigma = 2$, symmetric about the $t_0=0$ axis for all three boundary conditions. We find the concurrence is highest in the region around $t_0=0$ and low-to-moderate detector separation.  This maxima is present for all boundary conditions, but most pronounced for Dirichlet boundary conditions $(\zeta=1)$

When the AdS length is small, $\ell/\sigma=1$, we note that there are two secondary maxima present in the case of transparent boundary conditions $(\zeta = 0)$ around $t_0/\sigma\approx6.5$.  Additionally in this regime, entanglement harvesting is possible at large detector separation
when there is no relative time delay in the detectors' switching functions, most notably for Dirichlet conditions $(\zeta=1)$. At larger AdS length, $\ell/\sigma=5$ and $\ell/\sigma=20$, entanglement harvesting is possible for larger detector separations when the time delay is non-zero, noted by the two ``peninsulas'' of concurrence in the two rightmost columns in Fig.~\ref{fig:GeoTimeDelay}.  Again, this effect is present for all three boundary conditions, but most exaggerated for the Dirichlet case $(\zeta=1)$ .

Comparing Fig.~\ref{fig:GeoTimeDelay} to Fig.~\ref{fig:TimeDelay}, we note that when the AdS length is large $\ell/\sigma$, $\ell/\sigma=20$, the plots look very similar for all boundary conditions as a consequence of both the static and  circular trajectories having the same large $\ell$ (flat space) limit.  When the AdS length is smaller, the effect of the relative redshift between the detectors is striking. In addition to the asymmetry in the static case, we also find that for small AdS length, when $t_0$ is tuned correctly, entanglement harvesting is possible for much larger detector separations as compared to the circular geodesic case.  Additionally, the time dependent oscillations found when $\ell/\sigma = 1$ for static trajectories are no longer present for the circular geodesic trajectories.

%========================================
%========================================
\section{conclusion}
\label{sec-conc}

To investigate and quantify field entanglement in an operational manner, one must make measurements of the field by interacting an appropriate measuring apparatus with the field, and then analyze the measurement outcomes for indications of field entanglement. The entanglement harvesting protocol constructs such a measurement: two detectors interact locally with the field and after their interaction become entangled. As the detectors do not interact directly, and assuming any interaction mediated by the field is negligible, the amount of entanglement present in the final state of the detectors is entanglement that has been extracted from the localized regions in which the detectors interacted with the field. In this way the entanglement harvesting protocol can be used to probe the entanglement structure of the vacuum state of a quantum field theory.

In this article we performed a detailed study of both the transition probability of a single detector and the entanglement harvesting protocol for detectors in AdS$_3$ space interacting with to a real, massless, conformally  coupled scalar field beginning  in the vacuum state. We examined how this transition probability and protocol depend on the detectors' locations and trajectories in AdS space, spacetime curvature (in AdS$_3$ the scalar curvature is $R = - 6/\ell^2$), and the boundary conditions satisfied by the field at spatial infinity. The parameter space is a rather large one to explore, and so we considered identical detectors with equal energy gaps and switching widths in their own rest frame. Our results complement the earlier work done in flat spacetime \cite{Pozas-Kerstjens:2015,Pozas-Kerstjens:2016}, and provide a useful benchmark for further investigations of quantum information and detector physics in other asymptotically AdS spacetimes.

Beginning with a  study of the transition probability $P_D$ of a single detector, we found that it is most sensitive to changes in parameters at small values of the AdS length $\ell/\sigma$.   For Dirichlet boundary conditions ($\zeta =1$), the  transition probability of a detector at any position approaches zero as $\ell/\sigma \to 0$, whereas for  Neumann boundary conditions ($\zeta = -1$),  the transition probability likewise decreases to zero for a detector fixed at the origin, but increases a little with decreasing $\ell/\sigma$ if the detector is located elsewhere. For transparent boundary conditions ($\zeta = 0$),  the transition probability of a detector positioned far away from the origin remains constant as a function of $\ell/\sigma$.

For entanglement harvesting, we found that for static detectors there is an optimal AdS length $\ell/\sigma$ and detector energy gap $\Omega \sigma$ at which the concurrence reaches a maximum value.  One unexpected phenomenon is the appearance of ``separability islands"  for a range of small but finite $\ell/\sigma \approx 2.5 $.  In this region the detector transition probability $P_D$ remains approximately constant as the proper separation of the detectors changes, but the non-local correlation $|X|$ attains a local minimum whose origin remains to be understood.

We also observed a strong effect on the efficacy of entanglement harvesting if the detectors switch on at different times (in other words, the peak of their switching functions is offset by a time $t_0$).  For $t_0 < 0$ entanglement
harvesting is significantly suppressed compared to $t_0 > 0$, an effect due to locally different detector gaps induced by differing location-dependent redshifts.  A similar effect is seen in AdS$_4$ \cite{KRE} and we expect it to be present whenever the proper times of the detectors differ and a time delay is present in the switching functions.

Finally, we considered a scenario in which both detectors move along circular geodesic orbits about the origin. This has the effect of removing any relative redshift between the two detectors and so constitutes an important comparative setting.  We found a number of similarities with entanglement harvesting by static detectors:  there is a maximum in the concurrence as a function of both the detector energy gap and the AdS length in both cases, though the quantitative details differ slightly.

The entanglement harvesting protocol provides an operational way in which to probe the entanglement structure of a quantum field. We hope that our investigations here will inspire further studies of entanglement harvesting in other curved spacetimes to better understand how field entanglement depends on spacetime structure. In addition, we hope that connections between entanglement harvesting and other methods used to study entanglement in field theory, such as algebraic and path integral methods, will be made in the near future.

%========================================
%========================================
\begin{acknowledgments}
We would like to thank Jorma Louko, Hugo Marrochio, Eduardo Martin-Martinez, Keith Ng, and Erickson Tjoa for useful discussions and comments on various aspects of this work. This work was supported in part by the Natural Sciences and Engineering Research Council of Canada, the Ontario provincial government through the Ontario Graduate Scholarship, and the Dartmouth College Society of Fellows. R. A. Hennigar acknowledges the support of the NSERC Banting Postdoctoral Fellowship programme. The author J. Zhang thanks for the support from the National Natural Science Foundation of China under Grants No.11435006 and No.11690034.
\end{acknowledgments}

%========================================
%========================================
\appendix

\onecolumngrid
%========================================
%========================================
\section{Derivation of $P_D$ and $X$}
\label{Derivation of PD and X}

In this appendix we derive the numerical form of the transition probability  $P_D$  and matrix element $X$ defined in Eqs.~\eqref{PJ} and \eqref{defX} for detectors on static and circular geodesic trajectories in AdS$_3$ considered respectively in Secs. \ref{sec3} and \ref{Detectors on circular geodesics}.

%========================================
%========================================
\subsection{Static detectors}

Beginning with the definition of the transition probability $P_D$ in Eq.~\eqref{PJ}, we can express $P_D$ in terms of the integration variables $u \ce \tau_D$ and $s \ce \tau_D-\tau'_D$ and evaluate the integral over $u$
\begin{align} \label{PD-1}
P_D &\ce \lambda^2 \int d\tau_D  d \tau_D' \, \chi_D(\tau_D) \chi_D(\tau_D') e^{-i \Omega \left(\tau_D-\tau_D'\right)} W\!\left(x_D ,x_D'\right) \nn \\
&=2\lambda^2\int_{-\infty}^{\infty}du\chi_D(u)\Re \left[\int_{0}^{\infty}ds\, \chi_D(u-s) e^{-i \Omega s}
 W\!\left(x_D ,x_D'\right) \right ] \nn \\
 &=2\lambda^2\sqrt{\pi}\sigma \gamma_D \Re\int_{0}^{\infty}d(\Delta{t})\, e^{-\gamma_D^2\Delta{t}^2/(4\sigma^2)}
e^{-i \Omega \gamma_D \Delta{t}}W\!\left(x_D ,x_D'\right),
\end{align}
where in the last equality we have expressed the remaining integral in terms of the integration variable $\Delta{t} \ce  t-t'=s/\gamma_D$. Upon substituting the AdS$_3$ Wightman function given in Eq.~\eqref{wightmanf} into Eq.~\eqref{PD-1}, it is seen that the transition probability can be expressed as a difference of two terms
\begin{align}\label{PD-2}
P_D=P_D^- -\zeta P_D^+
\end{align}
where
\begin{align}
P_D^- \ce\lambda^2 \frac{\gamma_D \sigma }{2 \sqrt{2\pi} \ell} \Re \int_{0}^{\infty} d(\Delta{t})\, e^{-\gamma_D^2\Delta{t}^2/(4\sigma^2)}
\frac{e^{-i \Omega \gamma_D \Delta{t}}}{\sqrt{\sigma(x_D,x'_D)}} ,
\end{align}
and
\begin{align}
P_D^+ \ce \lambda^2 \frac{\gamma_D \sigma }{2 \sqrt{2\pi} \ell} \Re \int_{0}^{\infty} d(\Delta{t})\, e^{-\gamma_D^2\Delta{t}^2/(4\sigma^2)}
\frac{e^{-i \Omega \gamma_D \Delta{t}}}{\sqrt{\sigma(x_D,x'_D)+2}}.
\end{align}
Using Eqs.~\eqref{deltasigma} and \eqref{coordinate1}, and the detector's trajectory given in Eq.~\eqref{traj-static}, we may express the denominators appearing in the integrands defining $P_D^\pm$ as
\begin{align}\label{rhopm1}
\sqrt{\sigma(x_D,x'_D)} &= \gamma_D \left[\alpha_D^-+\cos(\Delta{t}/\ell-i\epsilon)\right]^{1/2}, \\
\sqrt{\sigma(x_D,x'_D)+2} &= \gamma_D \left[\alpha_D^++\cos(\Delta{t}/\ell-i\epsilon) \right]^{1/2},
\end{align}
where we have made explicit the $i\epsilon$ dependence indicating the appropriate branch cut \cite{Lifschytz:1994} and defined \mbox{$\alpha^{\pm}_D \ce  \left[- (R_D/\ell)^2 \pm 1\right]/\gamma_D^2$}; note that from the definition of $\gamma_D$ below Eq.~\eqref{traj-static} it is seen that $\alpha_D^- = -1$.

Let us first  express $P_D^-$ in terms of the dimensionless integration variable $y \ce \Delta{t}/\ell$
\begin{align}\label{PDminus}
P_D^- &= \frac{\lambda^2 \sigma}{2\sqrt{2\pi}} \Re \int_{0}^{\infty} d y \frac{ e^{-a_D y^2  } e^{-i \beta_D y} }{\sqrt{ -1 + \cos (y - i \epsilon)}}
\end{align}
where $\beta_D \ce \gamma_D\ell \Omega$ and  ${a_D} \ce\ell^2\gamma_D^2 / 4\sigma^2$.

Before we conitnue, we note that if we only consider the principal value of the square root, then integrals of the form
\begin{equation}
  \int_{-1}^{1}dy\frac{f(y)}{\sqrt{\sin^2(y)}} = \int_{-1}^{1}dy\frac{f(y)}{|\sin(y)|} = \int_{-1}^{0}dy\frac{f(y)}{-\sin(y)} + \int_{0}^{1}dy\frac{f(y)}{\sin(y)}
\end{equation}
are infinite, and the Cauchy principle value of integration cannot be applied to correct this.  However, $P_D$ is calculated using the Wightman function, which is a tempered distribution (i.e. $P_D^-$ must be finite).  In order to correct this, we require that
\begin{equation}
  \int_{-1}^{1}dy\frac{f(y)}{\sqrt{\sin^2(y)}} \ce \int_{-1}^{1}dy\frac{f(y)}{\sin(y)}
  \label{eq:Sqrt}
\end{equation}

Now under this condition, the denominator may be simplified
 \begin{equation}\label{sinfunction}
\sqrt{ -1 + \cos (y - i \epsilon)}=\sqrt{ -2\sin^2\left(\frac{y}{2}- i \epsilon \right)}=i \sqrt{2}\sin \left(\frac{y}{2}- i \epsilon\right),
\end{equation}
 Direct application of Sokhotsky's formula yields the identity
\begin{equation}
\frac{1}{\sin \left(x- i \epsilon \right)} = \PV  \frac{1}{\sin{x}} + i \pi \sum_{n\in{\mathbb{Z}}}(-1)^n\delta(x-n\pi), \label{trigIdenityepsilon}
\end{equation}
which when combined with  Eq.~\eqref{PDminus} allows for the simplification of $P_D^-$ to
\begin{equation}
P_D^-= \frac{\lambda^2\sigma}{4\sqrt{\pi}} \left [ - \PV \int_{0}^{\infty} dy \, \frac{e^{-{a_D} y^2} \sin(\beta_D{y})}{\sin(y/2)}+ \pi\sum_{n \in \mathbb{Z}} (-1)^{n}\cos(2n\pi\beta_D)e^{-4n^2\pi^2{a_D}} \right]. \label{PD-final}
\end{equation}
Turning our attention to $P_D^+$, we note that it may also be rewritten in the form
\begin{align}\label{PDplus}
P_D^+:&= \frac{\lambda^2 \sigma}{2\sqrt{2\pi}}  \Re \int_{0}^{\infty} d y \frac{ e^{-y^2 a_D } e^{-i \beta_D y} }{\sqrt{ \alpha_D^+ + \cos (y - i \epsilon)}}.
\end{align}
Note that  $|\alpha_D^+| \neq 1$ for finite $\ell$, in which case the singularities appearing in the above integrand are integrable and may take $\epsilon \to 0$. Again, by requiring that the Wightman function is a tempered distribution, we use Eq.~\eqref{eq:Sqrt}, which implies
\begin{align}
\label{cosy}
 \sqrt{\cos y + \alpha^{+}_D}=
 \begin{cases}
|\sqrt{\cos y + \alpha^{+}_D}|,  & \quad y\in(0,\pi-\Theta_D^+)\\
{i}|\sqrt{-\cos y - \alpha^{+}_D}|, & \quad y\in(\pi-\Theta_D^+,\pi+\Theta_D^+)\\
-|\sqrt{\cos y + \alpha^{+}_D}|, & \quad  y\in(\pi+\Theta_D^+, 3\pi-\Theta_D^+)\\
-{i}|\sqrt{-\cos y - \alpha^{+}_D}|, & \quad  y\in(3\pi-\Theta_D^+, 3\pi+\Theta_D^+)\\
\quad \vdots &\quad \quad  \vdots
\end{cases}
 \end{align}
where $\Theta_D^+ \ce \arccos\alpha_D^+$.
Using Eq.~\eqref{cosy},$ P_D^+$ can be written in the form
\begin{align}
P_D^+= \frac{\lambda^2 \sigma}{2\sqrt{2\pi}}  \Re \left[  \int_{0}^{\pi+\Theta_D^+} dy \, \frac{e^{-a_D y^2} e^{-i\beta_D{y}}}{\sqrt{\cos y  + \alpha_D^{+}}}+ \sum_{n \in \mathbb{Z}^+} (-1)^{n} \int_{\Theta_D^++(2n-1)\pi}^{\Theta_D^++(2n+1)\pi} dy \, \frac{e^{-a_D y^2} e^{-i\beta_D{y}}}{\sqrt{\cos y + \alpha_D^{+}}} \right]. \label{PD+final}
\end{align}
Combining Eqs.~\eqref{PD-final} and \eqref{PD+final} yields the transition probability stated in Eq.~\eqref{PAPB}.

We now evaluate the matrix element $X$ defined in Eq.~\eqref{defX}. Taking the switching function to be the Gaussian functions given in Eq.~\eqref{eq:TimeDelay}, $X$ may be simplified to
\begin{align}
X &= - \lambda^2  \gamma_A \gamma_B \int_{-\infty}^{\infty}  dt
\int_{-\infty}^{t}  dt' \, \bigg[
e^{-(t-t_0/2)^2\gamma_B^2/2\sigma^2} e^{-(t'+t_0/2)^2\gamma_A^2/2\sigma ^2} e^{-i \Omega( \gamma_B  t+  \gamma_A t')}  W\!\left(x_A( t' ), x_B(t )\right) \nonumber \\
 &\quad \hspace{2 in} +e^{-(t+t_0/2)^2\gamma_A^2/2\sigma^2} e^{-(t'-t_0/2)^2\gamma_B^2/2\sigma^2}e^{-i\Omega(\gamma_A t+  \gamma_B  t' )} W\!\left(x_B( t'),x_A( t) \right)
 \bigg]  \nn \\
&=- \lambda^2 2 \sqrt{2\pi}\sigma \frac{ \gamma_A
 \gamma_B}{\sqrt{\gamma_A^2+\gamma_B^2}}   e^{- \frac{\sigma^2\Omega^2}{2} \frac{\left(\gamma_A+\gamma_B \right)^2}{\gamma_A^2+\gamma_B^2 }-\frac{t_0^2}{2\sigma^2}\frac{\gamma_A^2\gamma_B^2}{\gamma_A^2+\gamma_B^2}+i\frac{\Omega t_0}{2}\frac{(\gamma_A+\gamma_B)^2(\gamma_A-\gamma_B)}{\gamma_A^2+\gamma_B^2}} \nn\\
 &\quad \times\int_{0}^{\infty} ds \,    \cosh \left[\left(i\frac{ \gamma_A\gamma_B (\gamma_A
-\gamma_B)}{\gamma_A^2+\gamma_B^2}  \Omega - \frac{t_0}{\sigma^2}\frac{\gamma_A^2\gamma_B^2}{\gamma_A^2+\gamma_B^2}\right) s\right]
e^{-\frac{s^2}{2 \sigma^2} \frac{ \gamma_A^2\gamma_B^2}{\gamma_A^2+\gamma_B^2}}W\!\left(x_A(
t' ), x_B(t )\right) ,\label{integrateX1}
 \end{align}
where in arriving at the last equality we have introduced the integration variables $u \ce  t$ and $s \ce t-t'$ and carried out the integration over $u$.

Upon substituting the spacetime trajectories of the static detectors given in Eq.~\eqref{traj-static} into the Wightman function in Eq.~\eqref{wightmanf}, the denominators become
\begin{align}\label{rhopm2}
\sqrt{\sigma(x_A,x'_B)} &=\sqrt{\gamma_A\gamma_B}\Big[\alpha_X^-+\cos(s/\ell-i\epsilon)\Big]^{1/2} ,\nonumber\\
\sqrt{\sigma(x_A,x'_B)+2} &=\sqrt{\gamma_A\gamma_B}\Big[\alpha_X^++\cos(s/\ell-i\epsilon)\Big]^{1/2},
\end{align}
where $ \alpha^{\pm}_X :=   \left[-R_AR_B/\ell^2 \pm 1\right]/\gamma_A \gamma_B$. Using Eqs.~\eqref{rhopm2} and \eqref{wightmanf}, $X$ simplifies to
\begin{align}\label{integrateX2}
X = - \frac{\lambda^2  \sigma}{2 \sqrt{\pi}}K_X \int_0^\infty dy \, e^{-a_X y^2}\cosh \big((\Delta_T+i\beta_X) y]\big) \left[ \frac{1}{\sqrt{\alpha^-_X + \cos (y - i \epsilon) }} - \zeta \frac{1}{\sqrt{\alpha^+_X + \cos (y - i \epsilon) }}\right],
\end{align}
where we have introduced the integration variable $y \ce s/\ell$ and defined %$K_X$, $a_X$, $\Delta_T$ and $\beta_X$ are defined in Eq.~\eqref{Kxeq}.
\begin{align}
K_{X} &\ce \sqrt{\frac{\gamma_A\gamma_B}{\gamma_A^2+\gamma_B^2}}\exp\left(-\frac{\Omega^2\sigma^2}{2}\frac{(\gamma_A+\gamma_B)^2}{\gamma_A^2+\gamma_B^2}-\frac{t_0^2}{2\sigma^2}\frac{\gamma_A^2\gamma_B^2}{\gamma_A^2+\gamma_B^2}+i\frac{\Omega t_0}{2}\frac{(\gamma_A+\gamma_B)^2(\gamma_A-\gamma_B)}{\gamma_A^2+\gamma_B^2}\right) \nn\\
{a_X} &\ce \frac{\gamma_A^2 \gamma_B^2}{\gamma_A^2 + \gamma_B^2} \frac{\ell^2}{2 \sigma^2 } \\
\Delta_T &\ce -\frac{t_0\ell}{\sigma^2}\frac{\gamma_A^2\gamma_B^2}{\gamma_A^2+\gamma_B^2} \nn\\
\beta_X &\ce \frac{\gamma_A \gamma_B \left( \gamma_A -\gamma_B\right)}{\gamma_A^2 + \gamma_B^2} \ell \Omega. \nn
\end{align}

Through the methods similar to those used to obtain $P_D^+$, the matrix element $X$ can be brought into the form given in Eq.~\eqref{X-equation}.

%========================================
%========================================
\subsection{Detectors on circular geodesics}

As discussed in Sec.~\ref{Detectors on circular geodesics}, we evaluate the transition probability of detector $B$ orbiting around the origin on a circular geodesic. Similar to Eq.~\eqref{PD-1}, the transition probability can be expressed as
\begin{align} \label{PD-21}
\tilde{P}_B &= 2\lambda^2\sqrt{\pi}\sigma_B \Re\int_{0}^{\infty}d(\Delta{t})\, e^{-\Delta{t}^2/(4\sigma^2)}
e^{-i \Omega \Delta{t}}W\!\left(x_B ,x_B'\right) \nn \\
&=\frac{\lambda^2 \sigma}{2 \sqrt{2 \pi}} \Re \left[ \int_{0}^{\infty} d y \frac{ e^{-\tilde{a} y^2 } e^{-i  \ell \Omega y} }{\sqrt{ -1 + \cos (y - i \epsilon)}} - \zeta\int_{0}^{\infty} d y \frac{ e^{-\tilde{a} y^2    } e^{-i \ell \Omega y} }{\sqrt{ 1 + \cos (y - i \epsilon)}} \right],
\end{align}
where in the last equality we have used the integration variable $y \ce \Delta t/\ell$ and defined $\tilde{a}=\ell^2/4\sigma^2$. Again, the appropriate branch of the square root function appearing in Eq.~\eqref{PD-21} is dictated by the requirement that the Wightman function is a tempered distribution, and so just as in Eq.~\eqref{sinfunction} we have
\begin{subequations}
\begin{align}
\sqrt{ -1 + \cos (y - i \epsilon)} &=i \sqrt{2}\sin\big(y/2- i \epsilon\big), \\
 \sqrt{ 1 + \cos (y - i \epsilon)} &= \sqrt{2}\cos\big(y/2- i \epsilon\big).
\end{align}
\end{subequations}
Finally, using Eq.~\eqref{trigIdenityepsilon} the transition probability of a detector on a circular geodesic can be brought to the form stated in Eq.~\eqref{transitionCircular}.

We now compute the matrix element $X$ defined in Eq.~\eqref{defX} for the case where detector $A$ and detector $B$ orbit around the origin  on a circular geodesic given by Eq.~\eqref{traj}. Through the same manipulations leading Eqs.~(\ref{integrateX1}) and \eqref{integrateX2}, $\tilde{X}$ may be written as
\begin{align}
\tilde{X} &=- 2\lambda^2{\sqrt{\pi}\sigma }  e^{-\sigma^2\Omega^2-\frac{t_0^2}{4\sigma^2}} \int_{0}^{\infty} ds \, e^{-s^2/(4\sigma^2)}\cosh\left(\frac{t_0}{2\sigma^2}s\right)W\!\left(x_A( t' ), x_B(t )\right) \label{integrateX21} \nn \\
&= - \frac{\lambda^2 \sigma}{2 \sqrt{2\pi}} \tilde{K}_X\int_0^\infty dy \, e^{- \tilde{a} y^2}\cosh(\tilde\Delta_T y) \left[ \frac{1}{\sqrt{ \cos (y - i \epsilon) - \tilde\alpha_X}} -  \frac{\zeta}{\sqrt{ \cos (y - i \epsilon) + \tilde\alpha_X}}\right]
 \end{align}
where in the last equality we have used the integration variable $y \ce s/ \ell$ and defined
\begin{align}
%\tilde{K}_X &\ce \frac{\ell}{\sqrt{\sqrt{R_A^2+\ell^2}\sqrt{R_B^2+\ell^2}-R_AR_B}}e^{\sigma^2\Omega^2} \nn\\
\tilde{K}_X &\ce \sqrt{\tilde\alpha_X}\ \exp\left(-\sigma^2\Omega^2-\frac{t_0^2}{4\sigma^2}\right) \nn\\
\tilde\Delta_T &\ce \frac{\ell t_0}{2\sigma^2} \label{eq:tildealphaX}\\
\tilde\alpha_X &\ce \frac{\ell^2}{\sqrt{R_A^2+\ell^2}\sqrt{R_B^2+\ell^2}-R_AR_B}. \nn
\end{align}
Employing the same methods to treat square root function used in arriving at Eq.~\eqref{PDplus}, $\tilde{X}$ can be brought to the form given in Eq.~\eqref{Xcircular}.

%========================================
%========================================
\section{Entanglement Harvesting in flat spacetime}
\label{EH in flat spacetime}

To facilitate comparison between the entanglement harvesting protocol for static detectors in AdS$_3$ with the flat space limit $\ell \to \infty$, we evaluate both the transition probability of a single detector and the matrix element $X$ appearing in the joint state of two detectors interacting with a real massless scalar field in (2+1)-dimensional Minkowski space.  The Wightman function associated with such a field is given by
\begin{equation}
W_{\rm flat}(x,x') = \frac{1}{4\pi}\frac{1}{\sqrt{\Delta x^2-\Delta t^2 +i \sgn \left(\Delta t\right)\epsilon}}.
\label{eq:flatWF}
\end{equation}

For static detectors and with Gaussian switching functions, as considered in Sec.~\ref{sec3}, we can evaluate the transition probability of a single detector and matrix element $X$ directly from their definition in Eqs. \eqref{PJ} and \eqref{defX}, with the result
\begin{align}
P_D &= \frac{\lambda^2\sigma\sqrt{\pi}}{4} \big(1-\text{erf}(\sigma\Omega)\big) \label{eq:flatPD},\\
X &= - \frac{\sigma \lambda^2}{4 \sqrt{\pi}} \exp\left[-\frac{d^2}{8 \sigma^2} - \sigma^2 \Omega^2 \right] \left(\pi I_0\left(d^2/(8\sigma^2)\right) - i K_0\left(d^2/(8\sigma^2)\right) \right) \label{eq:flatX}
\end{align}
where $d \ce |R_B - R_A|$ and  $K_0$ and $I_0$ are zeroth order modified Bessel functions of the first and second kind, respectively. These expressions were in producing the flat space limits depicted in  Figs~\ref{Negativity-ell}, \ref{Negativity-omega}, \ref{Cvsell-Om}, and \ref{CvsOm-ell}.
\twocolumngrid

\bibliography{AdSpaper}
\end{document}